\documentclass[journal]{IEEEtran}
\usepackage{cite}
\UseRawInputEncoding
\usepackage{booktabs}
\usepackage{amssymb}
\usepackage{graphicx}
\usepackage{subfigure}
\usepackage{epstopdf}
\usepackage{multicol}
\usepackage{textcomp}
\usepackage{multirow}
\usepackage{algorithm}
\usepackage{algorithmic}
\usepackage{amsthm,amsmath,amsfonts}
\usepackage{mathrsfs}
\usepackage{courier}

\usepackage{bm}
\usepackage{subfigure}

\usepackage{color}
\usepackage[hyphens]{url}
\usepackage[backref]{}
\def\BibTeX{{\rm B\kern-.05em{\sc i\kern-.025em b}\kern-.08em
		T\kern-.1667em\lower.7ex\hbox{E}\kern-.125emX}}

\begin{document}
	\title{Survey on Near-Space Information Networks: Channel Modeling, Transmission, and Networking Perspectives}
 %\title{Near-Space Information Networks}
	%
	%
	% authors names and IEEE memberships
	\author{
 Xianbin Cao,~\IEEEmembership{Senior Member,~IEEE}, Xiaoning Su, Peng Yang,~\IEEEmembership{Member,~IEEE}, \\ Yue Gao,~\IEEEmembership{Fellow,~IEEE}, Dapeng Oliver Wu,~\IEEEmembership{Member,~IEEE}, and Tony Q. S. Quek,~\IEEEmembership{Member,~IEEE}
 
 %Tony Q. S. Quek,~\IEEEmembership{Fellow,~IEEE}
\IEEEcompsocitemizethanks{\IEEEcompsocthanksitem X. Cao and P. Yang, and X. Su are with the School of Electronic and Information Engineering, Beihang University, Beijing 100191, China.
\IEEEcompsocthanksitem X. Cao and P. Yang are also with the Department of Mathematics and Theories, Peng Cheng Laboratory, Shenzhen, Guangdong 518055, China.
\IEEEcompsocthanksitem Y. Gao is with the School of Computer Science, Fudan University, Shanghai 200433, China.
\IEEEcompsocthanksitem D. O. Wu is with the Department of Computer Science, City University of Hong Kong, Kowloon, Hong Kong.
\IEEEcompsocthanksitem T. Q. S. Quek is with the Information Systems Technology and Design Pillar, Singapore University of Technology and Design, Singapore 487372, Singapore.
}
}

	\maketitle
	
	\begin{abstract}
 Near-space information networks (NSINs) composed of high-altitude platforms (HAPs) and high- and low-altitude unmanned aerial vehicles (UAVs) are a new regime for providing {quick, robust, and cost-efficient} sensing and communication services. Precipitated by innovations and breakthroughs in manufacturing, materials, communications, electronics, and control techniques, NSINs have been envisioned as an essential component of the emerging sixth-generation of mobile communication systems. This article {reveals some critical issues needing to be tackled in NSINs through conducting experiments} and discusses the latest advances in NSINs in the research areas of channel modeling, networking, and transmission from a forward-looking, comparative, and technical evolutionary perspective. In this article, we highlight the characteristics of NSINs and present the promising use cases of NSINs. The impact of airborne platforms' unstable movements on the phase delays of onboard antenna arrays with diverse structures is mathematically analyzed. The recent advances in HAP channel modeling are elaborated on, along with the significant differences between HAP and UAV channel modeling. A comprehensive review of the networking techniques of NSINs in network deployment, handoff management, and network management aspects is provided. Besides, the promising techniques and communication protocols of the physical (PHY) layer, medium access control (MAC) layer, network layer, and transport layer of NSINs for achieving efficient transmission over NSINs are reviewed, {and we have conducted experiments with practical NSINs to verify the performance of some techniques}. Finally, we outline some open issues and promising directions for NSINs deserved for future study and discuss the corresponding challenges. 
	\end{abstract}
	%(i.e., low altitude unmanned aerial vehicles (UAVs) equipped with base stations)
	%  so as to maximize the reward
	%, the use of which as radio access platforms in 5G and beyond-5G wireless communication networks has received substantial attentions in research community.
	
	
	% \fi
	\IEEEpeerreviewmaketitle

	\section{Introduction}
\IEEEPARstart{R}{ecent} advances in manufacturing, materials, communications, electronics, and control techniques have witnessed an unprecedented increase in the application of various types of non-terrestrial platforms in the military and civil fields. 
	In the wireless communication research field, non-terrestrial networks ({NTNs}) have been included as a part of the third generation partnership project (3GPP) Rel-17 specifications.
 {3GPP Rel-18 aims to standardize the integration of NSINs into the terrestrial networks (TNs). By ensuring interoperability and efficient integration, 3GPP focuses on unlocking the full potential of NTN components in enhancing connectivity and extending the capabilities of the fifth-generation of mobile communication systems (5G) ecosystem \cite{NR20223GPP}.} 
 %and discussions of {NTNs} are scheduled in the 3GPP Rel-18 work plan to support the fifth-generation of mobile communication systems (5G) advanced use cases \cite{NR20223GPP}. 
 Additionally, the {integrated} networks of {NTNs} and {TNs} have been globally considered a promising proposal for the sixth-generation of mobile communication systems (6G) \cite{DBLP:journals/chinaf/YouWHGZWHZJWZSW21}. 
 {3GPP Rel-19 further discussed the topics of NTN evolution for new radio (NR) and internet of things (IoT).}
%	Key benefits of HAP: 
%	Area of Interest Persistence
%	Rapid Deployment
%	Predictive Flight Modeling
%	vStorm Operating System
%	Payload Recovery
%	Wide Area Coverage
%	Low-Cost Stratospheric Access
%	Navigational Capabilities
%	Solar Regeneration
%	Payload Integration
%	Diverse Payload Compatibility
%
%	MISSION POSSIBILITIES:
%	Augment Existing Network
%	Research and Development
%	Tailored Missions
%	Reconstitute Lost Aerial Assets
%	Communications
%	Satellite Alternative
%	Position-Navigation-Time (PNT)
%	Intelligence, Surveillance, and Reconnaissance (ISR)
%
%	Military Communications
%	Military surveillance
%	Commercial Communications
%	Maritime Surveillance
%	Civil and Border Security
%	Earth Observation
%	Environmental Monitoring
%	Agricultural Monitoring
%		
	As a crucial component of NTNs, near-space information networks (NSINs) are a new regime of situational awareness, processing, and serving networks composed of high-altitude platforms (HAPs, including aircraft, airships, and balloons), high-altitude unmanned aerial vehicles (HAUAVs), and low-altitude unmanned aerial vehicles (LAUAVs). 
	Compared with space networks, NSINs have some unique advantages in terms of deployment flexibility and agility, deployment and operation cost, transmission latency, and responsiveness. HAPs in NSINs, deployed at an altitude ranging from 17 km to 25 km, 
 have the ability to extend the communication coverage quickly and provide wide-area communication services. The unmanned aerial vehicles (UAVs) in NSINs, deployed at a relatively low altitude ranging from several meters to thousands of meters, can achieve service enhancements in critical areas. 
	Further, NSINs overwhelm TNs in terms of quick response to flash crowd traffic, providing coverage in unserved or underserved areas cost-effectively, and being resilient to natural and man-made disasters. 
	Summarily, NSINs have some outstanding advantages in achieving wide-area and three-dimensional (3-D) coverage as well as local on-demand service enhancement. 
	
	As a result, NSINs have become an essential segment of the emerging space-air-ground integrated networks (SAGINs) and have wide application prospects in such fields as regional surveillance, emergency communications, intelligent transportation systems (ITS), and computational offloading. 
	Although NSINs have many attractive advantages and application prospects, there are many key issues, such as channel modeling, networking, and transmission, waiting to be solved. After solving these key issues, NSINs are able to seamlessly integrate with space networks and TNs and really contribute to the emerging 6G technological standards. 
	
	\subsection{Existing Surveys and Reviews} 
	During the past few years, the research on {SAGINs as well as the particular} NSINs has attracted extensive attention from both academia and industry. In this subsection, we overview the surveys and reviews on three fundamental issues, i.e., channel modeling, networking, and transmission of SAGINs and NSINs in the recent five years. Table \ref{table_survey} summarizes and compares the related surveys and reviews. 
 %And the reader is recommended to refer to these articles and the references therein for more detailed information on NSINs. 
%Next, this article discusses the existing surveys on the channel modeling, networking, and transmission of SAGINs and NSINs. 
% 	\begin{table*}[!t]
% 	\renewcommand{\arraystretch}{1.2}
% 	\caption{An overview of surveys \& reviews on NSINs in recent ten years}
% 	\label{table_survey}
% 	\newcommand{\tabincell}[2]{\begin{tabular}{@{}#1@{}}#2\end{tabular}}
% 	\centering
% 	\begin{tabular}{l}
% %\begin{minipage}{0.18\textwidth}
% 			\centering
% 			\includegraphics[width=7.0 in]{Table_survey.pdf}
% 		%\end{minipage} 
% 	\end{tabular}
% \end{table*}
\begin{table*}[!t]
		\newcommand{\tabincell}[2]{\begin{tabular}{@{}#1@{}}#2\end{tabular}}
		\renewcommand{\arraystretch}{1.3}
		% if using array.sty, it might be a good idea to tweak the value of
		%	 \extrarowheight as needed to properly center the text within the cells
		\caption{{Survey structure and content comparison with the related surveys \& reviews in recent five years (partially covered: $\partial $, covered: \checkmark) }}
		\label{table_survey}
		\centering
		% Some packages, such as MDW tools, offer better commands for making tables
		% than the plain LaTeX2e tabular which is used here.
  %\tabincell{l}{circuits \\ with }
		\begin{tabular}{|c|c|c|c|c|c|c|c|c|c|c|}
			\hline
			{Refs.} &  {Netw.  category} &  Year & Satellite & HAP & UAV & \tabincell{c}{HAP \& UAV \\ comparison} & \tabincell{c}{Channel \\ modeling} & {Networking} & {Transmission} & \tabincell{c}{Experiments/ \\ Simulations} \\ \hline
			{\cite{ Cheng2022ChannelNA }} &   {}  & 2022  & \checkmark  & No &  \checkmark &  No &  \checkmark & No & No & Simulations \\ \cline{1-1} \cline{3-11}
			\cite{ Niu2020SpaceairgroundIV } &   {}  & {2020}  & {\checkmark} & {\checkmark} & \checkmark & $\partial$ & No & \checkmark & No & Simulations \\	\cline{1-1} \cline{3-11}
		\cite{ Saeed2020PointtoPointCI } &  {} &  {2021}  & {\checkmark}  & {\checkmark} & {\checkmark} & No & $\partial$ & $\partial$ & No & Simulations \\ \cline{1-1} \cline{3-11}
		\cite{ DBLP:journals/tits/ShengCLWAWKY22 } &   {}  & 2022  & \checkmark & {\checkmark} & No & No & No & \checkmark & No & No \\ \cline{1-1} \cline{3-11}
		\cite{ Cheng20216GSS }  &  {} & 2022  & \checkmark  & \checkmark & \checkmark & No & No & \checkmark & No & No \\ \cline{1-1} \cline{3-11}
		\cite{Qiu2022MobileEC} &  {}  & {2022}  & \checkmark  & \checkmark & \checkmark & No & No & \checkmark & No & No \\ \cline{1-1} \cline{3-11}
   \cite{Shen2023ASO}  &  {}  & {2023}  & \checkmark  & \checkmark & \checkmark & No & No & \checkmark & No & No \\ \cline{1-1} \cline{3-11}
   \cite{ Arani2022UAVAssistedSI}  &  {}  & {2022}  & \checkmark & {No} & \checkmark & No & No & \checkmark & No & Simulations \\	\cline{1-1} \cline{3-11}
   \cite{Ye2021NonterrestrialCA}  &  SAGINs  & 2022  & \checkmark & \checkmark & \checkmark & $\partial$ & No & No & \checkmark & Simulations \\ \cline{1-1} \cline{3-11}
   \cite{ DBLP:journals/comsur/BaltaciDOACS21 }   &  {} & 2021 &  \checkmark  & \checkmark  & \checkmark &  $\partial$ &  No & No & \checkmark & No\\ \cline{1-1} \cline{3-11}
   \cite{Dai2022ASO} & {} & 2022 &  \checkmark  & No  & \checkmark &  No &  No & $\partial$ & $\partial$ & No \\ \cline{1-1} \cline{3-11}
   \cite{Zhang2022SpectrumSI} & {} & 2022 &  \checkmark  & \checkmark  & \checkmark &  No &  No & $\partial$ & $\partial$ & No \\ \cline{1-1} \cline{3-11}
   \cite{Ray2021ARO} & {} & 2022 &  \checkmark  & \checkmark  & \checkmark &  No &  No & $\partial$ & \checkmark & No \\ \cline{1-1} \cline{3-11}
   \cite{Azari2021EvolutionON} & {} & 2022 &  \checkmark  & \checkmark  & \checkmark &  $\partial$ &  $\partial$ & $\partial$ & $\partial$ & Simulations \\ \cline{1-1} \cline{3-11}
   \cite{Zhou2023AerospaceIN} & {} & 2023 &  \checkmark  & \checkmark  & \checkmark &  $\partial$ &  $\partial$ & $\partial$ & $\partial$ & No \\ \cline{1-1} \cline{3-11}
   \cite{ Bakambekova2024OnTI } & {} & 2024 &  \checkmark  & \checkmark  & \checkmark &  $\partial$ &  $\partial$ & $\partial$ & $\partial$ & No \\ \cline{1-1} \cline{3-11}
   \cite{ Mahboob2023RevolutionizingFC} & {} & 2024 &  \checkmark  & \checkmark  & No &  No &  $\partial$ & $\partial$ & $\partial$ & No \\ \hline
   \cite{DBLP:journals/comcom/ArumGM20} & \multirow{4}{*}{{NSINs}}  & 2020 &  No  & \checkmark  & No &  No &  $\partial$ & $\partial$ & $\partial$ & No \\ \cline{1-1} \cline{3-11}
   \cite{DBLP:journals/comsur/KurtKAIDAYY21} & {}  & 2021 &  No  & \checkmark  & No &  No &  $\partial$ & $\partial$ & $\partial$ & Simulations \\ \cline{1-1} \cline{3-11}
  % \cline{1-1} \cline{3-11}
   %\cite{Belmekki2024CellularNF} & {}  & 2024 &  No  & \checkmark  & No &  No &  No & No & No & Experiments \\ \hline
   \tabincell{c}{This \\ article} & {}  & 2024 &  No  & \checkmark  & \checkmark &  \checkmark &  \checkmark & \checkmark & \checkmark & \tabincell{c}{Experiments} \\ \hline
		\end{tabular}
	\end{table*}

{Channel models that can fully emulate the underlying characteristics and features of NTNs are indispensable for the successful design of NTNs. Nevertheless, wide-area and dynamic communication scenarios result in two key concerns for NTNs’ channels, i.e., channel non-stationary and channel consistency \cite{ Cheng2022ChannelNA }. The survey \cite{ Cheng2022ChannelNA } reviewed the recent advances in the topic of capturing channel non-stationarity and channel consistency for SAGINs.} 

{The networking of NTNs is a hot research topic. Plenty of surveys \cite{Niu2020SpaceairgroundIV,Saeed2020PointtoPointCI,DBLP:journals/tits/ShengCLWAWKY22,Cheng20216GSS,Qiu2022MobileEC,Shen2023ASO,Arani2022UAVAssistedSI,Ye2021NonterrestrialCA,DBLP:journals/comsur/BaltaciDOACS21,Dai2022ASO,Zhang2022SpectrumSI,Ray2021ARO,Azari2021EvolutionON,Zhou2023AerospaceIN,Bakambekova2024OnTI,Mahboob2023RevolutionizingFC,DBLP:journals/comcom/ArumGM20,DBLP:journals/comsur/KurtKAIDAYY21} elaborated on the network architecture design, the establishment of airborne communication links, and the network management of SAGINs.  
For example, the authors in \cite{ Niu2020SpaceairgroundIV } proposed a novel space-air-ground integrated vehicular network (SAGiven) architecture to gracefully integrate information and network resources from heterogeneous networks. Then, the up-to-date solutions to tackle the challenges of this new architecture were reviewed. 
The authors in \cite{ Saeed2020PointtoPointCI } surveyed various communication techniques for establishing communication links in the same network segment and across different network segments of SAGINs. 
The authors in \cite{ DBLP:journals/tits/ShengCLWAWKY22 } discussed the core scientific problems and reviewed the existing key network management techniques for achieving SAGINs-assisted high-speed railway communications. 
The authors in \cite{ Cheng20216GSS } reviewed the network slicing, software-defined networking (SDN), and network function virtualization (NFV) techniques for implementing the interoperability of different network segments of SAGINs and cooperatively orchestrating heterogeneous resources across different domains according to mission requirements. 
The authors in \cite{Qiu2022MobileEC} surveyed the key approaches (e.g., reinforcement learning (RL), mathematical programming, and game theory) of solving the resource management issue for SGAINs, including the deployment of airborne computing platforms, traffic offloading, and distribution of airborne caching platforms. 
The computing resource management approaches that supported the utilization of SAGINs for infrastructure-less environments were also comprehensively surveyed in \cite{Shen2023ASO}. 
Besides, the authors in \cite{ Arani2022UAVAssistedSI} reviewed the recent learning-based algorithmic approaches to tackle the platform deployment and resource allocation problems in UAV-assisted SGAINs.} 

{In addition to surveying the networking issue for SAGINs, the transmission issue of SAGINs was also discussed. For example, the authors in \cite{Ye2021NonterrestrialCA} gave a comprehensive literature review on reconfigurable intelligent surfaces (RIS)-assisted SAGINs from the perspectives of performance analysis and optimization, followed by the widely used methodologies. 
The authors in \cite{ DBLP:journals/comsur/BaltaciDOACS21 } provided comprehensive wireless communication approaches for enabling connectivity to aerial vehicles, including satellites, HAPs, and UAVs, and reviewed recent findings from the literature toward the possibilities and challenges of employing wireless communication standards.} 

{Several recent surveys also simultaneously discussed the networking and transmission issues for SAGINs. 
For example, the authors in \cite{Dai2022ASO} surveyed the collaboration methods of multiple segments of SAGINs and the recent progress on the topic of RIS-enabled SAGINs. 
The authors in \cite{Zhang2022SpectrumSI} summarized the existing spectrum utilization rules, spectrum sharing modes, and beamforming techniques in aerial/space networks and TNs. 
The authors in \cite{Ray2021ARO} reviewed key transmission techniques related to UAV and satellite-based communications and discussed key enablers of SAGINs from network management and security perspectives. The role of UAVs in augmenting the comprehension of 6G-enabled SAGIN was also elaborated on. 
The authors in \cite{Azari2021EvolutionON} mainly surveyed the resource allocation and beamforming approaches for SAGINs operation in mmWave and elaborated on candidate 6G techniques and advances in media access control (MAC) and network layers of SAGINs. 
Besides, the authors in \cite{Zhou2023AerospaceIN} discussed the potentially promising methodologies and models in SAGINs from the perspectives of system architecture, networking design, and enabling techniques (e.g., multiple-input multiple-output (MIMO) and artificial intelligence (AI)), to enable flexible and scalable management and control of SAGINs and ensure network stability. 
The authors in \cite{ Bakambekova2024OnTI } comprehensively surveyed the recent advances in the topic of utilizing AI techniques to solve networking and transmission issues in SGAINs, including orchestration and topology management, scheduling and collaborative resource management, routing, and flexible mobility management. 
The authors in \cite{ Mahboob2023RevolutionizingFC} surveyed the recent progress on utilizing AI techniques to address many critical issues of SAGINs, including channel estimation, resource allocation, handover optimization, and multiple access.} 

Considering that a HAP is an integral component in the realization of the vision of SAGINs, a number of recent surveys discussed the advancement for the HAP networks of the future. For example, a review of using HAPs for wireless communications in rural communities was conducted in \cite{DBLP:journals/comcom/ArumGM20}. The authors reviewed HAP resource and interference management techniques, where the potential of utilizing radio environment maps and AI for HAP interference and resource management was discussed. Some coverage extension methods were also summarized in this review. 
The work \cite{DBLP:journals/comsur/KurtKAIDAYY21} presented a comprehensive literature review of channel modeling, radio resource management, interference management, waveform design, handoff management, and network management in HAP networks of the future. Besides, a HAP communication system and two HAP onboard subsystems (exactly, energy management and communication payload subsystems) were elaborated on in this work.
 \subsection{Motivation and Contributions}
	{Different from the above research works, this survey will explicitly distinguish HAPs and UAVs, highlight the design considerations of NSINs, and comprehensively review the research on channel modeling, networking, and transmission {techniques/methods} of NSINs. The comparison of this article with the existing surveys and reviews on SAGINs and NSINs in the past five years is presented in Table \ref{table_survey}.} 

{It can be observed from Table \ref{table_survey} that most research works recognize the importance of distinguishing the role of HAPs and UAVs in SAGINs. Yet, a holistic distinction was missing in these works. 
Many studies \cite{ Cheng2022ChannelNA, Saeed2020PointtoPointCI,DBLP:journals/tits/ShengCLWAWKY22, Cheng20216GSS,Qiu2022MobileEC,Shen2023ASO,Arani2022UAVAssistedSI,Dai2022ASO,Zhang2022SpectrumSI,Ray2021ARO,Mahboob2023RevolutionizingFC} did not distinguish the HAP networks and UAV networks in SAGINs and considered HAP networks and UAV networks as a single air segment. 
Some research works even ignore the integral role of HAPs in SAGINs, considering only that SAGINs consist of space, UAV, and terrestrial networks. 
Actually, HAPs and UAVs are quite different in terms of many aspects, such as footprints, endurance, and service capacity. It is crucial to distinguish between a HAP and a UAV, and the integration of HAPs and UAVs will greatly enhance the network resource utilization and performance of NSINs. For example, a HAP and a UAV have different footprints. Deploying a HAP to accomplish what a UAV can do indicates a great waste of network resources. And their integration will simultaneously enable wide-area coverage and local-area enhancement coverage, which will achieve the efficient utilization of heterogeneous network resources. {Therefore, this article will provide a holistic discussion of the differences between a HAP and a UAV, which is unique.}
}

{Meanwhile, as presented in \cite{ Zhou2023AerospaceIN,Bakambekova2024OnTI}, the integration of multi-tier networks can increase the overall network robustness and will be a significant paradigm shift for the evolution of aerospace networks. Therefore, the investigation into the integration of HAP and UAV networks is imperative. Yet, the discussion of their integration is missing.} 

{Last but not least, most of the above research works focused on looking forward to the vision of SAGINs or NSINs. They did not conduct experiments to validate the effectiveness of the discussed approaches and techniques in practical applications. 
Although the authors in \cite{Belmekki2024CellularNF} conducted an experiment to verify the potential of HAPs, it only provided an overview of the HAP itself and did not survey the key techniques required to be tackled behind the HAP networks. {This article will support the discussion of some approaches and techniques with experimental evidence, which is unique.}
}

	%\subsection{Contributions and Organization}
The goal of this article is to present a comprehensive overview of the technical research on NSINs in the literature in the past ten years.
Although NSINs achieve a series of technical innovations and major breakthroughs in the research fields of seamless communications, wide-area sensing, ubiquitous computing, and flexible control, this article mainly surveys and discusses the latest advances in NSINs from the perspectives of channel modeling, networking, and transmission.
The main contributions are as follows:
{\begin{itemize}
\item \textbf{Distinguishing HAPs and UAVs:} We characterize the NSINs and holistically distinguish the HAPs and UAVs from diverse perspectives of mobility model, energy sources and endurance, payload and power capability, {as well as deployment altitude and} footprint. 
%{We characterize} the NSINs and introduce the NSINs as a promising, responsive, and cost-effective solution for wide-area coverage services and local-area service enhancement. We highlight the differences between HAP and UAV from the diverse perspectives of mobility model, energy sources \& endurance, payload \& power capability, {as well as deployment altitude \&} footprint. 
%We introduce the design consideration of NSINs and discuss the impact of heterogeneity of airborne platforms in the protocol design and resource allocation of NSINs. The cooperation among heterogeneous airborne platforms is also advocated. 
%We elaborate on the size (or space), weight, mobility, and power (SWMAP) constraints of airborne platforms in NSINs. The concept of the situation (information) of NSINs is introduced, and the significance of the network and security situation in the design of NSINs is discussed. 
\item \textbf{HAP and UAV integration:} We introduce the integration design consideration of NSINs and discuss the impact of the characteristics (e.g., the heterogeneity) of airborne platforms on the design of many key techniques for NSINs. 
%Besides, based on the overview of the state-of-the-art methods in channel modeling, networking, and transmission perspectives, we summarize the lessons learned about their application to NSINs as well as their benefits and limitations. 
\item \textbf{Holistic study:} We comprehensively survey the recent studies on NSINs from the holistic perspectives of channel modeling, networking, and transmission techniques. Detailed and systematic insights are provided. 
\item \textbf{Experiments:} We summarize some key issues encountered when experimenting with actual NSINs and discuss some techniques in the context of experimental evidence. 
\end{itemize}}

\subsection{{Literature Review Protocol and Organization}}
{This article conducted a comprehensive overview of the articles published in many digital libraries, including IEEE Xplore, ACM, ScienceDirect, SpringerLink, arXiv, and Web of Science. To search for relevant articles from the vast amount of articles in the above digital libraries, we take some literature review initiatives. Select the keywords, e.g., near-space information network, near-space communication, high-altitude platform, airship, aerial heterogeneous network, and space-air-ground integrated network. Choose the articles published in the last ten years. Filter the articles according to their types (e.g., review articles and research articles), languages (exactly, English), and subject areas (e.g., computer science and engineering). The publication title is also an important consideration. The articles published in high-level conferences, journals, and magazines will be cited. Besides, the article resorts to Google Scholar and the Semantic Scholar to search for relevant articles.}

\begin{figure*}[!t]
		\centering
		\includegraphics[width=6.2 in, height = 3.9 in]{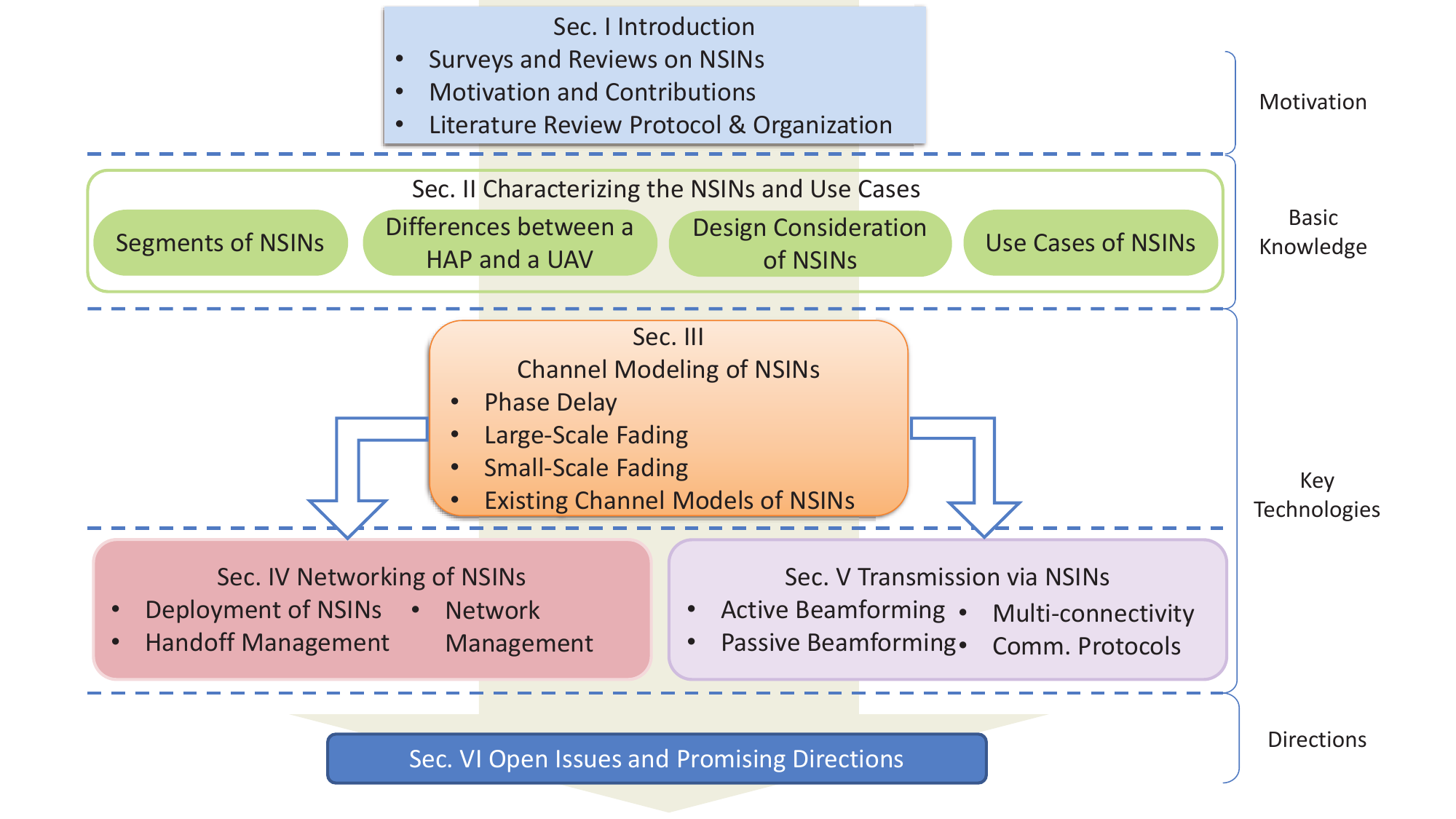}
		\caption{Organization of this article.}
		\label{fig:fig_org}
	\end{figure*}

We illustrate the outline of this article in Fig. \ref{fig:fig_org}.
Specifically, in Section II, we characterize the NSINs in detail and present the promising use cases of NSINs that aim to enable the reader to deepen their knowledge and understanding of NSINs. In Section III, we elaborate on the components of the channel models of NSINs while highlighting the impact of unstable platform movement and meteorological factors on the channel modeling of NSINs. 
Section IV describes key issues in the networking of NSINs, including network deployment, handoff management, and network management. The objectives, solutions, and key techniques of networking in NSINs are summarized and discussed in detail.  
In Section V, we comprehensively present the key techniques at the physical (PHY) layer, MAC layer, network layer, and transport layer of NSINs for achieving effective transmission. The main ideas and performance of these techniques are discussed and compared.
In Section VI, some open issues that need to be tackled and future directions that deserve attention are presented.
%Finally, we conclude by summarizing the open issues and promising directions for future study. 
Finally, we conclude this article in Section VII. 
For ease of reading, some significant acronyms in this article are shown in Table \ref{table_modular}.
 \begin{table*}[!t]
		\renewcommand{\arraystretch}{1.05}
		\caption{LIST OF ABBREVIATIONS}
		\label{table_modular}
		\newcommand{\tabincell}[2]{\begin{tabular}{@{}#1@{}}#2\end{tabular}}
		\centering
		\begin{tabular}{c c| c c}
			\hline
Acronyms &    Description & Acronyms &   Description \\ \hline
NSINs & Near-Space Information Networks & 
HAPs & High-Altitude Platforms \\
3GGP  &Third Generation Partnership Project& 
A-GR & ADS-B system aided Geographic Routing \\
5G & {\tabincell{c}{The Fifth-Generation of mobile \\ communication systems}} &
6G & {\tabincell{c}{The Sixth-Generation of mobile \\ communication systems}} \\
ACO & Ant Colony Optimization &
ADS-B & Automatic Dependent Surveillance-Broadcast \\
AI & Artificial Intelligence &
ALOHA &  Additive Links On-line Hawaii Area \\
AoA/AoD & Angle of Arrival/ Angle of Departure &
ARIS & Aerial-RIS \\
ATM &  Air Traffic Management &
CDMA & Code Division Multiple Access\\
{LEO} & {Low-Earth-Orbit} &
CIR & Carrier-to-Interference Ratio \\
CNPC  &Control and Nonpayload Communication &
CSMA/CA & {\tabincell{c}{Carrier Sense Multiple Access \\ with Collision Avoidance}}  \\
CTS  &Clear to Send &
DTN & Delay-Tolerant Networks \\
E2E & End-to-End &
EM  & Electromagnetic \\
FDMA  &Frequency Division Multiple Access &
FSO & Free Space Optical \\
FSPL & Free Space Path Loss &
FWA & Fixed Wireless Access \\
HGHR & {\tabincell{c}{Hybrid Time-Space Graph Supporting \\ Hierarchical Routing}} &
IMT & International Mobile Telecommunications\\
IoT & Internet of Things&
IRS & Intelligent Reflecting Surfaces\\
ISR  &Intelligence, Surveillance, and Reconnaissance &
ITS & Intelligent Transportation Systems \\
ITU & International Telecommunication Union &
LAP  &Low-Altitude Platform \\
LAUAVs & Low-Altitude Unmanned Aerial Vehicles &
LMS  &Least Mean Square \\
LSTM & Long Short-Term Memory Network &
ML  &Machine Learning\\
NFV & Network Function Virtualization&
NLoS& Non-Line-of-Sight\\
NOMA & Non-Orthogonal Multiple Access&
NTNs & Non-Terrestrial Networks \\
OFDM & Orthogonal Frequency Division Multiplexing &
PDR & Packet Delivery Ratio\\
PNT & Positioning, Navigation, and Timing&
PSNR & Peak Signal-to-Noise Rate \\
QoE & Quality-of-Experience &
QoS &Quality-of-Service \\
RIS & Reconfigurable Intelligent Surfaces&
RL &Reinforcement Learning\\
RMS&  Root Mean Square &
RNN & Recurrent Neural Network \\
RSS & Received Signal Strength &
RTS & Request to Send \\
SAGIN & Space-Air-Ground Integrated Networks &
SDMA&  Spatial Division Multiple Access \\
SDN & Software-Defined Networking &
SFC & Service Function Chaining \\
SNR&  Signal-to-Noise Ratio &
SWMAP & Sizes (or Space), Weights, Mobility, and Power \\
TDL & Tapped Delay Line &
TDM & Time Division Multiplexing \\
TDMA & Time Division Multiple Access&
TNs & Terrestrial Networks\\
UAVs&  Unmanned Aerial Vehicles&
HAUAVs&  High-Altitude Unmanned Aerial Vehicles \\
UtG & UAV-to-Ground&
VNF & Virtualized Network Function \\
WRC & World Radiocommunication Conference&
XR & Extended-Range
 \\ \hline
		\end{tabular}
	\end{table*}
 
\section{Characterizing the NSINs and USE CASES}
	The purpose of this section is to provide a detailed characterization of NSINs, focusing on its structure, specificity, and design considerations. Additionally, promising use cases for NSINs will be presented. 

\subsection{{Review of Network Definitions}}
\begin{figure*}[!t]
		\centering
		\includegraphics[width=5.6 in]{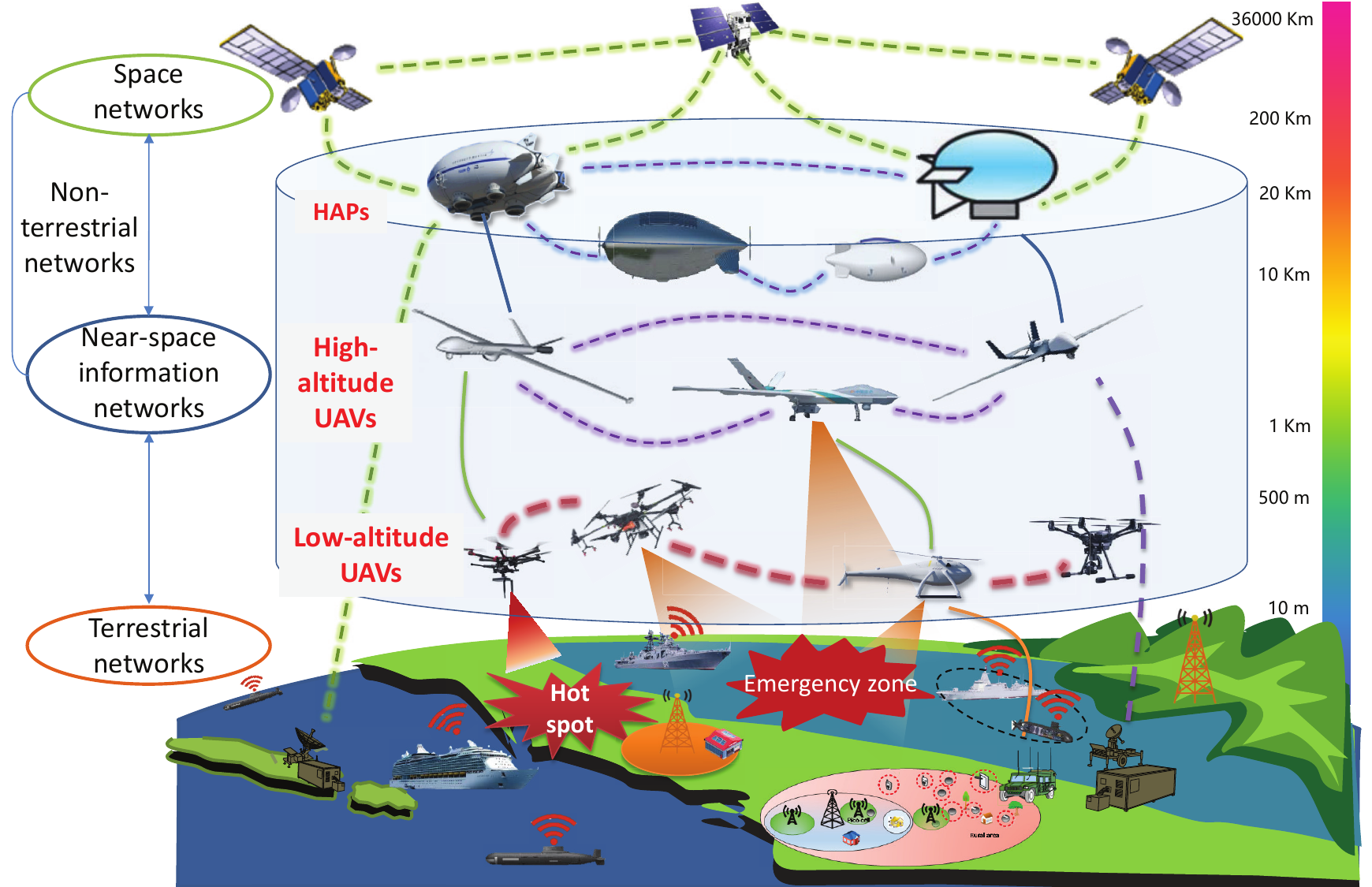}
		\caption{An architecture of integrated NTNs and TNs.}
		\label{fig:fig_sagin_Archi}
	\end{figure*}

 {NSINs are composed of HAPs, HAUAVs, and LAUAVs with diverse types of onboard devices. NSINs can seamlessly connect to space networks and TNs to construct NTNs- and TNs-integrated networks, the architecture of which is shown in Fig. \ref{fig:fig_sagin_Archi}.}
{In addition to the definition of NSINs presented at the beginning of this article, there are several different, albeit slightly different, definitions of such aerial heterogeneous networks in the literature \cite{DBLP:journals/iotj/PhamRFNYLDH22,DBLP:journals/jsac/CaoYAXWY18,Cheng20216GSS, DBLP:journals/comsur/GuoLLTK22, DBLP:journals/comsur/WangSNZS22,DBLP:journals/comsur/BaltaciDOACS21}. 
\begin{itemize}
    \item Aerial (radio) access network (AAN): The authors in \cite{DBLP:journals/iotj/PhamRFNYLDH22} considered the AAN as a combination of low-altitude platforms (LAPs), HAPs, and low-earth-orbit (LEO) satellites. An AAN is deployed to fully complement TNs to create a future access network in 6G. 
    \item Airborne communication networks (ACNs): ACNs are engineered to utilize various aircraft (including LAPs, HAPs, and satellites), which are equipped with transceivers and sensors, to build communication access platforms \cite{DBLP:journals/jsac/CaoYAXWY18}. 
    \item Air/Aerial network: In \cite{Cheng20216GSS, DBLP:journals/comsur/GuoLLTK22, DBLP:journals/comsur/WangSNZS22}, an air/aerial network is defined as a network made up of aircraft, UAVs, airships, and balloons to provide connectivity from the sky. This type of network can transmit information cooperatively with TNs and space networks.
    \item Future aerial communication (FACOM) network: The authors in \cite{DBLP:journals/comsur/BaltaciDOACS21} regarded FACOM as a connectivity ecosystem incorporating all connectivity from the sky (e.g., air-to-air (A2A), HAP communications, and satellite communications).
    %these looming aerial connectivity use cases and their potential connectivity solutions.
\end{itemize}
}

	\subsection{Segments of NSIN}
	As shown in Fig. \ref{fig:fig_sagin_Archi}, NSIN composed of HAPs, HAUAVs, and LAUAVs with diverse types of onboard devices are a novel and promising network regime. 
	From the perspective of platform deployment altitude, it can be categorized into two-layer vertical aerial networks: HAP networks and UAV networks. 
 
	HAP networks that consist of one or more connected HAP platform(s) are deployed in the near space of 17 km to 25 km. Connected HAP platforms can form a mesh network topology, which will greatly enhance the service capability of HAP networks.
 Further, it can be concluded that the role of HAP in NSIN is fundamental and irreplaceable. The design and manufacture of HAPs have attracted extensive attention from industry, and many recent HAP projects or products have been developed in recent ten years, as summarized in Table \ref{table_project}.
 	\begin{table*}[!t]
		\renewcommand{\arraystretch}{1.2}
		\caption{{{Some representative HAP projects/products that are still under construction or maintenance}}}
		\label{table_project}
		\newcommand{\tabincell}[2]{\begin{tabular}{@{}#1@{}}#2\end{tabular}}
		\centering
		\begin{tabular}{|l|l|l|}
			\hline
			{\tabincell{l}{Project/product}} & {\tabincell{l}{Company/\\organization}} & {\tabincell{l}{Important descriptions}}  \\\hline
			
			{\tabincell{l}{Zephyr S \\ High-Altitude \\ Pseudo-Satellite \cite{Airbus2021Zephyr}}} & {\tabincell{l}{Airbus and \\ NTT DOCOMO}} & {\tabincell{l}{$\bullet$ It demonstrated the feasibility of providing communication \\
					 services from the stratosphere to smartphones. \\ $\bullet$ Data transmissions across various speeds up to a distance \\ of 140 km were successfully demonstrated. \\
     $\bullet$ It loaded with systems for diverse communication \\ and sensing services such as TV, radio, \\ mobile telephony, VoIP, and remote sensing. \\ $\bullet$ It supports a payload of 100 kg and a flight \\ duration of up to one year. \\
   $\bullet$ It covers an area of up to 1000 km in diameter.}}  \\ \hline

			{\tabincell{l}{AVEALTO Wireless \\ Infrastructure \\ Platform \cite{AVEALTOAVEALTO2021}}} & {\tabincell{l}{Avealto Ltd.}} & {\tabincell{l}{$\bullet$ Remain on-station in the stratosphere at an altitude of \\ 20 km for missions
					of several months’ duration. \\ $\bullet$ Each HAP will provide services to a circular terrestrial \\ footprint of 240 km in diameter. }}  \\ \hline
			
			{\tabincell{l}{Thunderhead Balloon  \\System \cite{RAVENRAVEN2021}}} & {\tabincell{l}{Raven Aerostar}} & {\tabincell{l}{$\bullet$ It delivers long term evolution (LTE) from 70,000 ft \\ and demonstrates a solution for delivering internet to \\ underserved, denied or disaster-affected populations. \\ $\bullet$ It redefines stratospheric missions and provides \\ cost-effective persistence. }}  \\ \hline
			
			{\tabincell{l}{Yuanmeng \cite{Peter2022This,Peter2022China}}} & {\tabincell{l}{ Beihang University \\ \& Beijing Aerospace \\ Technology Company}} & {\tabincell{l}{$\bullet$ A volume of approximately 18,000 cubic metres. \\ $\bullet$ Payload (mission equipment): 300 kg \\ $\bullet$ Photovoltaic cells on the hull for power during the day \\ and battery or fuel cell for power at night \\ $\bullet$ Fly for up to six months at a stretch, with a large array \\ of solar panels covering the aircraft's top side. \\ $\bullet$ Loaded with systems for wideband communication, relay, \\ high-definition observation, and spatial imaging.  }} \\ \hline
			
			{\tabincell{l}{Stratobus \cite{Thales2017what}}} & {\tabincell{l}{Thales Alenia Space}} & {\tabincell{l}{$\bullet$ A solution for local missions out to a terrestrial horizon of \\ up to 500 km \\ $\bullet$ Providing security (the fight against terrorism, \\ drug trafficking), environmental monitoring \\ (forest fires, soil erosion, pollution), telecommunications \\ (Internet, 4G/5G) and navigation (GPS local reinforcement). \\ $\bullet$ It can carry up to 450 kg of payload, with a power \\ rating of eight kW for a five-year mission  }}  \\ \hline
			
			{\tabincell{l}{HAWK30 \cite{HAPSMobileHAWK30}}} & {\tabincell{l}{HAPSMobile}} & {\tabincell{l}{$\bullet$ It aims at connect mobiles, UAVs, \\ and the Internet of Things (IoT) devices around the world. \\ $\bullet$ It has a wingspan of 78 meters and provides a 100 km \\ coverage radius for several months. }}  \\ \hline
			
			{\tabincell{l}{PHASA-35 \cite{BAE2021phasa}}} & {\tabincell{l}{BAE Systems and \\ Prismatic}} & {\tabincell{l}{$\bullet$ It aims at providing some many services, e.g., 5G. \\ $\bullet$ It has a payload capacity of 15 kg with \\ a power capacity of 300-1000 W. \\ $\bullet$ It can cover a radius of up to 200 km. }}  \\ \hline
   	{\tabincell{l}{Elevate \cite{Zero2infinityElevate}}} & {\tabincell{l}{Zero 2 Infinity}} & {\tabincell{l}{$\bullet$ It aims at providing transportation services. \\ $\bullet$ It has  launched a payload of almost 500 kg to an altitude of 32 km. }}  \\ \hline
			
			{\tabincell{l}{X-station \cite{StratXXX}}} & {\tabincell{l}{StratXX}} & {\tabincell{l}{$\bullet$ It loaded with systems for diverse communication \\ and sensing services such as TV, radio, \\ mobile telephony, VoIP, and remote sensing. \\ $\bullet$ It supports a payload of 100 kg and a flight \\ duration of up to one year. \\
   $\bullet$ It can cover up to a 1000 km area in diameter.}}  \\ \hline
		\end{tabular}
	\end{table*}
	UAV networks consisting of a number of HAUAVs and LAUAVs are deployed in tropospheric space ranging from a few meters to ten km. 
 
	With the integration of the above two-layer networks, NSIN is able to achieve wide-area and 3-D coverage along with local on-demand service enhancement and has promising application prospects in many critical applications, including ITS, emergency communications, and security protection of energy, territory, homeland, sovereignty, and so on. 

	\subsection{Differences between a HAP and a UAV}
	Nevertheless, there are many significant and distinct differences between a HAP and a UAV in terms of mobility model, energy source and endurance, payload and power capability, and footprint. These differences are summarized in Table \ref{table_difference} and elaborated on in the following subsections. 
	\begin{table*}[!t]
		\renewcommand{\arraystretch}{1.2}
		\caption{{{Some significant differences between a HAP and a UAV}}}
		\label{table_difference}
		\newcommand{\tabincell}[2]{\begin{tabular}{@{}#1@{}}#2\end{tabular}}
		\centering
		\begin{tabular}{|c|l|l|l|l|l|}
			\hline
			{\tabincell{l}{Platform}} & {Type} & {Mobility} & {\tabincell{l}{Energy source \& endurance}} & {\tabincell{l}{Payload \& power capability}} & {Footprint}  \\\hline
		
			{} & {Aircraft} &	{\tabincell{l}{Trajectory re-plan \\ enable}} & {\tabincell{l}{$\bullet$ Fuel and solar cells \\ $\bullet$ Several months {\cite{BAE2021phasa}}}} & {\tabincell{l}{$\bullet$ Many tens of kg \\ $\bullet$ A power rating of 300 W -\\  1000 W \cite{BAE2021phasa}}} & {Cover a radius of up to 200 km \cite{BAE2021phasa}}  \\ \cline{2-6}
			
			{HAP} & {Balloon} &  {\tabincell{l}{Stratospheric winds\\  dependent}}  & {\tabincell{l}{$\bullet$ Mainly solar cells \\ $\bullet$ Several months {\cite{DBLP:journals/comsur/KurtKAIDAYY21}}}} & {\tabincell{l}{$\bullet$ Up to 100 kg {\cite{qiu2019air}} \\ $\bullet$ A power rating of 100 W}} & \tabincell{l}{Providing coverage over a radius of \\ 64 km {\cite{DBLP:journals/comsur/KurtKAIDAYY21}}} \\ \cline{2-6}
			
			{} & {Airship} &  {\tabincell{l}{Quasi-stationary or \\ path re-plan enable}}  & {\tabincell{l}{$\bullet$ Fuel and solar cells \\ $\bullet$ Several years {\cite{DBLP:journals/comsur/KurtKAIDAYY21}}}} & {\tabincell{l}{$\bullet$ Up to 450 kg \cite{Thales2017what} \\ $\bullet$ A power rating of 8 kW \cite{Thales2017what}}} & {Cover up to 500 km in diameter \cite{Thales2017what}} \\ \hline
			
			\multirow{3}{*}{UAV} & {\tabincell{l}{Fixed-\\wing}} &  {\tabincell{l}{Fast movement with \\ limited steering ability}}  & {\tabincell{l}{$\bullet$ Mainly fuel cells \\ $\bullet$ Tens of hours {\cite{qiu2019air}}}} & {\tabincell{l}{$\bullet$ Tens of kg \\ $\bullet$ Power of tens of watts \cite{DBLP:journals/wcl/LiuZSSCL21}}} & {\tabincell{l}{Up to many tens of km in radius {\cite{DBLP:journals/wcl/LiuZSSCL21}} \\ (Depending on the coverage region)}} \\ \cline{2-6}
			{} & {\tabincell{l}{Rotary-\\wing}} &  {\tabincell{l}{Flexible and \\ fast movement}}  & {\tabincell{l}{$\bullet$ Battery-powered \\ $\bullet$ Within an hour {\cite{Izydorczyk2020ExperimentalEO}}}} & {\tabincell{l}{$\bullet$ Usually a few kg {\cite{Izydorczyk2020ExperimentalEO}} \\ $\bullet$ Enable a few watts of power}} & {\tabincell{l}{Up to several km in radius {\cite{Izydorczyk2020ExperimentalEO}} \\ (Depending on the coverage region)}} \\ \hline			
		\end{tabular}
	\end{table*}
	
	\subsubsection{Mobility model}
	%Node mobility is one of the most apparent differences between NSIN and many diverse types of ad hoc networks, such as the vehicle ad hoc network (VANET) and mobile ad hoc network (MANET). 
 %In UAV networks, the extent of mobility may be much higher than that in both VANET and MANET.
The maximum flight speed of a UAV is much greater than that of a HAP.
	In accordance with vertical applications, the flight speed of a UAV can be in the range of 0-460 km/h \cite{DBLP:journals/adhoc/BekmezciST13,DBLP:journals/comsur/GuptaJV16}, and the flight speed of a HAP can be {in the range of 0-200} km/h \cite{Ultra2017NCNA}.
	A mobility model is a fundamental mathematical expression supporting network connectivity analysis, routing protocol design, and network performance evaluation. It mathematically characterizes how the locations, velocities, and acceleration of network nodes change over time. 
	Next, we discuss the mobility models designed for UAVs and HAPs, respectively.
 
	\begin{table*}[!t]
	\renewcommand{\arraystretch}{1.2}
	\caption{A list of some mobility models of UAVs}
	\label{table_mobility_model}
	\newcommand{\tabincell}[2]{\begin{tabular}{@{}#1@{}}#2\end{tabular}}
	\centering
	\begin{tabular}{l}
%\begin{minipage}{0.18\textwidth}
			\centering
			\includegraphics[width=7.0 in]{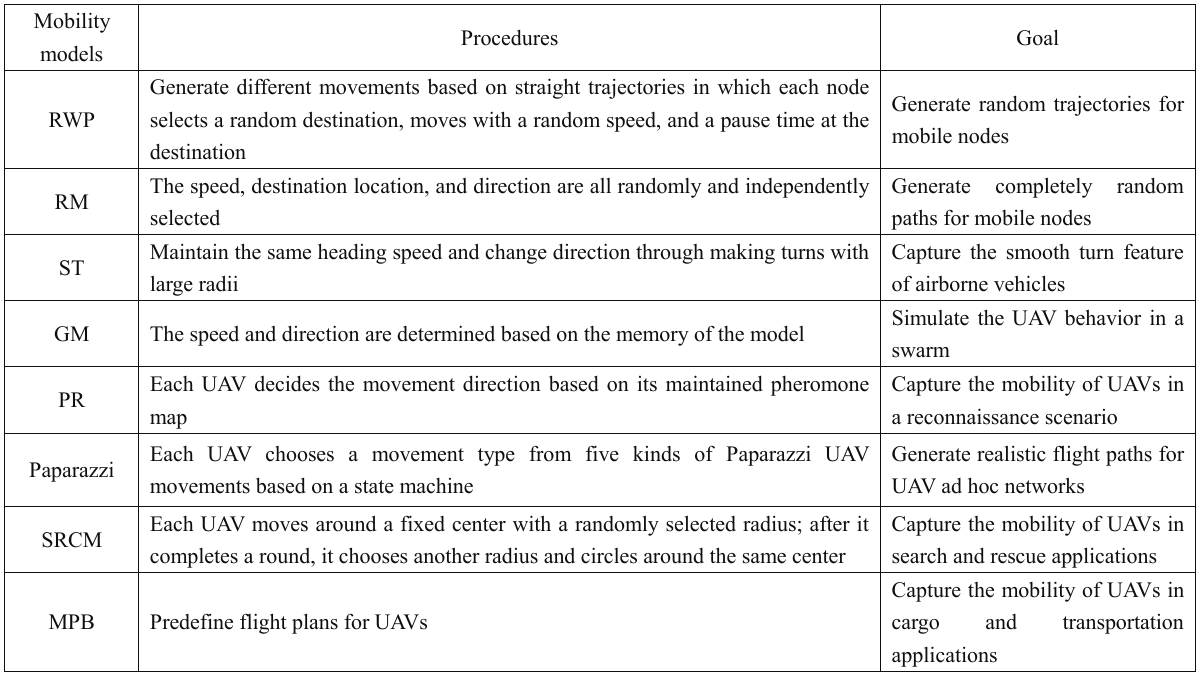}
		%\end{minipage} 
	\end{tabular}
\end{table*} 
	Table \ref{table_mobility_model} summarizes seven types of UAV mobility models, namely, random way point (RWP) \cite{Yassein2016FlyingAN}, random movements (RM) \cite{DBLP:conf/icwmc/KuiperN06}, smooth-turn (ST) \cite{Wan2012ASM}, Gauss–Markov (GM) \cite{Naser2022ImplementationOR}, pheromone repel (PR) \cite{DBLP:conf/icwmc/KuiperN06}, Paparazzi \cite{Bouachir2014AMM}, semi-random circular movement (SRCM) \cite{DBLP:journals/isci/WangGWW10}, and mission plan-based (MPB) \cite{DBLP:journals/wicomm/CampBD02}. 
	From this table, we observe that the selection of the mobility model is vertical application-related. 
	For instance, RWP and RM mobility models can well capture the random activities of aerial platforms. As a result, many researchers in the research community leverage them to model the mobility of UAVs.
	However, in practical applications, UAVs will choose neither the RWP model nor the RM model due to restrictions on maneuverability, aerodynamics, and energy consumption. 
	As for the ST, GM, and PR mobility models, they are suitable for patrolling applications as they characterize the memory-equipped movement of UAVs. 
	The flight path generated by the SRCM model for a UAV is a circular trajectory with a fixed circling center and a diverse range of turning radii. Hence, the SRCM mobility model is well-suited for UAVs performing some search and rescue tasks.
	The MPB mobility model is well-suited for UAVs engaged in cargo delivery and transportation tasks where the flight paths are pre-planned. 
    Although uncertainties, such as unknown dangerous airspace and departure delays, may occasionally cause flight deviations from the intended paths, UAVs have the ability to promptly re-plan their flight paths based on the pre-determined routes \cite{DBLP:journals/iotj/YinXCXYW18}. 
	
 {A HAP is usually considered to be quasi-stationary in the stratospheric.} The relatively weak stratosphere airflow will cause unstable HAP movements such as horizontal drift, vertical motion, rotation, and attitude angle disturbance. 
	A HAP is then regarded as stationary if it does not escape from a small stereoscopic airspace. ITU suggests that the diameter of the airspace is 500 m \cite{seriesF2000Preferred}. 
	However, accurate attitude control of a HAP with a huge volume {that requires a refined design of the propulsion system, including improved motors and propeller structures adaptive to the near-space environment} is hard to achieve, and a HAP may escape from the airspace with a large probability. 
 %{To achieve this goal, }
 \begin{figure}[!t]
		\centering
		\includegraphics[width=3.2 in, height = 1.3 in]{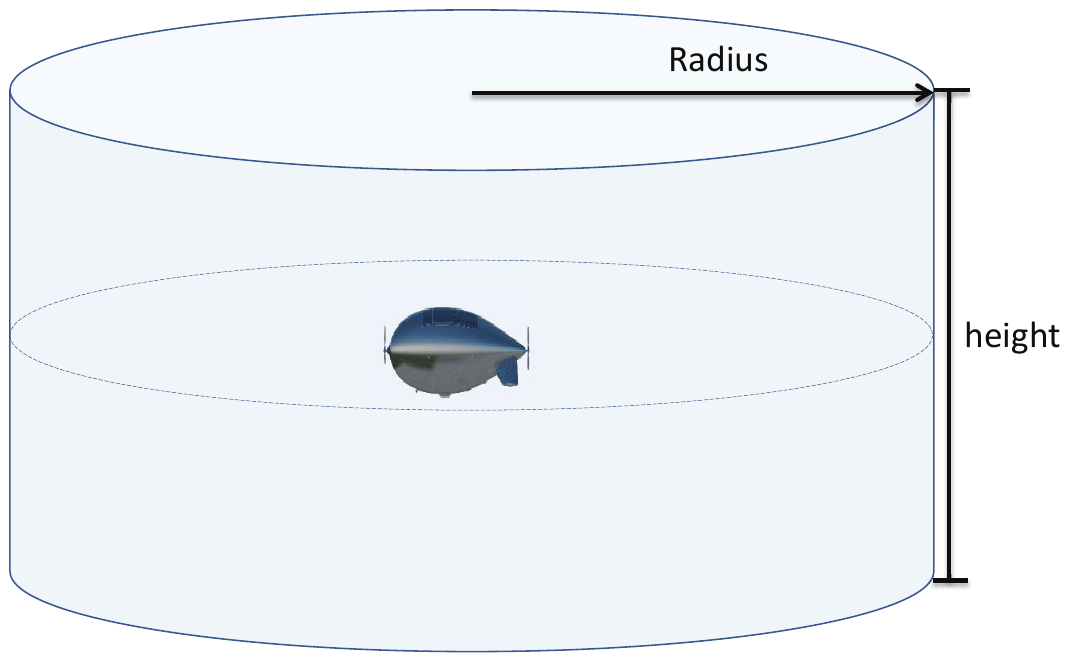}
		\caption{A stationary model (i.e., a cylindrical airspace) of HAP.}
		\label{fig:fig_stationary}
	\end{figure}
 %, i.e., a cylindrical airspace
	%Owing to the impact of breezes and gusts in the stratosphere on a HAP, a HAP may move in any direction, thereby resulting in escape from the airspace with a large probability. Therefore, this constraint is quite tight for a HAP with huge volumes. 
	Then, the HeliNet Project designated another stationary model for HAP \cite{Karapantazis2005BroadbandCV}. The stationary model is a cylindrical model with a diameter of four km and a height of three km, {as shown in Fig. \ref{fig:fig_stationary}.} 
 A HAP is considered to be stationary if it does not flee from this cylindrical airspace 99.9\% of the time. 
	There is another similar quasi-stationary model for HAP that allows a HAP to change its altitude over time, bounded by an upper limit (e.g., 26 km) and a lower limit (e.g., 20 km), with the cylinder’s radius ranging from one km to five km.
 The quasi-stationary model is an important and commonly used model for HAP, despite the fact that a HAP can fly at a speed of 110 km/h.
{However, in some practical applications requiring a HAP to remain stationary, the HAP (especially an airship) will fly in a circular trajectory with a small radius (e.g., one km).} 
% representing the spatial location of a HAP as a cylindrical volume rather than a deterministic point would be more appropriate.
		
	%{Use a table to collect the mobility differences among all types of NSIN platforms. 
	%From the perspective of speed range, mobility model, easy of control, atmosphere impact on the mobility of platforms, flexibility and agility}
	%Can describe two types of UAV mobility model in detail, one is just reflect the smoothness of UAV movement without considering the traffic-driven characteristic of UAV deployment; the other is the QoS-driven or QoE-driven movement control under communication and safety constrains. Give some examples to prove. 
 
	 \begin{figure*}[!t]
		\centering
		\includegraphics[width=5.9 in]{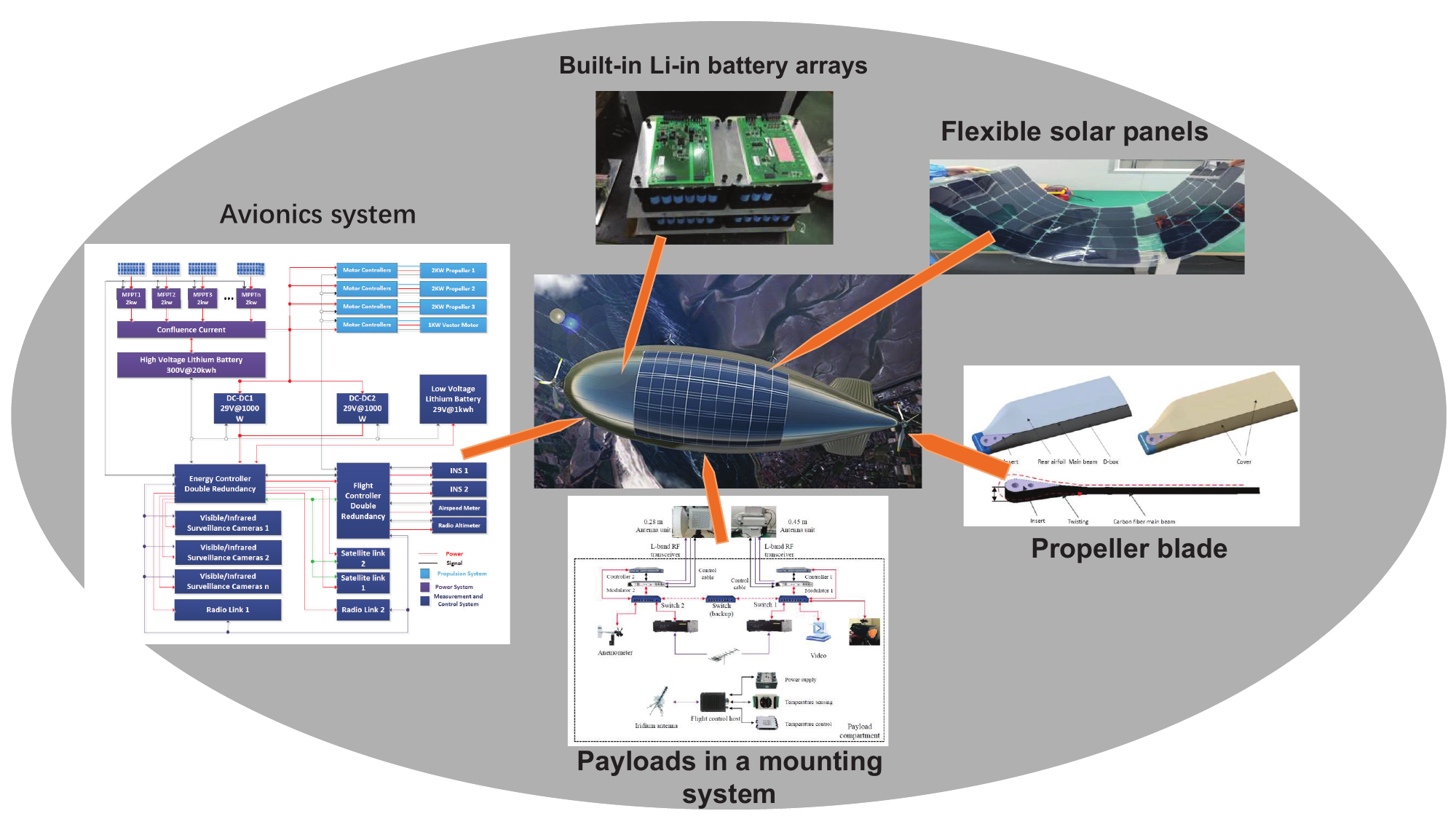}
		\caption{{A general view of some main components of an airship designed \& manufactured by Beihang University (abbreviated as Beihang airship).}}
		\label{fig:fig_beihang_airship}
	\end{figure*}
	\subsubsection{Energy sources and endurance}
A HAP is huge and has many mission systems. Fig. \ref{fig:fig_beihang_airship} presents a general view of some main components of a HAP (exactly, an airship), an energy management system, including an avionics system, a payload system, and a propulsion system, and so forth.
%Thus, it's indisputable that the avionics system of a HAP will be sophisticated. Fig. \ref{fig:fig_avionics} illustrates an avionics system of our airship, from which we can observe that a HAP will be extremely energy-consumption.	
 
 Up to now, HAPs have been able to utilize three types of energy sources, including fuel energy, beam energy, and solar energy. 
	Due to the limited capacity of fuel tanks, HAPs powered by conventional energy sources have a short flight duration. Additionally, the processes of takeoff and landing for a HAP are time-consuming and energy-intensive. Then, the cruise duration of conventional energy-driven HAPs will be significantly shortened, and their application scenario will be rather limited. 
	HAPs can be charged by energy beams from external energy sources. The concept of power transmission using radio waves has a long history, going back almost one hundred and forty years to Heinrich Hertz. In the 20th century, a stationary high-altitude relay platform (SHARP) project demonstrated such a process of transmitting microwave power from a large ground antenna system to a HAP \cite{Summary2013Jull}. 
	Except for microwave beam power, laser beams can also be explored as energy sources for HAPs. Nonetheless, microwave or laser beam power will bring the risk of high-power irradiation; thus, the practical application of beam-powered HAPs is severely restricted. 		
    HAPs are deployed in the stratosphere to take advantage of the abundant solar energy available. As a result, modern HAPs are designed with a focus on efficiently harnessing solar energy, taking into account operational costs and safety requirements. 
	HAPs are bulky, and then a plethora of solar panels can be attached to HAP surfaces to convert solar energy into direct current (DC) power. 
 %{Fig. \ref{fig:flexible_solar} shows the flexible solar panels attached to the surface of our airship.}
	Solar-powered HAPs will also utilize {batteries (e.g., li-ion battery)} to power HAP functions and maintain the platform's attitudes during dark nights. Besides, the fuel energy plays a crucial role in the flight altitude control of HAPs (exactly, airship). For instance, our airship will increase its buoyancy and flight altitude by producing hydrogen, and reduce its buoyancy and flight altitude by consuming hydrogen though hydrogen fuel cells. %Fig. \ref{fig:hydrogen_generation} illustrates the hydrogen generator and hydrogen fuel cells onboard our airship the related test experiments.}
 % \begin{figure}[!t]
	% 	\centering
	% 	\includegraphics[width=3.3 in, height = 1.6 in]{flexible_solar.pdf}
	% 	\caption{{Flexible solar panels of our airship.}}
	% 	\label{fig:flexible_solar}
	% \end{figure}
 %  \begin{figure}[!t]
	% 	\centering
	% 	\includegraphics[width=3.3 in]{li_battery.pdf}
	% 	\caption{{Li-in battery arrays onboard our airship.}}
	% 	\label{fig:liion_battery}
	% \end{figure}
 % \begin{figure}[!t]
	% 	\centering
	% 	\includegraphics[width=3.6 in]{Stationary_model_HAP.pdf}
	% 	\caption{A hydrogen generator, hydrogen fuel cells onboard our airship, and related tests.}
	% 	\label{fig:hydrogen_generation}
	% \end{figure}
 
	HAUAVs can utilize two types of energy sources, including fuel energy and solar energy. However, as HAUAVs fly in the atmosphere, the deployment of solar-powered HAUAVs is closely related to weather conditions. On the contrary, fuel-powered HAUAVs are relatively robust to weather conditions and can fly for many tens of hours. As a result, fuel-powered HAUAVs have a wide range of applications in the military field, such as used for ground attack, ground surveillance, and early warning. 
	LAUAVs can also utilize two types of energy sources, i.e., fuel energy and electrical batteries. Owing to the advantages in manufacturing and maintenance costs, battery-driven LAUAVs are much more popular than fuel-powered LAUAVs. However, the total battery capacity of a LAUAV is limited owing to the constraint on its payload capability. Currently, depending on applications, the limited energy can support a battery-driven LAUAV to fly for only a few minutes, or at most a few tens of minutes. 
	
	Based on the above analysis, it can be concluded that effective energy management is of utmost importance for NSIN as it directly impacts its overall lifespan.
	The energy consumption of NSIN consists of consumption for flight control and performing missions (e.g., communications). The energy consumed by flight control includes that for propulsion and flight attitude control. The types and characteristics (e.g., size and weight) of platforms in NSIN also affect flight control consumption. Communication activities will generate communication energy consumption that includes circuit consumption (e.g., energy consumed by mixers, frequency synthesizers, D/A converters, and so on) and transmission consumption. The types of communication payloads and adopted communication techniques will also significantly affect the amount of communication energy. 
	Many mechanisms related to flight control and transmissions can be adopted to improve the energy efficiency of airborne platforms in NSIN. For example, as the engine power of a UAV will decrease and then increase as the flight speed increases \cite{Zeng2018EnergyMF}, the flight speed of a UAV should be optimized to reduce engine power. Reasonable flight plans should be made to reduce engine power, and maneuvers like small-radius turns and rapid increases in the flight altitude and speed should be reduced. 
	Additionally, the issue of optimizing the deployment altitudes of UAVs should be studied. A low deployment altitude will lead to reduced engine power. However, the non-line-of-sight (NLoS) probability of communications may increase when UAVs are deployed at low altitudes. As a result, greater transmission power will be required to satisfy the QoS requirements of terrestrial serving terminals. 
	Certainly, UAVs can reduce energy consumption by offloading computational tasks to HAPs \cite{DBLP:journals/iotj/KangCMMFL23,DBLP:conf/wcnc/CaoYC23,DBLP:conf/iwcmc/LiuLYLLTH19,DBLP:conf/ictc/NguyenP22}. 
	
	Summarily, the energy constraint of NSIN is determined by that of HAPs, HAUAVs and LAUAVs. Owing to the stringent constraints on aspects of platform size, weight, and payload capability, LAUAVs in NSIN have the most stringent energy consumption constraints. 
	
	\subsubsection{Payload and power capability}
	Payload and power capability are significant differences between a HAP and a UAV. 
 The payload and power capability of a HAP are much greater than that of a UAV. For example, {Fig. \ref{fig:fig_beihang_airship} shows a diagram of the payload system of the Beihang airship, where many different types of payloads are mounted into the payload compartment to accomplish diverse missions}.
 The recent near-space airship “Stratobus” also can carry up to 450 kg of payload, which envisions unprecedented potential for integrating many diverse types of mission systems (e.g., computing, communication, sensing, positing, navigating, timing (PNT), and intelligence, surveillance, reconnaissance (ISR) systems) and simultaneously providing multiple kinds of services. In this way, the efficiency of completing missions by resorting to a HAP can be greatly improved. Further, owing to the advantages of high payload capability, a HAP can conduct many crucial tasks that cannot be completed by a UAV, e.g., the collection of sufficient measurement data inside a typhoon. To complete this task, an airborne platform needs to carry and drop a large number of communication and sensing nodes into the interior of the typhoon. Meanwhile, the airborne platform should enable the collection of data from these measuring nodes.
	For a UAV, it has limited payload capacity and thus cannot carry out many critical missions. 
	% \begin{figure}[!t]
	% 	\centering
	% 	\includegraphics[width=3.4 in]{Payload_compartment.pdf}
	% 	\caption{{A diagram of the payload system of our airship.}}
	% 	\label{fig:Payload_compartment}
	% \end{figure}
	%Therefore, a multi-UAV network with many cooperation UAVs is usually deployed to carry out tasks. 
	
	Besides, a HAP can generate a greater power rating than a UAV, e.g., the ``Stratobus'' has a 1000 ${\rm m}^2$ solar generator and high-energy density batteries, which are able to generate a power rating of eight kW. The great power generation capability indicates that a HAP can provide communication services for a lot of users with high bitrate requirements. Meanwhile, higher power indicates that a HAP has a larger footprint, and the degradation of signal quality caused by meteorological attenuation will be alleviated by increasing transmit power to some degree.

	\subsubsection{Footprint}
	The large footprint is one of the most significant features of a HAP, which popularizes both research and utilization of it. 	
	Thanks to a high deployment altitude, a HAP has a great probability of enabling line-of-sight (LoS) propagation of around 565 km in diameter. 
	Due to topological obstacles and radio propagation characteristics, a realistic coverage area for a HAP is around 240 km in diameter over most land areas and around 500 km over oceans or plains (where the topography is flat).
	Owing to their wide footprint, HAPs have been envisioned to connect remote areas where traditional Internet connectivity is unserved or underserved.
 Theoretically, only thousands of HAPs are required to establish global interconnectivity. 
	Besides, the wide footprint enables a HAP to be considered a satellite alternative. 
	
	Different from a HAP, a UAV has a relatively small footprint. Yet, like a HAP, the size of the UAV footprint is closely related to the deployment environment. For instance, when deployed in a rural area, a UAV will have a large footprint because of the absence of severe shadow fading or signal blockage caused by high-rise buildings. A UAV has a small footprint when deployed in a dense urban area. As a result, extensive attention has been paid to the topic of extending the UAV footprint by optimizing the UAV deployment altitude.

	\subsection{Design Consideration of NSIN}
	\subsubsection{Heterogeneity}
	As presented above, NSIN consist of many different airborne platforms that are quite different in terms of deployment altitude, footprints, computing power, and so on. Therefore, heterogeneity is the most important feature of NSIN and must be taken into account in the NSIN design process, especially in the protocol design of NSIN. 
	
	HAPs and UAVs collaboratively construct a network with a multi-tier structure. In this multi-tier network, different footprints of a HAP and a UAV may lead to footprint overlap and inter-tier interference. To this end, advanced beamforming schemes and spectrum sharing mechanisms \cite{DBLP:journals/tcom/WeiWGW0H023} should be designed to alleviate the interference so that the performance of NSIN will not be degraded. 
	
	As presented above, HAPs and UAVs have distinct abilities, resulting in the fact that HAP communication systems and UAV communication systems may adopt diverse communication paradigms. For instance, they may work at different frequency bands, explore different modulation and demodulation techniques, and adopt diverse signal exchange methods. To this end, the design of NSIN should achieve some technical breakthroughs in terms of the integrated protocol design, the integrated network framework design, the optimal deployment of gateways, and so on. 
	
	UAVs are widely used in various scenarios due to their flexible and agile deployment capabilities. In comparison, HAPs have larger sizes and heavier masses, making them less flexible than UAVs. Consequently, the mobility control of NSIN cannot be modeled and optimized using the same approaches as UAVs. 
	HAPs and UAVs also exhibit different communication persistence and lifespans, which should be taken into account when designing protocols and optimizing network resources for NSIN.

	\subsubsection{Cooperation}
	The cooperation between HAPs and UAVs can significantly enhance the performance of NSIN. In many applications, increasing the number of collaborative airborne platforms can greatly improve the service ability of NSIN. 
 %{Fig. \ref{fig:NSIN_ITs} illustrates a scenario in a practical program we have undertaken. We attempt to monitor complex ground traffic using multiple cooperative airships and UAVs in the program.} 
% \begin{figure}[!t]
% 		\centering
% 		\includegraphics[width=3.2 in]{fig2_use_case_ITS.pdf}
% 		\caption{{Ground traffic monitoring using NSIN.}}
% 		\label{fig:NSIN_ITs}
% 	\end{figure}
 
	On one hand, the footprint of a HAP is much larger than that of a UAV, so the collaboration of HAPs and UAVs will greatly extend the coverage of UAV networks \cite{Arum2023ExtendingCA}.  
	On the other hand, in some areas, such as mountainous and metropolitan regions, the coverage of HAP networks can be improved by deploying UAV relays. By acting as relays, UAVs can enhance the quality of HAP-UAV-user links in a flexible and proactive manner.
Airborne platforms in NSIN can act as aerial network nodes with diverse functionalities, including sensing nodes, computing nodes, communication nodes, and so on. 
Thus, through the cooperation among these airborne platforms, NSIN can complete some critical tasks (e.g., search, rescue, surveillance, and earth observation) faster and more accurately. In non-cooperative HAP and UAV networks, each airborne platform may be required to accomplish several tasks, the process of which will not only lead to great task latency but also occupy a significant amount of radio resources (e.g., frequency and power). 
	The robustness of NSIN can also be enhanced by airborne platform cooperation; as a result, the heterogeneous NSIN will keep organized even in the case of platform or link failure. 

	\subsubsection{Disruption and delay prone network}
	The link disruption is likely to occur in NSIN. 
	Link disruption will be caused by many factors, including airborne platform-related factors (e.g., platform mobility, failure/replacement, interference, and antenna pattern) and propagation environment-related factors (e.g., blockage and various types of propagation fading). 
	Delays in NSIN information transmission may be caused by long-distance transmission, poor link status, reduced network throughput, packet queueing, and network congestion, etc. 
	
	Therefore, some mechanisms should be designed to tackle the link interruption issue and reduce transmission delay in NSIN, which include beam alignment, mobility control, interference management, transmission protocol design, queue management, and congestion control. 
	The reader can refer to more details of the mechanism design for NSIN in Section X.

	\subsubsection{Airborne platform constraints}
\begin{figure}[!t]
\centering
\begin{minipage}[t]{0.45\textwidth}
\centering
\includegraphics[width=2.4 in]{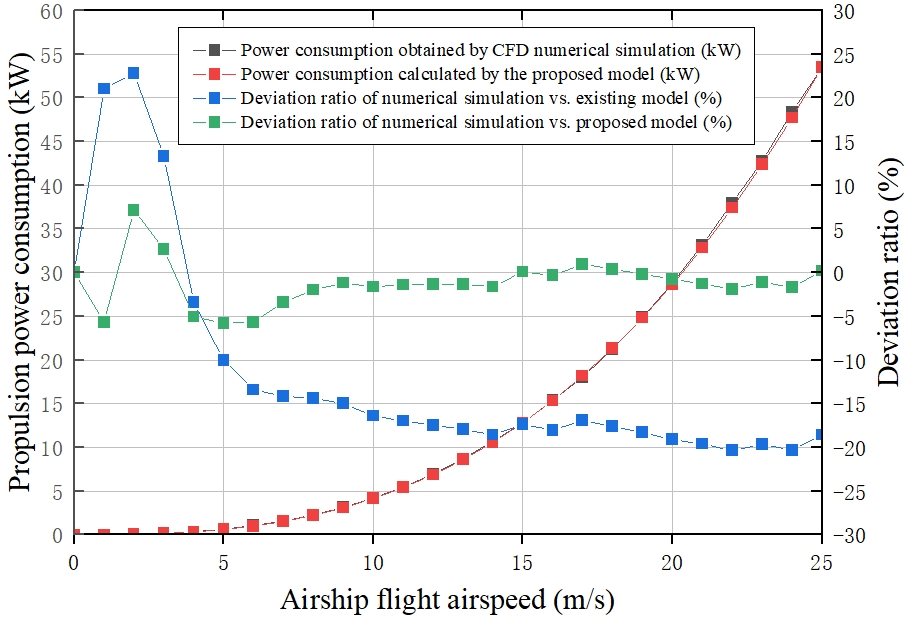}
\caption{Total required power and required power for propulsion versus speed of an airship, obtained through CDF numerical simulation.}
\label{fig:fig_HAP_power_vs_speed}
\end{minipage}
\hspace{0.1\linewidth}
\begin{minipage}[t]{0.45\textwidth}
\centering
\includegraphics[width=2.3 in]{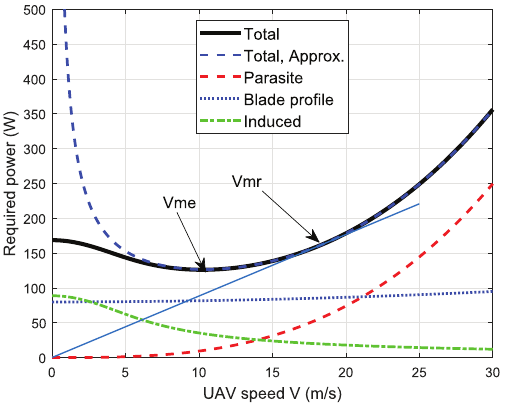}
\caption{Propulsion power consumption versus speed of a rotary-wing UAV \cite{Zeng2018EnergyMF}.}
\label{fig:fig_UAV_power_vs_speed}
\end{minipage}
\end{figure}

	When designing NSIN, one should simultaneously consider the constraints of HAPs and UAVs in terms of their sizes (or space), weights, mobility, and power. SWMAP constraints significantly affect the endurance, sensing, computation, and communication capabilities of NSIN. 
	\begin{itemize}
		\item \textbf{Size or space:}
		Airborne platforms with larger sizes or more space have a greater payload capacity and more mounting points. Thus, a larger airborne platform can carry more powerful sensors and peripherals, such as Internet of Things (IoT) devices, transceivers, hubs, cameras, three-axis accelerometers and gyroscopes, magnetometers, barometers, the global positioning system (GPS), and electro-optical pods. In this way, these platforms have the ability to complete preset tasks more accurately, quickly, and effectively. However, the larger the size of the platform, the heavier the platform. Correspondingly, the cost of manufacturing and operating the platform is higher, and its popularity is limited. 
		\item \textbf{Weight:}
		The weight of both the payload and the airborne platform itself will significantly affect its endurance. The weight of the payload is a crucial performance indicator for an airborne platform. For a HAP, it is closely related to the locations of the pods mounted on the HAP, and a HAP can carry more payloads by deploying decentralized pods. The lighter platform also enables it to mount more payloads and extends the platform's endurance. In fact, the design and manufacturing of lightweight airborne platforms is a significant trend. In recent years, advanced lightweight materials such as carbon fiber have been widely utilized in NSIN. For example, we have utilized carbon fiber materials to manufacture the propeller blades of our airship, as shown in Fig. \ref{fig:fig_beihang_airship}.
  %which greatly enhance the propeller structure, reduce the weight of the propeller, and lower energy consumption. 
  The Aquila drone engineered by Meta has a tremendous wingspan (141 feet) compared to a Boeing 737's 113 feet. 
  Furthermore, being built with carbon fiber, Aquila weighs only around 900 pounds, which is approximately half the weight of a smart car.
% \begin{figure}[!t]
% 		\centering
% 		\subfigure[Structure of a blade of a propeller]{
%     \label{fig:subfig:f} %% label for first subfigure
%     \includegraphics[width=3.3 in]{Propeller_blade.pdf}}
%     \hspace{3pt}
%     \subfigure[Ground testing]{
%     \label{fig:subfig:g} %% label for first subfigure
%     \includegraphics[width=3.3 in, height=1.4in]{Blade_testing.pdf}}
% 		\caption{{Structure of a carbon fiber propeller blade of our airship, and a ground scaling test experiment.}}
% 		\label{fig:fig_Propeller_blade}
% 	\end{figure}

		\item \textbf{Mobility:}
		Mobility is one of the most important characteristics of an airborne platform, which contributes a lot to its popularity. 
		Compared to stationary platforms, mobile airborne platforms have many merits, e.g., they can improve the fairness of network service, proactively improve channel quality, enhance network performance, and accomplish 360$^{\circ}$ panoramic sensing and monitoring of specific areas or targets. 
		The degree of performance improvement is also related to the flexibility of the platform. Airborne platforms have some constraints on their moving directions and speeds, and the types of airborne platforms affect their maneuverability. For instance, compared to a LAUAV, a HAP has greater maneuverability limitations owing to its larger size and inertia. To this end, NSIN designers need to separately enforce mobility constraints for airborne platforms in NSIN. 
		\item \textbf{Power consumption:}
	Power consumption 
        constraints are key consideration during the process of designing NSIN and play a crucial role in prolonging its lifetime. Different types of airborne platforms have diverse requirements for power consumption constraints. In particular, small and battery-driven platforms have stricter power constraints.
		Although solar-powered airborne platforms are not much more energy-sensitive than battery-driven platforms, developers have to design energy-aware protocols for them such that they can operate normally during hard and long nights. 
		%Power consumption of airborne platforms in NSIN mainly includes consumption for accomplishing some tasks (e.g., communications, sensing, and computing) and propulsion power consumption. 
  To save power consumption for accomplishing tasks, NSIN developers should also pay attention to the design of energy-efficient management mechanisms. 
		The propulsion power consumption of airborne platforms can be much greater than that for accomplishing tasks. The amount of propulsion power consumption is closely related to the maneuverability of the platforms. Eq. (\ref{eq:HAP_propulsion_energy}) presents a propulsion power consumption model (in W) of an airship derived by the aerodynamic theory and computational fluid dynamics (CFD) numerical simulation results.
  \begin{equation}\label{eq:HAP_propulsion_energy}
P=\frac{\text{0}\text{.56}\rho^{\frac{5}{6}} {{V}}^{\frac{17}{6}}{D}^{\frac{2}{3}}\frac{0.\text{172}{{\lambda }^{\text{1/3}}}+\text{0}\text{.252}{{\lambda }^{-\text{1}.2}}\text{+1}\text{.032}{{\lambda }^{\text{-2}\text{.7}}}}{{{{(l/\mu)}}^{\text{1/6}}}}}{-\text{0}\text{.17}{V^{\text{-0}\text{.4}}}+\text{0}\text{.6205}}
\end{equation}
where $V$, $D$, and $\lambda$ represent the flight speed, the volume, and the slenderness ratio of the airship, {$l$ the length of the airship, $\mu$ the coefficient of kinetic viscosity}, and $\rho$ the density of air.
  Fig. \ref{fig:fig_HAP_power_vs_speed} and Fig. \ref{fig:fig_UAV_power_vs_speed} also illustrate the propulsion power consumption versus the platform flight speed of a HAP and a rotary-wing UAV, respectively. {From these figures, we can know that the movement control of airborne platforms is extremely energy-consumption.} Therefore, designers should take into account the energy-efficient movement control of airborne platforms when deploying NSIN to accomplish tasks. 
	\end{itemize}

	\subsubsection{Situation assistance}
	The NSIN is deployed in a rather complex 3-D airspace. To effectively support network construction, management, utilization, and operation, NSIN designers should be aware of, analyze, and exploit situational information related to NSIN. 
	
	The situation of NSIN mainly includes the network situation and the security situation. 
	The network situation of NSIN characterizes such NSIN statuses as resources, capabilities, capacity, and traffic. In terms of resources, the network situation involves the status of spectrum, power, storage, and platforms. Network capability status describes the communication, computing, sensing, PNT, and ISR capabilities of NSIN. 
 Capacity status includes link capacity, sub-network capacity, and network capacity. Regarding the traffic situation, the network situation describes the status of network and node traffic congestion, as well as the types of traffic present in the network, such as control or payload traffic. It is noteworthy that different types of traffic have distinct service requirements on NSIN, and the necessity of merging multiple different types of traffic into a network has been widely recognized.
	
	In a designated time and space range, elements (e.g., neighboring flight platforms, terrain, and meteorology) of the airspace environment may threaten the safe flight of an airborne platform. 
 The security situation refers to the spatial and temporal distribution of the level of threat to the safe flight of platforms.
	It is an important prerequisite for the safe and efficient sharing of airspace by multiple heterogeneous airborne platforms. 
	For instance, airborne platforms in NSIN need to fly to a target airspace autonomously and share the limited and complex airspace when carrying out the mission. Then, when planning the trajectories of airborne platforms, developers should consider cooperative planning with civil airspace, the avoidance of obstacles, and the collision with other flight platforms in the neighboring area.

	\subsubsection{Aviation regulations}
	Owing to safety, privacy, and harmonious operation concerns, the aeronautical regulation activities of NSIN are regulated by national civil aviation authorities. It indicates that NSIN must operate under civilian laws and be subject to civilian regulations and licensing to guarantee safe and harmonious operations. Civil aviation authorities of different countries may formulate different rules for NSIN. However, relevant rules are formulated mostly considering the following factors: 
	1) The categories, sizes, and weights of airborne platforms (e.g., airships, balloons, and aircraft). For instance, airborne platforms weighing more than 0.25 kg must be registered in most countries. 
	2) The pilot or control methods of airborne platforms, i.e., unmanned or manned. For example, current aeronautical regulations restrict unmanned HAPs from flying in civilian airspace, but the deployment of manned HAPs may be a step towards allowing HAPs in such airspace.
	3) The flight altitude and safe distance towards other objects, e.g., LAUAVs are not allowed to fly above 120 m in China. 
	4) The flight regions of airborne platforms, whether they are above an urbanized area, restricted areas, near national borders, or near airports. For instance, airborne platforms are prohibited from flying within five km of the national border in China. 
	5) The urgency of flight missions, e.g., the flight plan approval process will be simplified in some countries during the period of executing anti-terrorism and emergency missions.

	\subsubsection{Spectrum regulation of NSIN}
	At the World Radiocommunication Conference in 1997 (WRC-97), the frequency bands, i.e., 47.2-47.5 GHz (downlink) and 47.9-48.2 GHz (uplink), further confirmed at WRC-19, were first allocated to HAPs for worldwide usage. 
	This proposal was promoted by the US as the federal communications commission (FCC) approved the application of the Sky Station International Corporation to utilize this 2 × 300 MHz band for stratospheric application of Internet service and interactive video communication service relay in 1996. 
	Yet, concerning the great rain fading and immature technical solutions in the upper frequency bands, some lower frequency bands were agreed to be occupied by HAPs at WRC-2000. 
	Thereafter, technological advances and the pressing requirement of providing global broadband applications resulted in a review of the current spectrum regulatory provisions. At WRC-19, delegates agreed to allocate a portion of millimeter-wave (mmWave) frequency bands to HAPs for worldwide usage. 
	An ITU-R “work in progress'' report under WRC-23 agenda item 1.4 will explain the sharing and compatibility studies of HAPs as International Mobile Telecommunications (IMT) base stations for mobile services in some frequency bands below 2.7 GHz. The detailed spectrum regulation for HAP networks is summarized in Table \ref{table_spectrum}. 
	
	\begin{table*}[!t]
	\renewcommand{\arraystretch}{1.2}
	\caption{Spectrum regulation of HAP networks}
	\label{table_spectrum}
	\newcommand{\tabincell}[2]{\begin{tabular}{@{}#1@{}}#2\end{tabular}}
	\centering
	\begin{tabular}{l}
%\begin{minipage}{0.18\textwidth}
			\centering
			\includegraphics[width=7.0 in]{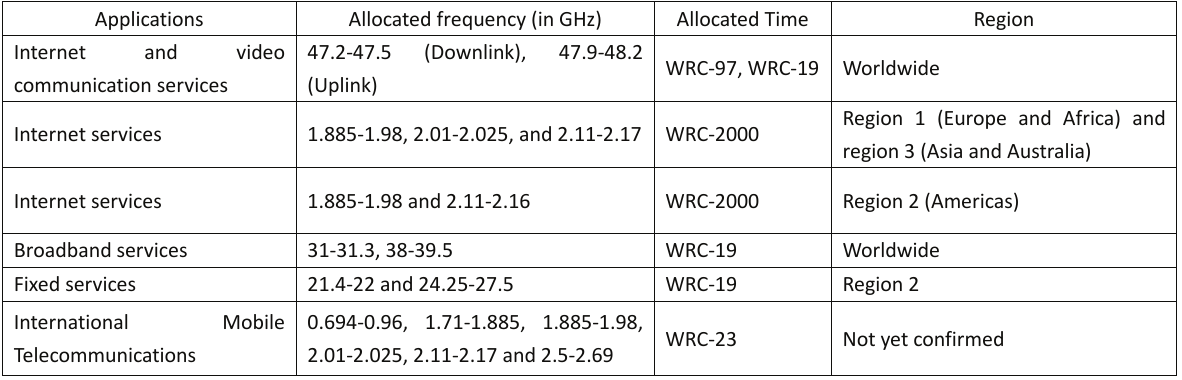}
		%\end{minipage} 
	\end{tabular}
\end{table*}

	As for UAV networks, the spectrum is allocated in accordance with usage.	
	The control and non-payload communication (CNPC) link is established to ensure the safe operation and efficient control of UAVs. 
	As shown in Table \ref{table_spectrum_UAV}, some frequencies for UAV CNPC links are allocated with some usage restrictions.
 Yet, partial frequencies fall in the bands allocated to fixed satellite services. Then, the sharing compatibility issue between UAV CNPC links and satellite-ground links should be tackled. 
	Payload links refer to the links used for other radio equipment than those for CNPC links. The payload devices (e.g., camera, repeater, and flying base station) are closely related to missions, which can transmit measurement and perception data from a UAV to the ground. 
	Table \ref{table_spectrum_UAV} also depicts the frequencies allocated to UAV payload links. 
\begin{table*}[!t]
	\renewcommand{\arraystretch}{1.2}
	\caption{Spectrum regulation of UAV networks}
	\label{table_spectrum_UAV}
	\newcommand{\tabincell}[2]{\begin{tabular}{@{}#1@{}}#2\end{tabular}}
	\centering
	\begin{tabular}{l}
%\begin{minipage}{0.18\textwidth}
			\centering
			\includegraphics[width=7.0 in]{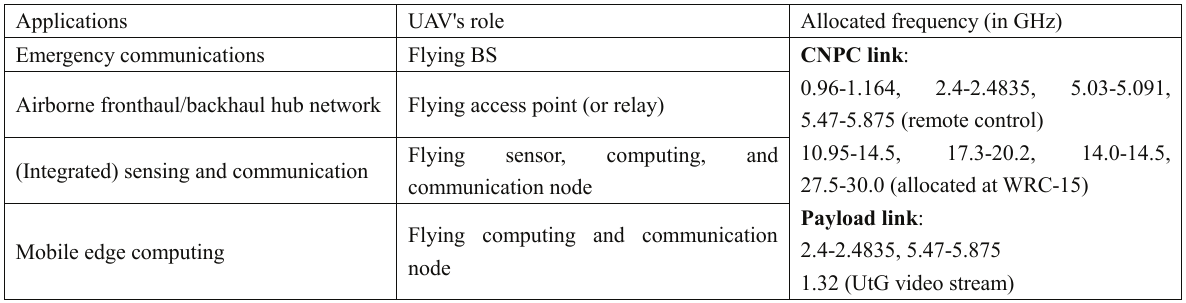}
		%\end{minipage} 
	\end{tabular}
\end{table*}
%	\begin{table*}[!t]
%	\renewcommand{\arraystretch}{1.2}
%	\caption{{{Spectrum regulation of UAV networks}}}
%	\label{table_1}
%	\newcommand{\tabincell}[2]{\begin{tabular}{@{}#1@{}}#2\end{tabular}}
%	\centering
%	\begin{tabular}{|l|l|l|}
%		\hline
%		{Applications} & {UAV's role} & {Allocated frequency (in GHz)}  \\\hline
		%% {\tabincell{l}{The minimum required data \\ rate of eMBB UEs in $s$}}
%		{Emergency communications} & {Flying BS} & {\multirow{5}{*}{{\tabincell{l}{\textbf{CNPC link:} \\ 0.96-1.164, 2.4-2.4835, 5.03-5.091, \\ 5.47-5.875, 10.95-14.5, and 17.3-30  \\ \textbf{Payload link:} \\ 2.4-2.4835, 5.47-5.875, and 1.32
%		}}}} \\ \cline{1-2}
%		{Airborne fronthaul/backhaul hub network} &  {Flying access point (or relay)}  & {} \\ \cline{1-2}
%		{(Integrated) sensing and communication} & {Flying sensor, computing, and} & {} \\ 
%		{} & {communication node} & {} \\ \cline{1-2}
%		{Mobile edge computing}  & {Flying computing and communication node}  & {} \\ \hline
		%\cline{1-4}
%	\end{tabular}
%\end{table*}
	
	\subsection{Promising Use Cases of NSIN}
	As presented above, airborne platforms in NSIN can fulfill various roles. By functioning as aerial macro-base stations, they have the ability to offer cost-effective communication services over a wide coverage area. Specifically, when the coverage radius of a HAP is 50 km and the maintenance cost is \$500 per hour, the communication cost will be \$1.325 million per gigabit per year. It indicates that the coverage cost per unit area of a HAP is only 6.7\textperthousand \  times that of a ground macro-base station. 
	They can act as mobile relays to connect isolated information islands, collect data, and deliver data back to TN or NTN gateways. 
	For instance, airborne relays can be deployed to facilitate communication between offshore vessels. Particularly, vessels within 200 km offshore can be interconnected by deploying two HAP relays when the coverage radius of a HAP is 50 km. Besides, by deploying 130 HAPs, any two vessels within 200 km offshore of China can be connected. 
	Airborne platforms can serve as aerial edge computing nodes \cite{DBLP:journals/iotj/ZhangWMGSH23,DBLP:journals/wcl/TraspadiniGGZ23,DBLP:journals/tvt/GongYWYDY23}, effectively reducing network latency by shifting computing capabilities from the cloud to the network edge. Enabling the ability of edge computing, one can regard NSIN as content delivery networks (CDN), which enable low-latency and high-interaction content delivery services. NSIN can provide network access services for rural or remote areas with inadequate communication infrastructures \cite{Zhang2023HAPenabledCI}.
	They can also act as caching nodes \cite{Abderrahim2023DataCH,Yuan2023JointME,DBLP:journals/cm/AbderrahimAS23}. 
	By caching content in NSIN, the network resource allocation and service abilities of data centers can be proactively optimized according to users' QoS requirements. 
		The aforementioned essential functions of airborne platforms provide NSIN with promising application prospects for various critical use cases, as discussed below. %, as depicted in Fig. \ref{fig:fig_use_case_NSIN}. 
	\textbf{Emergency communication:}
	Emergency communication is one of the most important goals of deploying NSIN. NSIN is resistant to natural disasters and geographical limitations. When TN are disrupted due to earthquakes, floods, mudslides, and other disasters, NSIN can be quickly deployed to recover communications and establish information transmission channels for emergency rescue. Additionally, NSIN can act as complement radio access networks to sustain services of TN when terrestrial flash crowd traffic occurs in densely populated areas \cite{DBLP:journals/cm/KementKJYSDZ23,DBLP:journals/twc/ZhouSLLH19}.%, as shown in Fig. \ref{fig:fig_emergency_comm}.
 % \begin{figure}[!t]
	% 	\centering
	% 	\includegraphics[width=1.5 in]{fig2_use_case_emergency_communication.pdf}
	% 	\caption{Emergency communication use case of NSIN.}
	% 	\label{fig:fig_emergency_comm}
	% \end{figure}
	
	\textbf{QoE enhancement:}
	One can deploy NSINs to assist TNs in enhancing users' service experience. Taking video transmission as an example, Intra-coded pictures (I frames) {are the base frames with a small compression ratio in the frame structure of video compression coding.}
 {The compression ratio of} Predictive-coded Pictures (P frames) {is greater than that of I frames.}
 Bidirectionally predicted pictures (B frames) {have the highest compression ratio, and the amount of information in P frames is usually about 2.5 times that of B frames. I frames and P frames are reference frames, the decoding failure for which affects other frames. 
However, decoding B frames that are not reference frames incorrectly will not cause the spread of decoding errors. 
Therefore, considering the diverse transmission performance requirements of video frames,} one can then choose to deliver I and P frames via TNs and transmit B frames through NSINs {when the proportion of B frames is not high}. 
 After combining I, P, and B frames at the terminal, users' service experience can be greatly enhanced.
		In addition, for applications involving video streaming and image rendering, real-time processes of such operations as video coding and decoding, orchestration, geometrical transformation, project transformation, and perspective transformation, are crucial. By shifting the rendering ability from the cloud to the network edge of NSINs, the process latency can be greatly reduced, and then the potential of enabling emerging time-critical applications such as extended-range (XR) or Metaverse via NSINs-assisted TNs may be unlocked \cite{DBLP:journals/network/SiQZL23, DBLP:conf/globecom/LuongSAFNK23}. 
	
	\textbf{Cargo-UAV applications:} 
	It is envisioned that numerous cargo-UAVs will fly in the airspace of densely populated cities. To ensure the safe operation of these cargo-UAVs, reliable and low-latency communication links among them and controllers (enforced by law) must be established, and a large amount of data about them should be collected and analyzed continuously and timely. HAPs in NSIN can be deployed to satisfy the above requirements perfectly. Owing to the high deployment altitude, HAPs can establish communication links of good quality with cargo-UAVs. Thus, controllers can control cargo-UAVs in a reliable and low-latency manner via HAPs. The on-board processors of HAPs also have strong computing and storage capabilities, which can handle the large amount of data generated by cargo-UAVs timely. 
	
	\textbf{ITS applications:}
	NSIN can provide improved coverage and surveillance services for intelligent transportation systems (ITS). UAVs in NSIN can achieve the goal of local service enhancement, and HAPs in NSIN can provide wide coverage services. Take railway transportation as an example, NSIN will become a new regime for continuously monitoring the surroundings of railways and detecting abnormal events around railway tracks. The service requests from high-speed cars and trains in ITS are increasing with the proliferation of ITS services. However, it is hard for stand-alone TN to provide low-latency and continuous coverage services for these transportation vehicles. To satisfy their requirements, these vehicles can establish connections with suitable network platforms in NSIN. For instance, high-speed cars and trains do not need to frequently hand over between access points by accessing HAPs. This reduces the signaling overhead and handover latency significantly. 
	
	\textbf{Communications among Non-Terrestrial Platforms:}
	As is known, 71\% of the earth’s surface is covered by the ocean. Humans have only explored and charted 5\% of the ocean, and there are rich resources waiting to be explored in the ocean. As a result, more and more fixed or mobile offshore platforms (e.g., gas and oil rigs, scientific expeditions, and surveillance platforms) are being established to exploit the abundant resources on the seabed and explore the unknown environments.  
 Nowadays, these platforms exchange data with their terrestrial controllers through renting satellite links, which inevitably results in long transmission latency. 
	NSIN, which is much closer to these platforms, can be deployed as relay networks to establish the connection between them and their controllers. 
	In this way, the end-to-end (E2E) transmission latency among these platforms and controllers can be significantly reduced. 
	
\begin{figure}[!t]
		\centering
		\includegraphics[width=2.8 in]{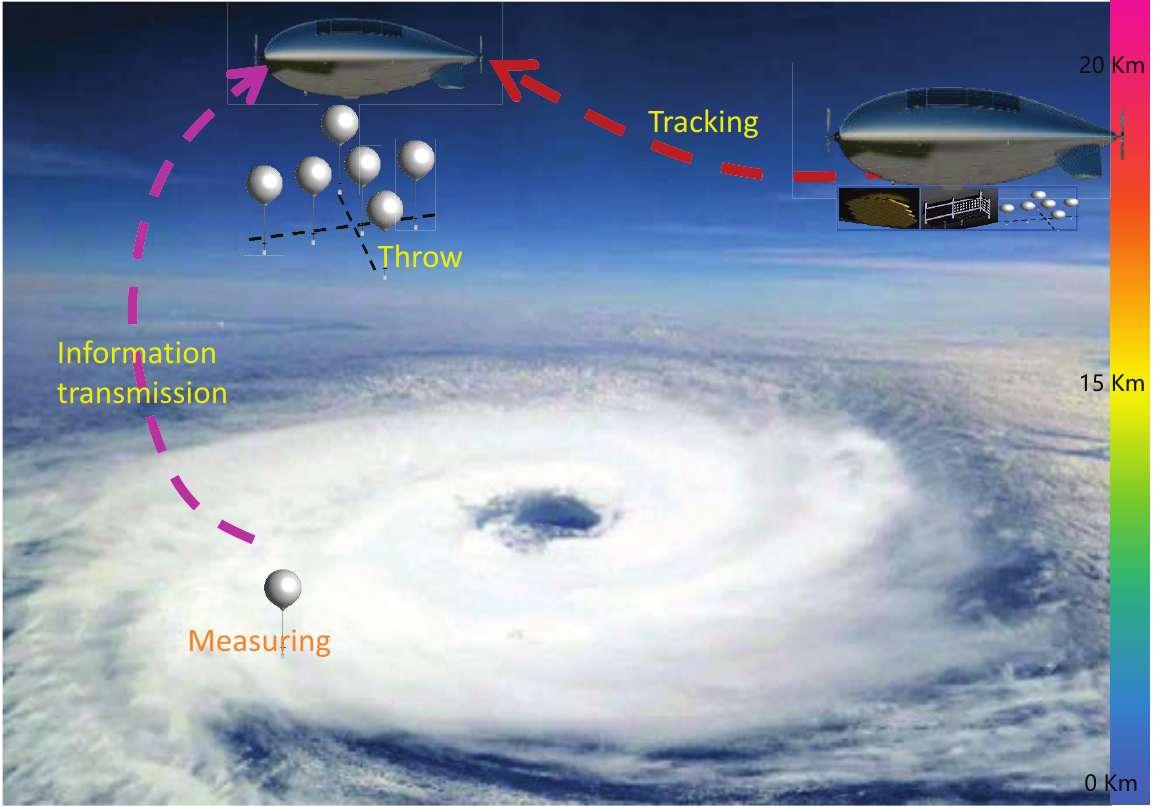}
		\caption{{A scenario of monitoring typhoons using NSIN.}}
		\label{fig:NSIN_typhoon}
	\end{figure}

	\textbf{Meteorological monitoring:}
	Meteorological monitoring (e.g., typhoon monitoring) is of great importance in both civilian and military research fields. Most existing meteorological monitoring systems are space- or ground-based. However, space-based systems face limitations in ensuring continuous monitoring due to satellite orbit constraints. Ground-based systems have limited detection ranges and poor flexibility. NSIN-based monitoring systems can overcome these shortcomings by offering continuous monitoring capabilities throughout the entire life cycle of climate phenomena like typhoons, snow, and rainfall. Therefore, the development of NSIN-based meteorological monitoring systems is expected to be a significant trend in the near future. {Fig. \ref{fig:NSIN_typhoon} shows a proposal of monitoring typhoons we are working on using NSIN. }
%{Fig. \ref{fig:NSIN_ITs} illustrates a scenario in a practical program we have undertaken. We attempt to monitor complex ground traffic using multiple cooperative airships and UAVs in the program.}
	
	\textbf{Proliferation of IoT services:} 
	Existing terrestrial infrastructures are limited and increasingly unable to support the highly demanding IoT services. For example, more and more IoT devices are expected to be deployed in such areas as forests, mountains, and oceans where there is no or rare TN coverage. In this context, NSIN is envisioned as a key infrastructure to complement existing terrestrial infrastructures to collect data or offload computationally intensive tasks from IoT devices and charge them in an efficient and cost-effective way \cite{DBLP:journals/iotj/JiaWDYH23,DBLP:conf/icoin/EiAPHH23,DBLP:journals/iotj/LakewTDC23,DBLP:journals/iotj/KangCMMFL23}.

	\subsection{Summary and Lessons Learned}
 \textbf{Summary:} NSINs are significant heterogeneous and multi-layered networks consisting of the HAP, HAUAV, and LAUAV network segments. 
 There are many notable differences between a HAP and a UAV, which prevent most protocols designed for UAV networks from being applied directly to NSINs. 
 Researchers and designers should conceive protocols and network management architectures for NSINs that take full account of the heterogeneity among airborne platforms as well as their integration. 
 Besides, there are many promising use cases for NSINs where HAP networks play an irreplaceable and fundamental role. 

 \textbf{Lessons learned:} However, the impact of the unstable movement of HAP on the performance of HAP networks is seldom studied. 
	%The monitoring data from our field experiments showed that an airship rotated around one revolution per minute, and its pitch angle could vary by up to about +/-30$^{\circ}$. 
 The unstable movement of the platform can result in position shifting and deformation of the HAP's footprint.
 As a result, users close to the boundary of the footprint will frequently trigger handoffs, and their QoS requirements are hard to be satisfied. 
	A HAP moves with six degrees of freedom, i.e., X-direction, Y-direction, Z-direction, yaw, roll, and pitch.
 The movement of the HAP along the X- and Y-directions will not affect the size of the footprint, but rather its location. 
 The movement along the Z-direction will result in the expansion or contraction of the covered area. 
 Further, the changes in its footprint are closely related to the installed antenna type of the HAP. 
 For instance, given a high-directional antenna, the area of the coverage area expands with the uplift of the HAP; however, in the case of a low-directional antenna (or omnidirectional antenna), the area of the coverage area shrinks with the uplift of the HAP. 
 Meanwhile, {the changes in the attitude angles of the airship may cause the changes in the beam.} Thus, novel beamforming techniques should be designed to withstand the adverse effects of {attitude angle variations.} 
  
	\section{Channel Modeling of NSINs}
A good understanding of the channel characteristics of NSINs, key physical parameters of the channel, and channel modeling issues lays the foundation for the design of NSINs. 
The signal propagation environment of NSINs is quite different from that of TNs, {space} networks, and UAV networks. 
%The NSINs' propagation environment goes through the stratosphere and troposphere, where the temperature, humidity, and atmospheric pressure vary greatly.
The platforms in NSINs also have special SWMAP characteristics. As a result, existing channel models of terrestrial, {space}, and UAV networks may not be directly applied to model channels of NSINs.

{In this section, we aimed at revealing some practical factors affecting the channel modeling of NSINs via experiments and discussing the existing channel models customized for NSINs.} 

%first discuss the phase delays of different types of antenna arrays onboard NSINs' platforms.
%Next, we discuss the channel propagation characteristics of NSINs from both large-scale fading and small-scale fading perspectives. 
%Third, we provide a detailed explanation of the existing channel models customized for NSINs.

\subsection{Problems Encountered in Experiments}
\begin{figure}[!t]
		\centering
		\subfigure[{Process of dropping sounders}]{
    \label{fig:subfig:f_w} %% label for first subfigure
    \includegraphics[width=1.6 in, height = 2.0 in]{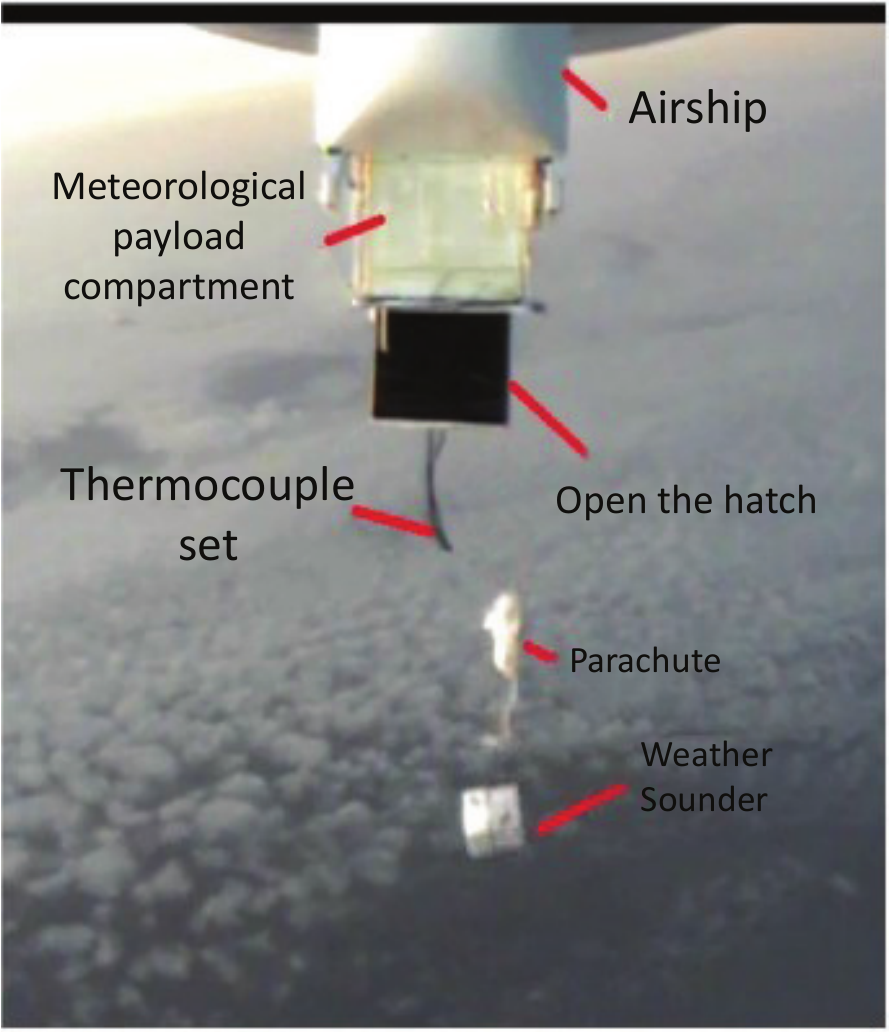}}
    \hspace{3pt}
    \subfigure[{Recorded humidity data}]{
    \label{fig:subfig:g_w} %% label for first subfigure
    \includegraphics[width=2.1 in]{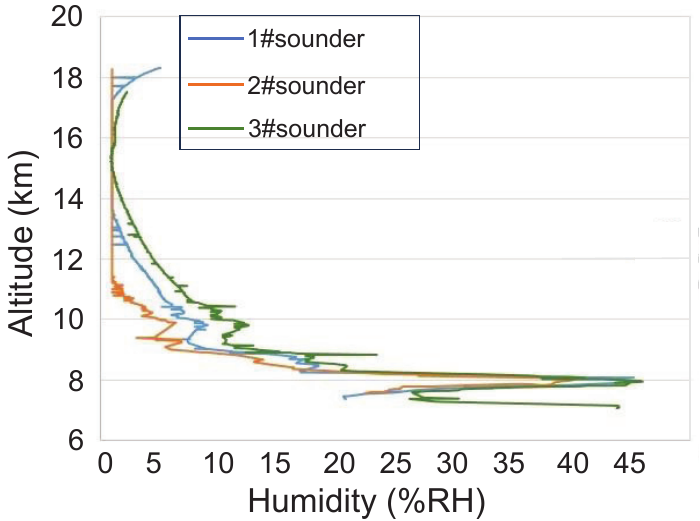}}
		\caption{{A  meteorological measurement experiment via an airship.}}
		\label{fig:meteorological_measurement}
	\end{figure}

\begin{figure}[!t]
		\centering
		\includegraphics[width=3.1 in]{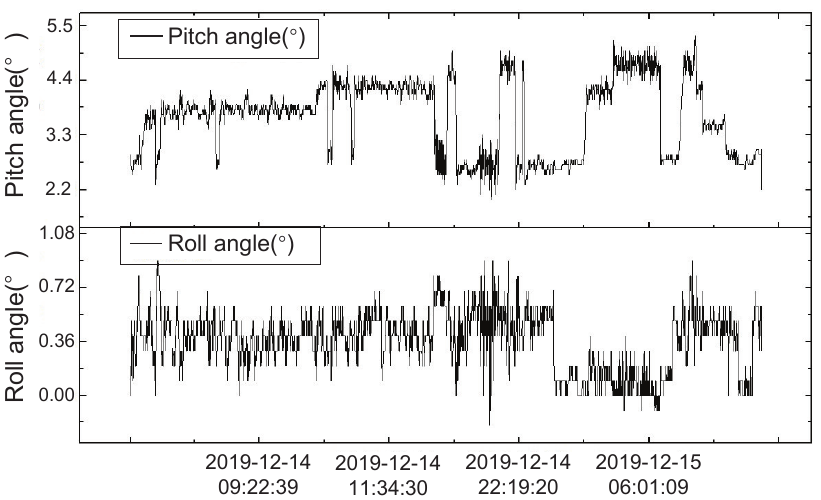}
		\caption{{Attitude changes of an airship.}}
		\label{fig:attitude_change}
	\end{figure}

{The authors in this article conducted an experiment to measure high-altitude meteorological data for a region of China using the Beihang airship, as shown in Fig. \ref{fig:subfig:f_w}. The measured high-altitude meteorological data supported the investigation of NSIN channels.
Some influence factors that should be considered in modeling channels of NSINs were revealed in the experiment. For instance, 
the humidity of NSINs' propagation environment affecting the signal fluctuation varies greatly, as depicted in Fig. \ref{fig:subfig:g_w}.
The attitude angles of the airship significantly affecting channel models of NSINs changed with time. Fig. \ref{fig:attitude_change} illustrates the attitude changes of the Beihang airship caused by mechanical vibrations and airflow over a period of time. From this figure, it can be observed that the pitch and roll angles of the airship can vary by 3.3$^{\circ}$ and 1$^{\circ}$.} 
%The NSINs' propagation environment goes through the stratosphere and troposphere, where the temperature, humidity, and atmospheric pressure vary greatly.
%Besides, Fig. \ref{fig:yaw_change} shows that sudden airflow has a significant affect on the yaw angle of airship, and the change is more than 5$^{\circ}$.
%This type of discovery will benefit for the study on the channel modeling of NSINs. 
{With the influence factors in mind, we next discuss how to model the channels of NSINs.} 
%In this experiment, Many problems were During the experiments, 
%we discuss the channel propagation characteristics of NSINs from both large-scale fading and small-scale fading perspectives. 
%We therefore conducted an experiment to test the realistic effects of these factors on an airship.
 %{The experimental results justify the necessity of investigating the phase delay of antennas mounted on a HAP.} 
 %Consequently, the locations or shapes of cells will change, as illustrated in Figure \ref{fig:footprint}. 
	\subsection{Phase Delay}
	Let ${\bm r}_{m,n} = [x_m, y_n, z_{mn}]^{\rm T}$ denote the location vector of the $(m, n)$-th array element with respect to (w.r.t) a reference array element and ${\bm d} = [\cos \varphi \sin \theta ,\sin \varphi \cos \theta ,\cos \theta ]^{\rm T}$ the DoA unit vector with $\theta$ and $\varphi$ representing the elevation angle and the azimuth angle, respectively, the phase delay of the $(m, n)$-th array element w.r.t the reference array element can then be given by {${\tau _{m,n}} = -\beta \left\langle {{\bm d},{{\bm r}_{m,n}}} \right\rangle $, where $\left\langle  \cdot  \right\rangle $ represents the inner product operation and $\beta = 2\pi /\lambda $ with $\lambda$ being the carrier wavelength. 	
	The $(m, n)$-th element of the array matrix can be expressed as $[\bm a(\theta ,\varphi )]_{m,n} = \exp (j{\tau _{m,n}})$.}
	
	Unfortunately, an airborne array antenna will jitter due to the unstable movement of the airborne platform, which challenges the analysis of the phase delay. 
	In this subsection, we discuss the impact of jittering on phase delay. 
	Denote $\bm R$ as a rotation matrix capturing the attitude of an airborne platform, which can be measured by yaw, pitch, and roll angles. %; we then have ${\bm R} = {\bm R_{yaw}}(\alpha ){\bm R_{roll}}(\beta ){\bm R_{pitch}}(\gamma )$,
	%The rotation matrix can be characterized by Euler angles, i.e., yaw, pitch, and roll, which are shown in Fig. and can be written as
	%where $\alpha$, $\beta$, and $\gamma$ represent the yaw angle, roll angle, and pitch angle of an airborne platform, ${\bm R_{yaw}}(\alpha )$, ${\bm R_{roll}}(\beta )$, ${\bm R_{pitch}}(\gamma )$ are rotation matrixes 
	% \begin{equation}
	% 	{{\bm R}_{yaw}}(\alpha ) = \left( {\begin{array}{*{20}{c}}
	% 			{ + \cos \alpha }&{ - \sin \alpha }&0\\
	% 			{ + \sin \alpha }&{ + \cos \alpha }&0\\
	% 			0&0&1
	% 	\end{array}} \right),
	% \end{equation}
	% \begin{equation}
	% 	{{\bm R}_{roll}}(\beta ) = \left( {\begin{array}{*{20}{c}}
	% 			{ + \cos \beta }&0&{ + \sin \beta }\\
	% 			0&1&0\\
	% 			{ - \sin \beta }&0&{ + \cos \beta }
	% 	\end{array}} \right),
	% \end{equation}
	% \begin{equation}
	% 	{{\bm R}_{pitch}}(\gamma ) = \left( {\begin{array}{*{20}{c}}
	% 			1&0&0\\
	% 			0&{ + \cos \gamma }&{ - \sin \gamma }\\
	% 			0&{ + \sin \gamma }&{ + \cos \gamma }
	% 	\end{array}} \right).
	% \end{equation}
	%Define the $(i,j)$-th element of ${\bm R}$ by $R_{ij}$ $\forall i, j \in \{1,2,3\}$. 
	{The impact of attitude changes on the array element can be modeled as ${\bm r}_{m,n}^{\rm true} = {\bm R} {\bm r}_{m,n}$, and denote $R_{ij}$ as the $(i,j)$-th element of ${\bm R}$.}
	%With ${\bm r}_{m,n,o}^{\rm true}$, the true phase delay under the platform jittering impact can be obtained. 
	%Besides, the effects of platform motion inherent to NSIN may induce Doppler effect. The Doppler shift can be classified into bulk doppler shift and doppler spread due to maneuvering (such as roll, pitch, and yaw of platforms), Doppler spread due to scattering. For high-velocity airborne vehicles, the bulk Doppler shift will be more than tens of times of the latter two types of Doppler spread \cite{DBLP:journals/taes/OgbeLRB19}. For simplicity, the bulk Doppler shift is assumed to dominate the Doppler effect In NSIN. 
	%Denote $f_d = f_c v/c$ as the frequency dispersion due to Doppler effect, where $f_c$ is the carrier frequency, $v$ is the relative speed, $c$ is the speed of light. Then, the Doppler effect on the channel modeling can be written as ${\rm e}^{j2 \pi f_dt}$ with $t$ being a time slot. 
	We can then summarize the expressions of phase delays and steering matrixes of three typical types of antenna arrays, including linear antenna arrays, rectangular antenna arrays, and circular antenna arrays in Table \ref{table_array_antenna_violation}.
	{The reader interested in studying NSIN channel models can refer directly to this table for the modeling of steering vectors/matrices.}
	\begin{table*}[htp]
		\centering
		\caption{A list of phase delays and steering matrices under three typical antenna arrays}
  \label{table_array_antenna_violation}
		\newcommand{\tabincell}[2]{\begin{tabular}{@{}#1@{}}#2\end{tabular}}
		\resizebox{\textwidth}{!}{
			\begin{tabular}{ | c | c | c | }
				\hline
				Array structure & Phase delay & Steering vector/matrix \\ \hline
				%  ULA antenna
				\begin{minipage}{0.25\textwidth}
					\centering
					\includegraphics[width=1.1 in,height= 0.8in]{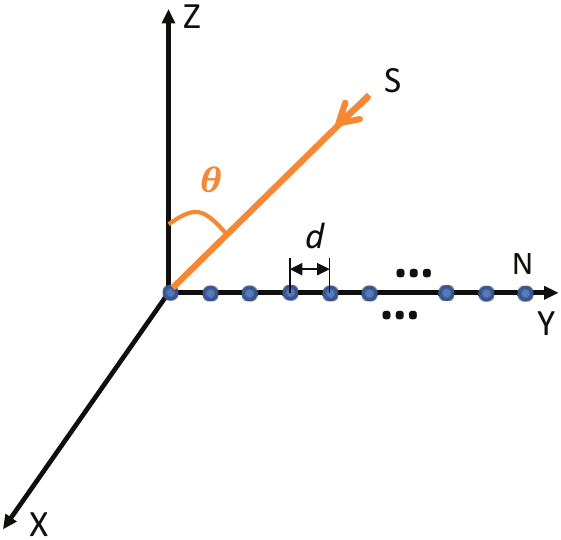}
				\end{minipage}
				& \tabincell{l}{ Case I (stable attitude) {\cite{an2024energy}}: \\ $\tau_n = -\beta(n-1)d \sin \theta $ \\ 
    Case II (attitude changes): \\ $\tau_n =  - \beta (n - 1)d({R_{22}}\sin \theta  + {R_{32}}\cos \theta )$} & $[{\bm a}(\theta)]_{n} = {\frac{1}{{\sqrt N }}}{\rm e}^{-j\tau_n}$
				\\ 
				\hline 
				%  UPA antenna
				\begin{minipage}{0.25\textwidth}
					\centering
					\includegraphics[width=1.4 in]{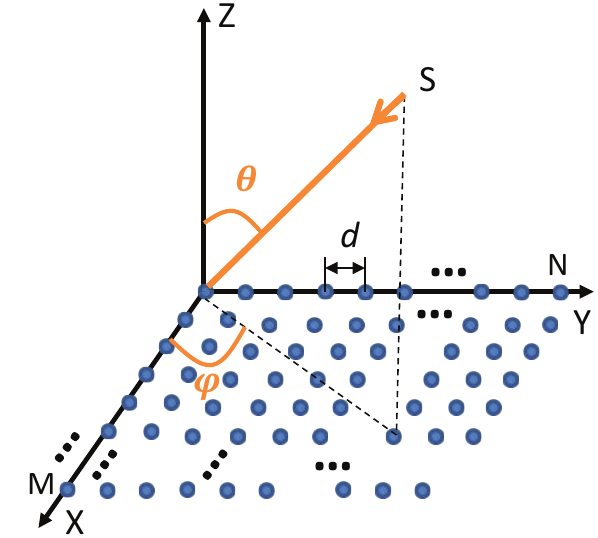}
				\end{minipage} &  \tabincell{l}{Case I (stable attitude) {\cite{DBLP:journals/tsp/LinLHCZ19}}: \\ $\begin{array}{l}
						\tau _m  =  - \beta (m - \frac{{M - 1}}{2})d\cos \varphi \sin \theta \\
						%\begin{array}{*{20}{c}}
							{\tau _n =  { - \beta (n - \frac{{N - 1}}{2})d\sin \varphi \sin \theta }}
						%\end{array}
					\end{array}$ \\ 
     Case II (attitude changes): \\ $\begin{array}{l}
						{\tau _m}  = {\begin{array}{*{20}{c}}
								\begin{array}{l}
									- \beta (m - \frac{{M - 1}}{2})d({R_{11}}\cos \varphi \sin \theta \\
									+ {R_{21}}\sin \varphi \sin \theta  + {R_{31}}\cos \theta )
								\end{array}
						\end{array}}\\
						% \begin{array}{*{20}{c}}
						% 	{}
						% \end{array}\begin{array}{*{20}{c}}
						% 	{}
						% \end{array}
      {\tau _n} = {\begin{array}{*{20}{c}}
								\begin{array}{l}
									- \beta (n - \frac{{N - 1}}{2})d({R_{12}}\cos \varphi \sin \theta \\
									+ {R_{22}}\sin \varphi \sin \theta  + {R_{32}}\cos \theta )
								\end{array}
						\end{array}}
					\end{array}$ }  & \tabincell{l}{$\begin{array}{*{20}{l}}
						{{\bm A}(\theta ,\varphi ) = }&{{{\bm a}_x}(\theta ,\varphi ) \times }\\
						{}&{{\bm a}_y^{\rm T}(\theta ,\varphi )}
					\end{array}$ \\ where, \\ ${[{{\bm a}_x}(\theta ,\varphi )]_m} = \frac{1}{{\sqrt M }}{e^{ - j{\tau _m}}}$ \\ ${[{{\bm a}_y}(\theta ,\varphi )]_n} = \frac{1}{{\sqrt N }}{e^{ - j{\tau _n}}}$ } \\
				\hline 
				% UCA antenna
				\begin{minipage}{0.25\textwidth}
					\centering
					\includegraphics[width=1.4 in]{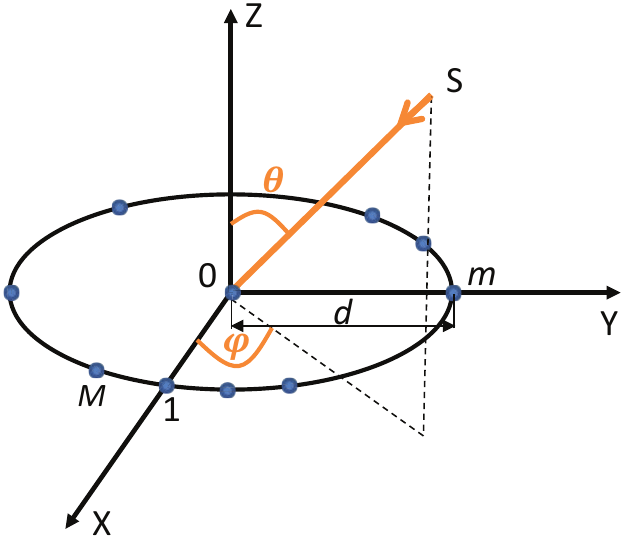}
				\end{minipage}
				& \tabincell{l}{Case I (stable attitude) {\cite{DBLP:journals/tsp/LinLHCZ19}}: \\ $\begin{array}{l}
						{\tau _1} = 0\\
						{\tau _m} =  - \beta d\sin \theta \cos (\frac{{2\pi (m - 2)}}{{M - 1}} - \varphi ),\\
						\begin{array}{*{20}{c}}
							{}&{\forall m \in \{ 2, \ldots, M - 1\} }
						\end{array}
					\end{array}$ \\ Case II (attitude changes): \\ $ \begin{array}{l}
						{\tau _1} = 0\\
						{\tau _m} =  - \beta d\cos (\frac{{2\pi (m - 2)}}{{M - 1}})({R_{11}}\cos \varphi \sin \theta \\
						\begin{array}{*{20}{c}}
							{}&{ + {R_{21}}\sin \varphi \sin \theta  + {R_{31}}\cos \theta )}
						\end{array}\\
						\begin{array}{*{20}{c}}
							{}&{ - \beta d\sin (\frac{{2\pi (m - 2)}}{{M - 1}})({R_{12}}\cos \varphi \sin \theta }
						\end{array}\\
						\begin{array}{*{20}{c}}
							{}&{ + {R_{22}}\sin \varphi \sin \theta  + {R_{32}}\cos \theta ),}
						\end{array}\\
						\begin{array}{*{20}{c}}
							{}&{\forall m \in \{ 2, \ldots, M - 1\} }
						\end{array}
					\end{array}$ }  & \tabincell{l}{$[{\bm a}(\theta, \varphi)]_{1} = {\frac{1}{{\sqrt M }}}$ \\ $[{\bm a}(\theta, \varphi)]_{m} = {\frac{1}{{\sqrt M }}}{\rm e}^{-j\tau_m}$, \\ $\forall m \in \{2, \ldots, M-1\}$}
				\\ 
				\hline
			\end{tabular}
		}
	\end{table*}

	%1)	Large-scale fading, due to path loss of signal as a function of distance and shadowing by large objects such as buildings and hills. This occurs as the mobile moves through a distance of the order of the cell size, and is typically frequency independent.
	
	%2)	Small-scale fading, due to the constructive and destructive interference of the multiple signal paths between the transmitter and receiver. This occurs at the spatial scale of the order of the carrier wavelength, and is frequency dependent. 
	
	\subsection{Large-Scale Fading}
	Large-scale fading happens because electromagnetic (EM) waves between Tx and Rx are blocked and/or shadowed by large objects (e.g., buildings, hills, and trees). It therefore includes free space path loss (FSPL) and shadow fading. The strength of large-scale fading is closely related to the distance between Tx and Rx and is typically frequency-independent. 
	Besides, rain, atmospheric gases, and scintillation may lead to large signal fluctuations.
	
	\subsubsection{Free space path loss}
	As the EM wave spreads from Tx to Rx over an increasing area, the transmitted signal attenuates with increasing distance.
    {The results of channel measurements, as shown in Fig. \ref{fig:channel_measurement}, indicate that the value of FSPL (in dB) can be calculated by \cite{DBLP:journals/wcl/LiuZSSCL21}
    \begin{equation}
		FSPL = 10 \alpha \log_{10}(d_{\rm 3D}) + EP_a
	\end{equation}
    where $\alpha$ is the path loss exponent, $d_{\rm 3D}$ the distance between a user and an airborne platform, $EP_a$ the extra path loss, depending on carrier frequency, flight altitude of the airborne platform, and the topography the propagation environment.}
 \begin{figure}[!t]
		\centering
		\subfigure[Fixed-wing UAV]{
    \label{fig:subfig:f_m} %% label for first subfigure
    \includegraphics[height = 1.3 in]{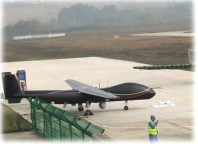}}
    \hspace{3pt}
    \subfigure[Measurement scenario]{
    \label{fig:subfig:g_m} %% label for first subfigure
    \includegraphics[height=1.4in]{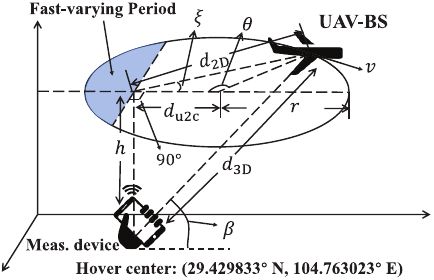}}
		\caption{{A channel measurement experiment \cite{DBLP:journals/wcl/LiuZSSCL21}.}}
		\label{fig:channel_measurement}
	\end{figure}
	%Given a propagation distance $d$ (in meters) and frequency $f_c$ (in GHz), FSPL (in dB) is computed by
	% For a ground user, the distance $d$ determined by HAP deployment altitude $h_0$ and an elevation angle $\alpha$ can be computed by
	% \begin{equation}
	% 	d = \sqrt {R_E^2{{\sin }^2}\alpha  + h_0^2 + 2{h_0}{R_E}}  - {R_E}\sin \alpha 
	% \end{equation}
	% where $R_E$ is Earth radius.
	
	\subsubsection{Shadow fading}
 {The
 shadow fading is closely related ground users' elevation angles towards an airborne platform.} 
Shadowing is the deviation of the received power of an EM signal from its averaged value. 
{Then, shadow fading is usually modeled using a statistical distribution. 
Experimental results in \cite{DBLP:journals/wcl/LiuZSSCL21} showed that depending on platform's altitude, it is suitable to model shadow fading using generalized Gamma distribution and Log-normal distribution. Besides, both the mean and the variance of shadow fading depend on the elevation angle, and a large angle results in a small mean and variance.}
	%One can model shadow fading (in dB) using a zero-mean normal distribution with a standard deviation $\sigma_{SF}^2$, i.e., $N(0, \sigma_{SF}^2)$, where $\sigma_{SF}$ depends on the elevation angle, carrier frequency, environment (e.g., urban and dense urban), and the probability of LoS or non-line-of-sight (NLoS) propagation.
	
	\subsubsection{Atmosphere attenuation}
	The atmospheric attenuation model characterizes the EM wave absorption of atmospheric gases. Its strength depends on the carrier frequency, elevation angle, transceiver deployment altitude above sea level, and absolute humidity.
	This type of attenuation should be considered at frequencies above ten GHz. Nevertheless, one should also calculate this type of attenuation at any frequency above one GHz when the elevation angle between Tx and Rx is below ten degrees. 
	ITU-R P.676 (in Annex 1) outlines in detail the computing procedure of atmospheric attenuation \cite{Attenuation2022ITU}.
	
	\subsubsection{Rain attenuation}
	Rain attenuation is a phenomenon relative to the EM wave frequency and the rainfall rate. Rain causes attenuation in EM waves through the processes of absorption and scattering.
    Rain attenuation is a critical factor to consider for frequencies above six GHz. As for signal transmission in typhoon environments, it is essential to calculate the rain attenuation for frequencies in the several hundred MHz range, as the rainfall rate can reach up to 72.8 mm/h \cite{Stories2020CWB}.

    ITU-R P.618 \cite{Propagation2017ITU} described a comprehensive approach for estimating the long-term statistics of rain attenuation, taking into account the factors of frequency, rainfall rate, elevation angle, and polarization. In addition to this, modeling rain attenuation can be achieved through the application of a synthetic storm technique and the utilization of raindrop size distribution models like lognormal and optimized models \cite{Afullo2021ANOO}.
	However, the mentioned rain attenuation models are independent of time series. As a result, it is not enough to design NSIN (especially the PHY and MAC layers of NSIN) using these models. Then, one should propose time series rain attenuation models. 
	ONERA $n$-state Markov chain and the German Aerospace Research Center (DLR) rain attenuation model are two types of well-known models of NSIN and satellite communications. 
	There are three components in an ONERA model. The $1^{\rm st}$ one describes a coarse attenuation time series with values of ``rain" and ``no rain". The $2^{\rm nd}$ one is a microscopic model describing attenuation time series with high resolution. The $3^{\rm rd}$ one combines the two obtained time series to generate a complete long-term attenuation time series \cite{fiebig2003review}. 
	The DLR model comprises a generic component and a specific set of parameters. These parameters enable the adjustment of the time series generator to simulate scenarios with different frequencies, elevation angles, and climatic zones \cite{fiebig2003review}.
	
	\subsubsection{Tropospheric scintillation attenuation}
	Tropospheric scintillation attenuation indicates rapid amplitude and phase fluctuations in received signals. 
 The NSINs' propagation environment goes through the troposphere, where the temperature, humidity, and atmospheric pressure vary greatly.
	Sudden changes in the refractive index caused by variations in temperature, humidity, and barometric pressure will result in signal fluctuations. 
	Tropospheric propagation should be considered for frequencies above six GHz, and the tropospheric scintillation effect increases with frequency, being especially significant above ten GHz. 
 %{Fig. \ref{fig:meteorological_measurement} depicts an experiment of dropping three meteorological sounders from an airship in a row and the recorded humidity data of them. Experimental results show the significant variation of water vapor content.} 
 Additionally, the scintillation effect becomes more pronounced at lower elevation angles due to longer propagation paths.
	The ITU-R P.618 presents an accurate prediction method for the scintillation attenuation amplitude \cite{series2017propagation}. It shows that 
	the tropospheric scintillation attenuation may be dozens of dB in the case of a low elevation angle. The scintillation attenuation can be negligible under a high elevation angle case (e.g., $> 10^{\circ}$).
	Nevertheless, 
 %when it rains one should also consider the tropospheric scintillation attenuation \textcolor{blue}{in the case of a high elevation angle during rain.} 
 when it rains, tropospheric scintillation attenuation at high elevation angles (e.g., $> 10^{\circ}$) should also be considered.
	
	\subsection{Small-Scale Fading}
	Small-scale fading is caused by the constructive and destructive interference of multi-path signals between Tx and Rx. It depends on the carrier frequency and occurs at a spatial scale approximately equaling to the wavelength of the signal. Small-scale fading characterizes the rapid fluctuation in received signal strength over short distances and short time periods. Small-scale fading includes flat fading, frequency selective fading, fast fading, and slow fading, which are affected by the EM propagation environment. 
	
	\subsubsection{Flat fading}
	In flat fading, the signal amplitude remains unchanged during each symbol duration, but the amplitude and phase of the symbol may change over short distances.
	To mitigate the impact of flat fading, some techniques, such as diversity reception and error correction coding, should be utilized.
	
	In flat fading, all frequency components of the received signal fluctuate simultaneously and in the same proportion. As a result, this type of fading is also called non-selective fading.
    One can model the flat fading channel using a two-state model. This model distinguishes two states: a good state corresponding to LoS and slightly shadowed channel conditions, and a bad state corresponding to severe shadowed channel conditions. A semi-Markov model can capture the state duration. 
    One can utilize a Loo distribution to characterize the fading in each state. The received signal consists of the direct path signal and multi-path components. 
    3GPP TR 38.811 provides system-level evaluations of flat fading \cite{Study20203GPP}.
	
	\subsubsection{Frequency selective fading}
    Frequency selective fading occurs when the symbol duration is shorter than the delay spread. 
	It may lead to inter-symbol interference (ISI). An equalization technique should be employed to mitigate the impact of this type of fading. This type of technique utilizes a digital filter at Rx to counteract the channel effect.
	Frequency selective fading, as the name indicates, affects diverse spectral components of signal with varying amplitudes.
	After modifying some crucial parameters related to NSIN scenarios, the procedure presented in subsection 7.5 of 3GPP TR 38.901 \cite{Study20171003GPP} should be utilized to model the frequency selection fading channel. The elevation angle between a ground user and a HAP is a crucial parameter. 
 NSIN scenarios with fewer clusters than those specified in 3GPP TR 38.901 require the inclusion of angular scaling factors in cluster generation \cite{Study20171003GPP}.% (Table 6.7.2-1aa in 3GPP TR 38.811 corresponds to Table 7.5-2 in 3GPP TR 38.901 and Table 6.7.2-1ab in 3GPP TR 38.811 corresponds to Table 7.5-3 in 3GPP TR 38.901). 
	
	\subsubsection{Fast fading}
	In a fast fading channel, the channel impulse response exhibits rapid variations within the symbol duration.
	Fast fading distorts the shape of the baseband pulse, resulting in linear distortion that causes inter-symbol interference (ISI). Adaptive equalization helps reduce ISI by eliminating the linear distortion induced by the channel.
	
	\subsubsection{Slow fading}
	In a slow fading channel, the CIR changes at a much slower rate than the transmitted baseband signal.
	One can utilize receiver diversity and error correction coding techniques to combat the effects of slow fading.

%	\subsection{Doppler Shift}
%	The effect of Doppler shift
%	What causes the Doppler shift
%	How to model
%	How to overcome or tackle Doppler shift
%	
%	\subsection{Scattering and Blockage Effect}
%	The effect of scattering and blockage effect
%	What causes the
%	How to model
%	How to tackle, as well as the challenges 

	\subsection{Existing Channel Models of NSIN}
	To fully comprehend the capabilities of NSIN, it is crucial to have a profound understanding of NSIN channel models.
    This paper focused on summarizing and analyzing NSIN radio-frequency (RF) channel models, and the reader can refer to \cite{DBLP:journals/jsac/CaoYAXWY18} for HAP free space optical (FSO) channel models. 
	Besides, NSIN channel modeling includes HAP channel modeling and UAV channel modeling, where UAV channel modeling has been extensively studied. In this subsection, we focus on the discussion of HAP channel modeling and analysis of the differences between HAP channel models and UAV channel models. 
	%Modeling and Parameter Estimation
	%Describe high accuracy channel modeling and universal channel modeling of NSIN.
	
	%3GPP-1 presents the comprehensive channel model for HAP communications. 
	%Describe the construction of the channel model
	
	%3GPP-2 also give comprehensively model the UAV channel under different communication scenarios such as urban, suburban, rural areas. 
	%Describe the construction of the channel model
	
	%The channel model can be classified into two types, i.e., deterministic channel model and stochastic channel model. 
	\subsubsection{HAP channel models}
\begin{figure}[!t]
		\centering
		\includegraphics[width=3.3 in]{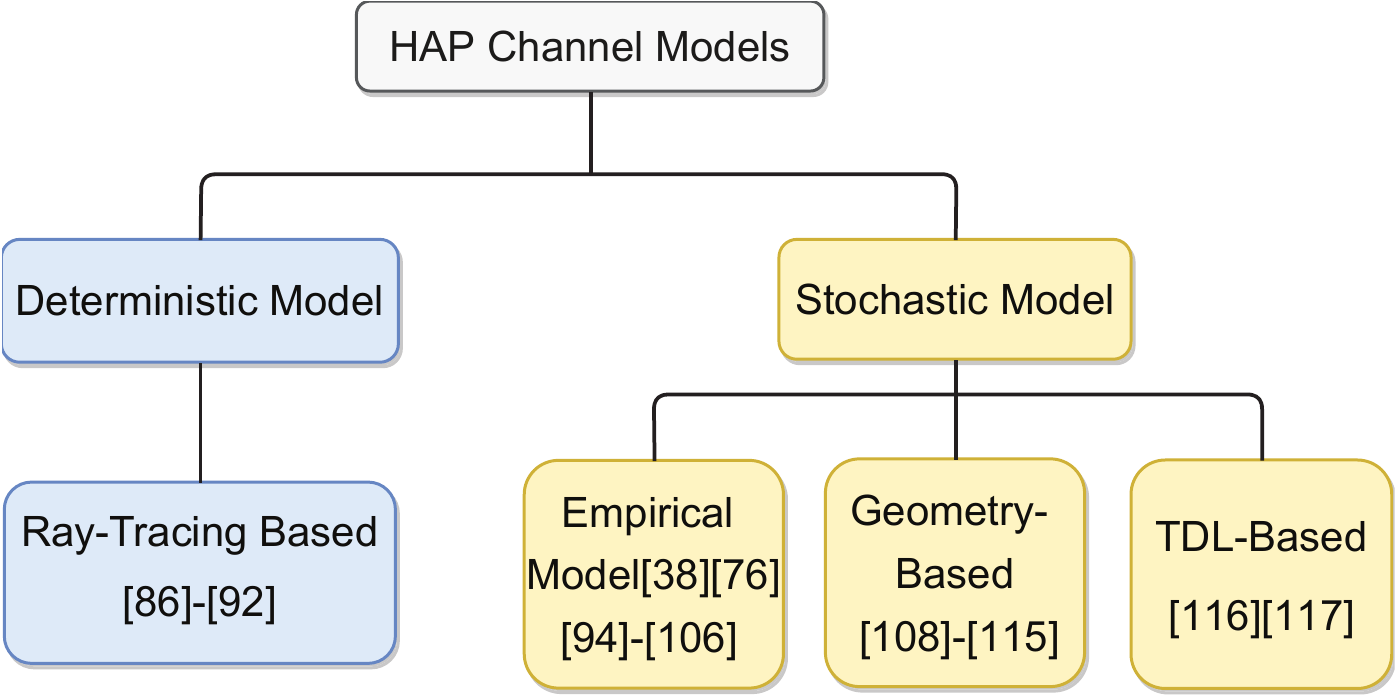}
		\caption{Classification diagram for existing HAP channel models.}
      
 \label{fig:fig_HAP_channel_model_classification}
	\end{figure}

	%However, the channel model constructed by 3GPP is so complicated that it hinders the study of researchers; as a result, researchers proposed many channel model for HAP and UAVs with different emphasis on the impact factors such as (jittering, rain attenuation, doppler, and so on). 
	%In summary, the channel model can be classified into five groups in the research community: 
	%From the aspect of different impact factors to discuss the channel model. (HAP and UAV are merged)
	In the research community, recent HAP channel models can be primarily categorized into two groups: deterministic models and stochastic models. These classifications are outlined in Fig. \ref{fig:fig_HAP_channel_model_classification}.
    %, and combination model. 
    
	\textbf{Deterministic model:} A deterministic channel model requires constant parameters (e.g., angle of arrival/angle of departure (AoA/AoD)) during simulations \cite{DBLP:conf/wcsp/LiC17} and leverages some methods (e.g., ray-tracing, digital map-assisted, and two-ray\cite{DBLP:conf/wcsp/LiC17}) to estimate the CIR or channel coefficients. 
	The accuracy of deterministic models relies on extensive and environment-specific databases, which can characterize the terrain topography, the electrical properties of buildings, and other obstacles. 
	Besides, model parameter adjustment according to propagation measurement data is required. 
	
	This type of HAP channel modeling method has attracted much attention. 
	For example, in \cite{DBLP:conf/icc/IskandarS06}, a ray-tracing-based propagation channel model between a HAP and a ground mobile station was developed. This model examined the effect of angle variations on propagation loss by utilizing a ray-tracing tool that accounted for changes in elevation and azimuth angles.
	In \cite{DBLP:conf/icc/IskandarS06}, the authors employed a ray-tracing technique to compute the power level and path loss experienced by a mobile user, as mentioned in \cite{DBLP:conf/globecom/ShimamotoI06}. By analyzing the outcomes, the propagation parameters, including path loss and Rice factor, were derived as functions of elevation and azimuth angles.
    In addition to evaluating angle variations, the authors investigated the effects of building height and street size on the propagation loss of downlink HAP channels in an urban environment. They utilized a ray-tracing tool to conduct this study and published their findings in \cite{kurniawan2011propagation,DBLP:conf/isspa/IskandarS05,iskandar2005ray}.
    The research presented in \cite{DBLP:journals/access/Hadidianmoghadam19} examined the propagation path loss of the downlink HAP channel using two-dimensional ray-tracing and the uniform theory of diffraction. The study evaluated the effects of wave reflection from building walls and the ground, as well as wave diffraction.
	Diverse from the above channel modeling methods using ray-tracing tools to simulate wave reflection and diffraction, the work in \cite{hsieh2019propagation} studied the impact of terrestrial clutter on downlink HAP propagation loss and developed a HAP propagation path loss model including free space path loss and clutter-induced attenuation using the ray-tracing data. 
	
	\textbf{Statistical model:} Without using measurement data, a statistical model utilizes statistical distributions and empirical parameters to capture the characteristics of propagation channels in a low-complexity manner and includes the empirical channel model, the geometry-based stochastic model, and tapped delay line (TDL)-based stochastic model. 
	
	In empirical channel model methods, some empirical channel fading models (e.g., Rician fading channel) are applied to compute the propagation loss of downlink HAP channels \cite{DBLP:conf/fgct/RaafatFE12}. 
	For instance, in \cite{holis2008elevation}, the downlink HAP propagation channel loss was calculated from the superposition of free space (or LoS) path loss and additional shadowing (or NLoS) path loss, where the LoS probability was determined by empirical parameters. 
	The work in \cite{DBLP:journals/wcl/LiuZSSCL21} modeled the downlink HAP propagation loss as the superposition of free space path loss and multipath and shadow fading, with the multipath and shadow fading being determined by empirical parameters. 
	Given the highly directional nature of EM wave propagation in the high-frequency band, the research presented in \cite{DBLP:journals/tsp/LinLHCZ19,DBLP:journals/iotj/XuLRK22} developed a downlink HAP channel model that included both dominant LoS and sparse NLoS propagation components.
	The work in \cite{DBLP:conf/vtc/PopoolaGC20} studied the mmWave downlink HAP channel modeling, where rain attenuation was a dominant propagation loss. Besides, as scattering was restricted at mmWave frequency bands, the path loss was considered to be dominated by LoS propagation loss. Therefore, the downlink HAP propagation channel was modeled as a superposition of rain attenuation and LoS path loss. 
	Some studies, such as \cite{DBLP:journals/twc/DingWZLL22,DBLP:journals/icl/JiJHLH21,DBLP:conf/wpmc/ZongJHJH19,azzahra2019noma,DBLP:journals/icl/SudheeshMMSM18}, have utilized the Rician fading channel to model the downlink HAP channel due to its ability to consider both LOS and NLOS components.
	Furthermore, as the LoS component is often obstructed by various obstacles such as buildings, trees, hills, and mountains, the effect of shadow fading has been highlighted in some studies. To model the downlink HAP channel, shadowed-Rician fading was utilized in \cite{nguyen2019study,DBLP:journals/taes/RamabadranSVM21}.
    Shadowed-Rician fading is a more generalized form of Rician fading, which assumes that the amplitude of the LoS component follows a distribution that is different from the Rician distribution.
	The multi-state HAP statistical channel model \cite{DBLP:conf/chinacom/LiYLG11,DBLP:conf/vtc/ZhaoWLZZ20} was developed based on the HAP propagation channel models mentioned above. This model could simulate more downlink HAP propagation states, such as LoS, shadowed, and blocked states. In this type of channel model, different fading models were utilized to simulate diverse propagation states, and a Markov chain was explored to theoretically model and analyze the state transitions. 
 Besides, the work in \cite{Ylmaz2023PathGA} analyzed the channel model between HAPs, which took into account the antenna radiation pattern, the effects of atmospheric factors, and the polarization mismatch of the Tr and Rx antennas.
	
	Geometry-based stochastic modeling methods capable of capturing spatial–temporal downlink HAP channel characteristics were derived from a predefined stochastic distribution of effective scatterers through applying the fundamental laws of EM wave propagation \cite{DBLP:journals/tvt/LianSWJ22}. The accuracy of this type of channel model is highly dependent on how the scatterers around the Tx and Rx are modeled. The scatterers are typically assumed to be distributed in either regular geometric shapes or irregular shapes.
	For example, multipath propagation models for downlink HAP channels based on (improved) circular straight cone geometry were developed in \cite{cuevas2009multipath,liu2012improved}, respectively. The propagation of reflected signals was modeled by taking into account all scatterers within the coverage area of a HAP system.
	3-D geometry-based reference models for downlink HAP channels were proposed in \cite{DBLP:journals/tvt/MichailidisK10,DBLP:journals/tvt/MichailidisTK13,michailidis20123} with the assumption that the scatters in the vicinities of Rx and or Tx were non-uniformly distributed within a regular geometry shape(s). 
	In contrast to the studies conducted in \cite{DBLP:journals/tvt/MichailidisK10,DBLP:journals/tvt/MichailidisTK13,michailidis20123}, which analyzed the impact of non-isotropic scattering environments, the approach presented in \cite{DBLP:journals/icl/LianJH18} introduced a 3-D geometry-based reference model that took into account both isotropic and non-isotropic scattering for HAP channels.
	Moreover, the dynamic behavior of scatterers and the potential for scatterers to reappear after they disappeared were explored in \cite{DBLP:journals/icl/LianJH17,DBLP:journals/tvt/MichailidisNTK20}. By analyzing the dynamic evolution of scatterers, 3-D geometry-based reference models were developed for HAP channels.
	
	Stochastic models for HAP channels can be created by utilizing a TDL method with varying numbers of taps. Each tap in this model can represent the fading statistics of the multipath components of HAP channels. Empirical measurements and numerical simulations can be used to analyze the fading statistics of individual taps.
	For instance, considering the multipath effect and the inherent dispersion characteristics of signals at different propagation paths, the works in \cite{guan2020channel,DBLP:journals/wpc/MichailidisK14} proposed to use TDL to model HAP mobile channels in a wide-area environment.

	\subsubsection{UAV channel models}
	UAV channel modeling is critical for the performance analysis and system design of UAV communication systems. Nevertheless, there are some challenges in UAV channel modeling; for example, the propagation features of UAV channels are under-studied for spatial and temporal variations in non–stationary UAV channels. The airframe shadowing and jittering are also under-explored. In recent years, there has been a significant increase in attention from both industry and academia towards the research on UAV channel modeling. 

 In industry, some technical standards for UAV channel modeling have been released by Nippon Telegraph and Telephone Corporation (NTT), Nokia, Ericsson, Intel, and Huawei, and so on \cite{new2018NTT170779,Enhancements2017ZTE1707264,WF2017NTT1714675,Large2017Nokia1714857,5G173GPP}. 
 
\begin{figure}[!t]
		\centering
		\includegraphics[width=1.9 in]{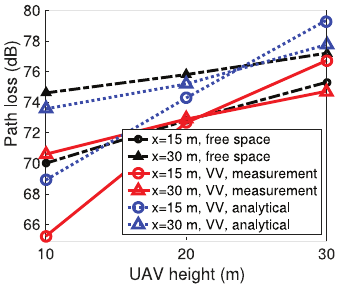}
		\caption{{Path loss vs. UAV-user horizontal distance and UAV height. VV: vertical–vertical Tx-Rx antenna orientation; free space: free space path loss model \cite{khawaja2020ultra}.}}
		\label{fig:path_loss_UAV_exper}
	\end{figure}

 {In academia, researchers have proposed a number of UAV channel models and conducted measurement campaigns to verify their validity \cite{khawaja2020ultra,chang20193d,chowdhery2018aerial,liu2019measurement}. For example, the work \cite{khawaja2020ultra} studied the UAV-to-ground (UtG) ultra-wideband channel through propagation measurements between 3.1 GHz and 4.8 GHz. Considering the LoS and ground-reflected component paths, the UtG path loss for the unobstructed UAV hovering scenario was modeled as \cite{khawaja2020ultra}
\begin{equation}
    \label{eq:UAV_exp_utg_channel}
    \begin{array}{l}
L({d_0},{d_1})[dB] = 10{\log _{10}}{\left( {{{4\pi {d_{\rm ref}}} \mathord{\left/
 {\vphantom {{4\pi {d_{\rm ref}}} \lambda }} \right.
 \kern-\nulldelimiterspace} \lambda }} \right)^2}\\
 \qquad \qquad + 10{\log _{10}}\frac{{{{\left( {{d_0}{d_1}} \right)}^2}}}{{{{\left( {{d_1}\sin \theta } \right)}^2} + d_0^2\sin \Omega \sin \Omega '{{\left| {{\Gamma _1}(\Psi) } \right|}^2}}}
\end{array}
\end{equation}
where $d_0$ and $d_1$ are related to the locations of UAVs and ground users, reference distance $d_{\rm ref} = 1$ m, $\lambda$ denotes the carrier frequency, $\Omega '$ and $\Omega$ denote the elevation angles of ground reflected components at the Tx and Rx, $\theta$ denotes the angle of the LoS component, and ${{\Gamma _1}(\Psi) }$ is the reflection coefficient with $\Psi$ being a pair of azimuth and elevation angles. 
Fig. \ref{fig:path_loss_UAV_exper} depicts the impact of the UAV height and the UAV-user horizontal distance (denoted by $x$) on path loss. This figure shows that the derived path loss at $x=15$ m is greater than that at $x=30$ m when the UAV height is 30 m. This is consistent with the measurements. It is because a higher elevation angle results in a smaller antenna gain \cite{khawaja2020ultra}.
}
 
 Besides, many surveys on UAV channel modeling have been published \cite{DBLP:journals/comsur/KhuwajaCZAD18,DBLP:journals/access/YanFZW19,DBLP:journals/comsur/KhawajaGMFS19}. Some surveys, including \cite{DBLP:journals/comsur/KhuwajaCZAD18,DBLP:journals/access/YanFZW19,DBLP:journals/comsur/KhawajaGMFS19}, have summarized the available UAV channel measurement campaigns, as well as various large- and small-scale UAV fading channel models. Some important issues (e.g., airframe shadowing, Doppler spread, blockage, scattering, and delay dispersion) in UAV channel modeling were also discussed. 
	
	\subsubsection{Differences between HAP and UAV channel models}
	There are many significant differences between HAP channel modeling and UAV channel modeling, for example, whether consider the impacts of meteorological factors and airframe shadowing on channel modeling. 
	Restricted by the deployment altitudes of UAVs and the allocated frequencies, few studies on UAV channel modeling \cite{DBLP:conf/iswcs/CuvelierH18,DBLP:journals/jsac/SaeedGBA21,DBLP:conf/vtc/SongHLDL20} have taken into account the influence of meteorological factors, such as atmospheric absorption, rain and cloud attenuation, and tropospheric scintillation.
	On the contrary, the investigation of the influence of meteorological factors is a highly researched topic in the field of HAP channel modeling.
	Particularly, at frequencies above 10 GHz, the atmospheric absorption must be calculated \cite{Kirik2023InterHAPBG}. Meanwhile, for elevation angles below 10 degrees, the atmospheric absorption is recommended to be calculated for frequencies above 1 GHz. At frequencies above 6 GHz, it is recommended to calculate rain and cloud attenuation. Similarly, at frequencies above 6 GHz, tropospheric scintillation corresponding to rapid amplitude and phase fluctuations of signals should be calculated. Furthermore, due to the longer transmission path, the influence of tropospheric scintillation becomes more pronounced at small elevation angles, especially for high carrier frequencies.
	
	The impact of airframe shadowing is seldom considered in HAP channel modeling. However, in UAV channel modeling, the research on the airframe shadowing impact has attracted much attention. In UAV communications, airframe shadowing occurs when the UAV-to-ground (UtG) path is blocked by UAV maneuvering, banking turns, or UAV structures such as wings, fuselage, and engines. 
	The shadowing results depend on some factors, like carrier frequency, antenna placement, and the exact shapes, sizes, and materials of UAVs. 
	
	\subsection{Summary and {Lessons Learned}}
{\textbf{Summary:}} In this section, we {discussed the modeling of phase delay of array antennas} under antenna jitter, the influence of atmospheric effects on NSIN channel models, and the small-scale fading of NSIN channels.
Besides, we revealed the significant differences between HAP and UAV channel models and reviewed recent efforts in the channel modeling of NSINs. 

{\textbf{Lessons learned:}} Nevertheless, the signal propagation environment of NSINs differs significantly from that of space networks, UAV networks, and TNs. As a result, the existing channel models for these types of networks may not be directly applicable to NSINs. 
The channel modeling of NSINs should consider the phase delay of deployed antennas (which may be greatly affected by unstable platform movement), atmosphere attenuation, rain attenuation, tropospheric scintillation attenuation, the effect of multipath, and Doppler frequency shift. 
Further, there is a lack of studies on the high-accurate channel estimation of NSINs.
Most existing channel models of NSINs are established based on ray-tracing tools or statistical/empirical distributions. 
There is a need to {improve} the accuracy of NSIN channel models by mining and analyzing field measurement data, which reflects the realistic behaviors of NSINs' propagation channels. However, due to the high cost of conducting extensive high-altitude channel measurement campaigns and certain restrictions imposed by aviation regulations, {it is challenging to obtain a large number of real channel samples.
A viable approach includes utilizing the generative adversarial network (GAN) and variational auto-encoder (VAE) to generate a large number of high-fidelity channel samples based on partially real samples.} 
%researchers have made limited contributions to measurement-based HAP channel modeling.
	
	% 	 \begin{figure}[!t]
	% 	\centering
	% 	\subfigure[{Transmit signals from our airship}]{
 %    \label{fig:subfig:f_c} %% label for first subfigure
 %    \includegraphics[width=1.4 in, height = 2.3 in]{Transmiter_airship.pdf}}
 %    \hspace{3pt}
 %    \subfigure[{A ground receiver}]{
 %    \label{fig:subfig:g_c} %% label for first subfigure
 %    \includegraphics[width=1.4 in, height=2.3in]{Receiver_airship.pdf}}
	% 	\caption{{A communication coverage and signal reception quality measurement experiment via our airship.}}
	% 	\label{fig:communication_measurement}
	% \end{figure}
	\section{Networking of NSINs}
\begin{figure}[!t]
		\centering
		\includegraphics[width=3.4 in]{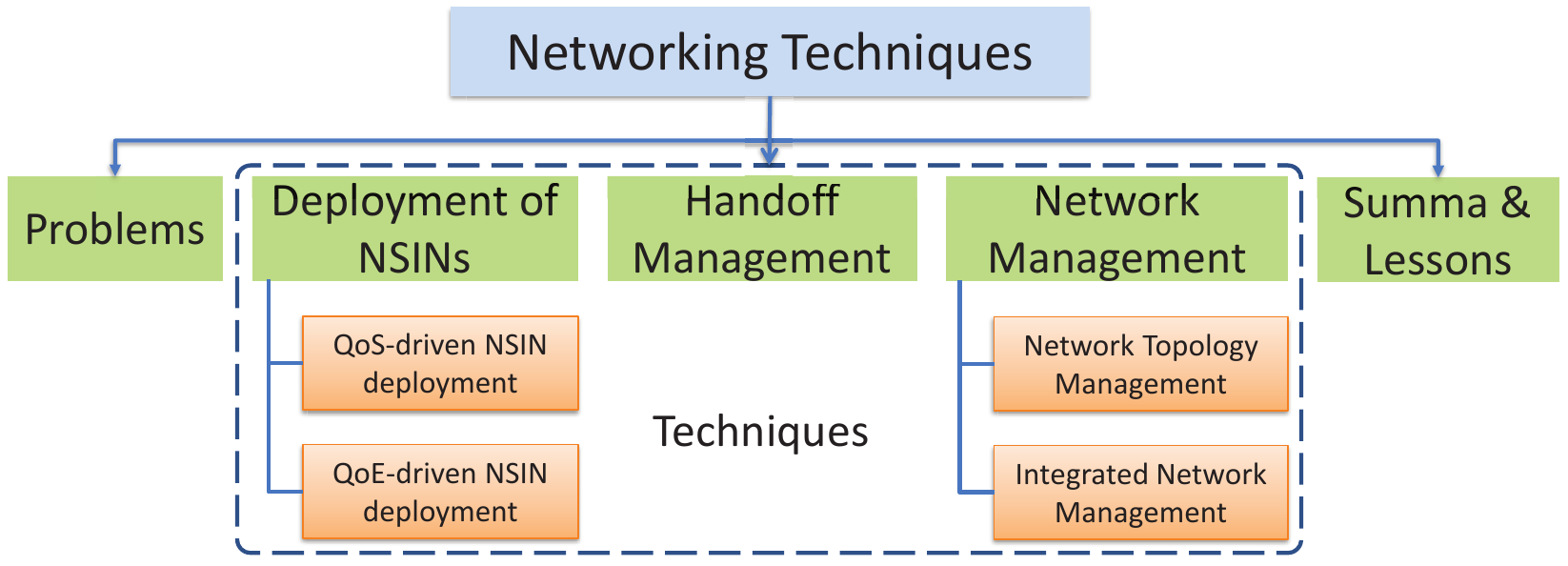}
		\caption{{Categorization of the networking techniques in this section.}}
		\label{fig:networking_network}
	\end{figure}

 {In this section, we reveal some influence factors that should be kept in mind when investigating the networking issues of NSINs via experiments.} Some networking techniques for NSINs will be elaborated on, mainly from the perspectives of constructing and managing NSINs.
 {Fig. \ref{fig:networking_network} illustrates the technical categorization of this section.}

\subsection{{Problems Encountered in Experiments}}
{The authors in this article conducted an experiment to test the coverage performance of NSINs using the Beihang airship, as shown in Fig. \ref{fig:communication_measurement}. 
		 \begin{figure}[!t]
		\centering
		\subfigure[{Transmit signals from Beihang airship}]{
    \label{fig:subfig:f_c} %% label for first subfigure
    \includegraphics[width=1.4 in, height = 2.1 in]{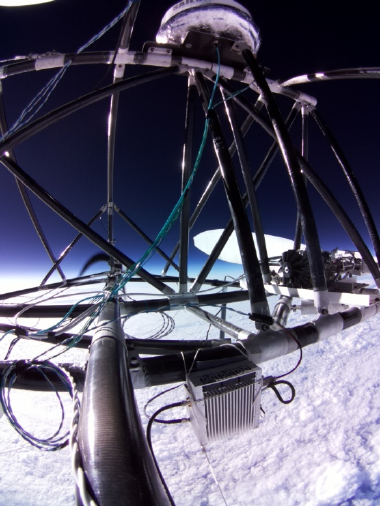}}
    \hspace{3pt}
    \subfigure[{A ground receiver}]{
    \label{fig:subfig:g_c} %% label for first subfigure
    \includegraphics[width=1.4 in, height=2.1in]{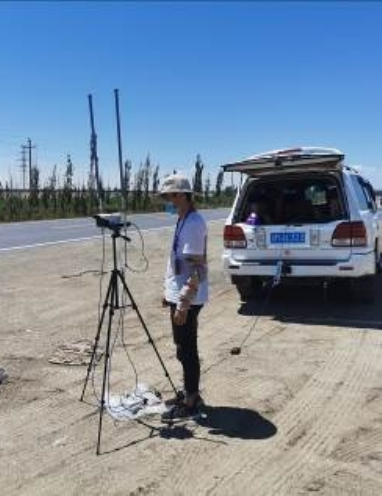}}
		\caption{{A communication coverage and signal reception quality measurement experiment via Beihang airship deployed at an altitude of 20 km.}}
		\label{fig:communication_measurement}
	\end{figure}
 Table \ref{table_communication_results} summarizes the test results. The experimental results show that the coverage area of NSINs can have a radius of several hundred km. Users at the edge of NSINs will experience a high packet loss rate. 
 Both the changes in the airship attitude and the great propagation attenuation may result in packet loss. Unstable airship attitude will cause changes in the locations or shapes of footprints projected by the airship.
 Therefore, it is essential to investigate the deployment and handoff issues of NSINs for improving service quality. 
 Besides, network management is a fundamental issue to be investigated for NSINs. We next provide an overview of the recent progress on these research topics. 
}
 \begin{table*}[!t]
		\newcommand{\tabincell}[2]{\begin{tabular}{@{}#1@{}}#2\end{tabular}}
		%\renewcommand{\arraystretch}{1.3}
		% if using array.sty, it might be a good idea to tweak the value of
		%	 \extrarowheight as needed to properly center the text within the cells
		\caption{{An experiment of testing communication coverage performance via Beihang airship.}}
		\label{table_communication_results}
		\centering
		% Some packages, such as MDW tools, offer better commands for making tables
		% than the plain LaTeX2e tabular which is used here.
  %\tabincell{l}{Promising technical advancements, including directional antenna and full-duplex radio circuits \\ with multi-packet reception, which can be adopted by MAC protocols of FANET.}
		\begin{tabular}{|c|c|c|c|c|}
			\hline
			{Airship location} & {Receiver location} &  {Distance (km)} & {Packet loss rate} & {Latency (ms)} \\
			\hline
			{(41.412629$^{\circ}$,88.167300$^{\circ}$)} & {(42.275686$^{\circ}$,87.076372$^{\circ}$)} &  132  & 0\%  & 3.656 \\
			\hline
			{(41.558200$^{\circ}$,87.975700$^{\circ}$)} & {(41.849713$^{\circ}$,86.329568$^{\circ}$)} &  141  & 0\%  & 3.688 \\
			\hline
		{(41.627400$^{\circ}$,87.817200$^{\circ}$)} & {(41.834235$^{\circ}$,85.923380$^{\circ}$)} &  159  & 0\%  & 3.676 \\
			\hline
		{(41.685900$^{\circ}$,87.662700$^{\circ}$)} & {(41.947432$^{\circ}$,85.409625$^{\circ}$)} &  190  & 0.029\%  & 3.997 \\
			\hline
		{(41.845100$^{\circ}$,87.541100$^{\circ}$)} & {(42.006960$^{\circ}$,84.597780$^{\circ}$)} &  245  & 0.05\%  & 4.132 \\
			\hline
		{(41.987700$^{\circ}$,87.503637$^{\circ}$)} & {(41.825335$^{\circ}$,83.861625$^{\circ}$)} &  302  & 0.08\%  & 4.167 \\
			\hline
   {(41.880320$^{\circ}$,87.542650$^{\circ}$)} & {(41.825335$^{\circ}$,83.861625$^{\circ}$)} &  306  & 0.12\%  & 4.525 \\
			\hline
		\end{tabular}
	\end{table*}
 
	\subsection{Deployment of NSINs}
	\subsubsection{QoS-driven NSIN deployment}
 %The deployment locations of airborne platforms in NSINs greatly affect the QoS of served users. 
 %For instance, we conducted an experiment to test the communication coverage performance of an airship, as shown in Fig \ref{fig:communication_measurement}. 
 %Table \ref{table_communication_results} summarizes test results. The experimental results show that users at the edge of the HAP footprint will experience a high packet loss rate. Therefore, it is essential to investigate the QoS-driven NSINs deployment problem.
 %In this subsection, we discuss the QoS-driven NSINs deployment problem.
	QoS-driven NSIN deployment refers to the optimization of the deployment locations or trajectories of aerial platforms in NSINs based on space-time distributions of serving users in a considered area, the goal of which is to meet users’ QoS requirements. 
 % \begin{figure}[!t]
	% 	\centering
	% 	\includegraphics[width=3.3 in, height= 1.9in]{QoS_driven_NSINs.pdf}
	% 	\caption{A scenario of deploying QoS-driven NSINs.}
	% 	\label{fig:fig_qos_driven_NSINs}
	% \end{figure}
	Typically, users’ QoS requirements can be characterized as their received signal-to-noise ratio (SNR) or achievable data rates.
 
	{\textbf{Discussion:}} Nevertheless, it is difficult to solve the QoS-driven NSIN deployment problem. First, NSIN consists of multiple sub-networks deployed at different altitudes. The sub-networks have many differences in capabilities, such as payload, communication, and maneuverability. Therefore, studying the QoS-driven NSIN deployment problem means a solution of a joint optimization and deployment problem of a stereoscopic, multi-layered, heterogeneous network. Secondly, the airspace environment in which NSIN are deployed is complex and constantly changing.  The airspace needs to be shared by the aerial platforms of NSIN and civil aviation aircrafts. There are also many unknown and uncertain risk areas in the airspace. Therefore, studying the QoS-driven NSIN deployment problem needs to explore how to ensure the flight safety of aerial platforms.
 
	To meet the above challenges, it is first necessary to mathematically model the stereoscopic, multi-layered, and heterogeneous characteristics of NSIN. A feasible approach to reflecting the stereoscopic and multi-layered characteristics of NSIN is to define different coverage or footprints for various platforms, design a time-varying sub-network selection scheme, and establish a multi-category signal interference regime. In order to characterize the heterogeneity of aerial platforms, a viable method is to constrain the service capabilities of platforms from the perspective of limiting available computing, storage, and network resources. Additionally, it is necessary to plan offline or online trajectories for NSIN platforms by detecting the surrounding environment.
	Next, formulating and solving the NSIN deployment problem to meet the user's QoS requirements, subject to the constraints on the stereoscopic, multi-layered, and heterogeneous characteristics and the flight safety requirement. 
	The formulated optimization problem is typically a multi-objective, multi-constrained, high-dimensional, and serialized decision-making problem. Therefore, it is challenging to use traditional optimization approaches to obtain the optimal or sub-optimal solution. Instead, intelligent optimization, reinforcement learning, and graph neural networks are better approaches to solve this type of problem.

\begin{table*}[!t]
	\renewcommand{\arraystretch}{1.2}
	\caption{A summary of objective, approach, and investigated network type of QoS-driven NSIN deployment problem}
	\label{table_qos_deployment}
	\newcommand{\tabincell}[2]{\begin{tabular}{@{}#1@{}}#2\end{tabular}}
	\centering
	\begin{tabular}{l}
%\begin{minipage}{0.18\textwidth}
			\centering
			\includegraphics[width=7.0 in]{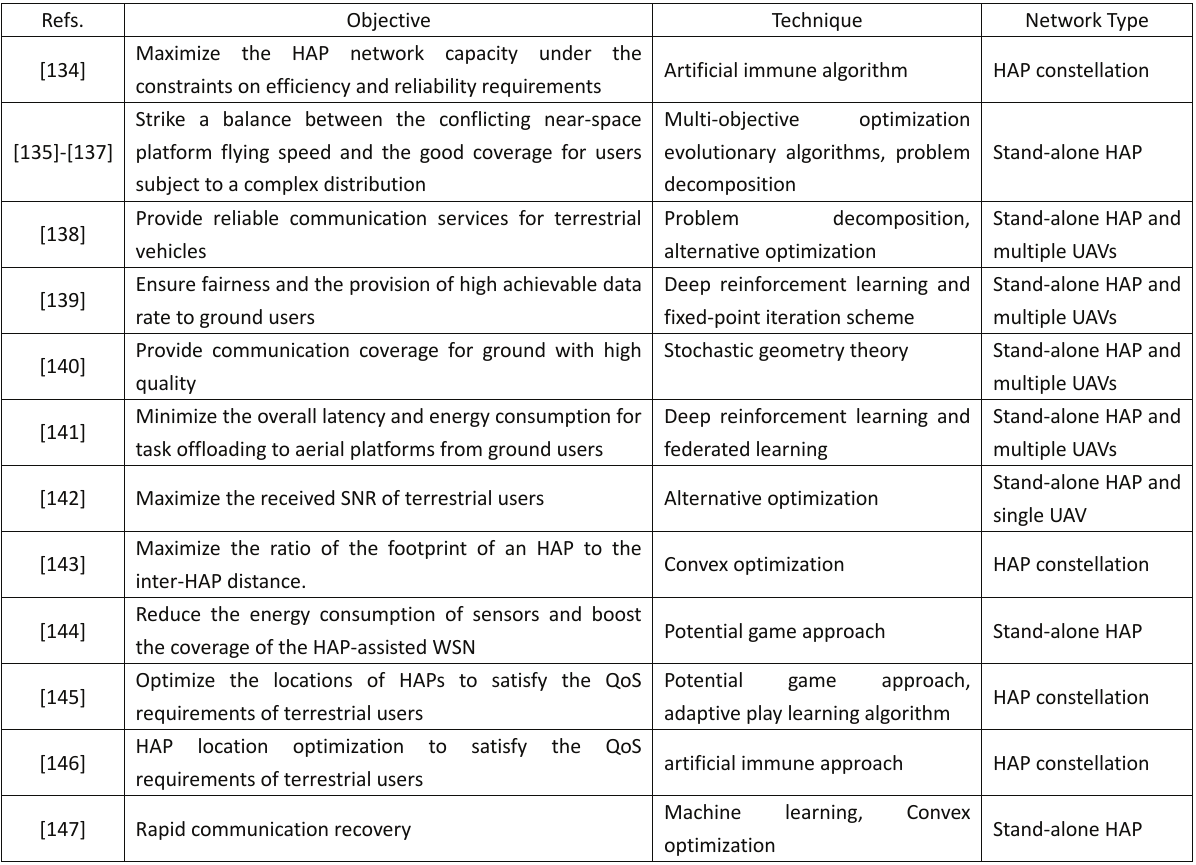}
		%\end{minipage} 
	\end{tabular}
\end{table*} 
	
 {\textbf{Existing studies:}} The QoS-driven NSIN deployment problem is currently under active research, and a number of related studies have been published in the literature during the past few years \cite{DBLP:journals/tmm/DongHGWL16,DBLP:conf/cec/WangGLWS16,DBLP:journals/tec/GongWZJ17,DBLP:conf/cec/WangGCMJ14,he2022noma,DBLP:journals/ojcs/AraniHZ23,DBLP:journals/icl/ZhaoZ000023,DBLP:conf/iwcmc/WangLX23,DBLP:journals/wcl/GaoJLM21,DBLP:conf/pimrc/XuYCLZX17,DBLP:conf/wcnc/Wang13,DBLP:conf/iccnc/ZongGWZ12,dong2015optimization,almalki2020deployment,tang2017multiobjective,DBLP:journals/ijdsn/YiDLC17}. 
 Table \ref{table_qos_deployment} summarizes the recent studies on the QoS-driven NSIN deployment problem in the literature. 
	For instance, the work in \cite{DBLP:journals/tmm/DongHGWL16} formulated a HAP constellation deployment problem with a goal of maximizing the network capacity under the constraints on QoS metrics that were characterized by the efficiency and reliability of communications. To efficiently search for deployment locations of HAPs, an improved artificial immune algorithm was designed.
	The works in \cite{DBLP:conf/cec/WangGLWS16,DBLP:journals/tec/GongWZJ17, DBLP:conf/cec/WangGCMJ14} proposed to deploy a near-space communication system to meet the QoS requirements of ground users. The following problem was studied: how to strike a balance between the flying speed of a near-space platform and the good coverage for users subject to a complex distribution? The authors transformed the conflict between flying speed and coverage into a multi-objective problem and developed various multi-objective optimization evolutionary algorithms based on decomposition to solve this kind of problem. 
 The issue of deploying a stand-alone HAP and multiple UAVs to provide communication coverage or partial computing services for ground users also attracts much attention from the research community \cite{he2022noma,DBLP:journals/ojcs/AraniHZ23,DBLP:journals/icl/ZhaoZ000023,DBLP:conf/iwcmc/WangLX23}.
	For instance, the authors in \cite{he2022noma} designed an NSIN architecture to provide reliable communication services for terrestrial vehicles. In this architecture, vehicles would receive both signals directly from a HAP and relayed signals from multiple UAVs. To enhance the spectral efficiency and downlink transmission rate of NSIN, researchers investigated the joint design of non-orthogonal multiple access (NOMA) and UAV location optimization. Furthermore, in \cite{DBLP:journals/wcl/GaoJLM21}, the joint optimization of UAV trajectory and phase shift of RIS was studied to maximize the received SNR of terrestrial users. An alternative optimization algorithm was designed to optimize the UAV trajectory and adjust the RIS phase shift. 
	In \cite{DBLP:conf/pimrc/XuYCLZX17}, researchers proposed a HAP deployment algorithm that aimed to maximize the ratio of the HAP footprint to the inter-HAP distance.
	The works in \cite{DBLP:conf/wcnc/Wang13} designed a HAP-assisted wireless sensor network (WSN) architecture to reduce the energy consumption of sensors and boost the coverage of the WSN. A potential game approach was developed to determine the deployment locations of HAPs so that the above goal could be achieved. 
	In \cite{DBLP:conf/iccnc/ZongGWZ12}, the deployment of HAPs for providing communication services to terrestrial users was formulated as a potential game. To identify the deployment locations of HAPs that could satisfy the QoS requirements of more users, a restricted spatial adaptive play learning algorithm was devised. 
	The work in \cite{dong2015optimization} investigated the methodology of HAP deployment to guarantee ground users’ QoS requirements.To optimize the locations of HAPs, an enhanced artificial immune approach based on immune review was developed.
	Additionally, the deployment scenario of a HAP system for quick communication recovery in a disrupted communication area was presented in \cite{almalki2020deployment}. To guarantee the QoS requirements of users in a given area, a machine learning-based propagation model was developed to optimize the deployment altitude of a HAP.

	%Proactive networking: Based on the predicted result, conduct proactive networking on the assistance of situational information. The movement control of UAVs with the assistance of HAP coverage. Resource allocation problem.
	
	\subsubsection{QoE-driven NSIN deployment}
	With the rapid growth of streaming services, it becomes increasingly important to focus on the subjective experiences of users. 
	QoS metrics can only measure network performance and cannot reflect users’ subjective feelings about service quality. 
	The concept of QoE was proposed to tackle this issue. QoE reflects the subjective feelings of users watching videos and becomes the main performance evaluation criteria for video service quality. 
	{Except for video services, QoE can be utilized to evaluate the service quality of such types of emerging services as digital twins (DTs) and Metaverse \cite{DBLP:journals/comsur/YangDLLXNCSM23}.}
 
	{\textbf{Discussion:}} Research on QoE in wireless communications is a hot and under-studied topic. There is a consensus that a unified QoE evaluation model cannot be achieved. This is because different video applications have diverse requirements for video quality.
	For instance, short video applications focus on the opening rate per second, high-definition long video applications have stringent requirements for high bandwidth, video conferences, and online games pursue low latency and high smoothness. 
	
	However, it is widely accepted that users cannot accept three categories of feelings during video transmission, i.e., freeze, blur, and large delay. 
	Therefore, some underlying layer and upper layer mechanisms need to be carefully designed so that users’ QoE requirements, including low latency, high video quality, and smooth playback, can be satisfied. 
	
	It is easy to know that the QoE evaluation indicators are subjective. In this case, one should analyze the relationship between users’ QoE and its influencing factors and establish a mapping model between them. 
	For example, one can utilize the peak signal-to-noise rate (PSNR) to objectively measure the quality of demodulated videos. To objectively measure the smoothness of played videos, a viable approach is to design and optimize a logarithmic utility function that takes into account factors such as the user's achievable data rate, desired playback rate, and application type. 
 {A typical smoothness measurement model can take the following form, $\phi \left( {\bm{\bar r}\left( t \right)} \right) = \alpha \sum\limits_{i = 1}^M {{{\log }_2}\beta (1 + \frac{{B{{\bar r}_i}\left( t \right)}}{{{R_i}}})}$, where $ {\bm{\bar r}\left( t \right)} = \left( {{{\bar r}_1}\left( t \right), \ldots ,{{\bar r}_M}\left( t \right)} \right)$, ${\bar r}_i(t)$ is the time-averaged data rate of user $i \in \{1, \ldots, M\}$, ${{R_i}}$ represents the desired playback bitrate of user $i$, {$B$ represents the total bandwidth,} and $\alpha $, $\beta$ are both positive values that are different for various types of applications. 
 }
 
	It is not difficult to optimize a certain QoE evaluation indicator alone. Nevertheless, it is highly challenging to simultaneously optimize all the indicators owing to their mutual restrictions. 
	
	\begin{table*}[!t]
		\newcommand{\tabincell}[2]{\begin{tabular}{@{}#1@{}}#2\end{tabular}}
		%\renewcommand{\arraystretch}{1.3}
		% if using array.sty, it might be a good idea to tweak the value of
		%	 \extrarowheight as needed to properly center the text within the cells
		\caption{The recent studies on the QoE-driven UAV deployment in the literature.}
		\label{table_qoe_deployment}
		\centering
		% Some packages, such as MDW tools, offer better commands for making tables
		% than the plain LaTeX2e tabular which is used here.
		\begin{tabular}{|c|c|c|c|c|c|c|c|c|c|c|c|c|}
			\hline
			Refs. & \tabincell{l}{Location/ \\ trajectory}
			 & \tabincell{l}{Transmit \\ power}	& \tabincell{l}{Band- \\ width} & Caching	& MEC	& Latency	& \tabincell{l}{Smooth \\ playback} & \tabincell{l}{High \\ quality} &	\tabincell{l}{Single \\ UAV} & \tabincell{l}{Multi- \\ UAVs} & \tabincell{l}{Optimi- \\ zation} & \tabincell{l}{AI/ \\ ML} \\
			\hline
      \cite{DBLP:journals/tnse/WuCYYWQ24} & 
            \checkmark & \checkmark & {} & {} & {} & \checkmark & \checkmark & \checkmark & {} & \checkmark & \checkmark & {} \\
			\hline
			\cite{DBLP:journals/tnse/HuZXFJC21} & \checkmark & \checkmark & \checkmark & {} & \checkmark & \checkmark & {} & {} & \checkmark & {} & \checkmark & {} \\
			\hline
			\cite{DBLP:journals/tvt/BeraMC20} & \checkmark & {} & {} & \checkmark & {} & \checkmark & {} & {} & {} & \checkmark & \checkmark & {} \\
			\hline
			\cite{DBLP:conf/infocom/Shen22} & \checkmark & {} & {} & {} & \checkmark & \checkmark & {} & {} & \checkmark & {} & \checkmark & {} \\
			\hline
			\cite{DBLP:journals/tvt/ZhanH20,DBLP:conf/icc/HuZAW19} & \checkmark & \checkmark & \checkmark & {} & {} & {} & \checkmark & {} & \checkmark & {} & \checkmark & {} \\
			\hline
			\cite{DBLP:journals/iotj/ZhouMHZCL22} & \checkmark & \checkmark & {} & {} & {} & {} & \checkmark & {} & \checkmark & {} & \checkmark & {} \\
			\hline
			\cite{DBLP:journals/tmm/ZhanHWFN20} & \checkmark & {} & {} & {} & {} & {} & \checkmark & {} & {} & \checkmark & {} & \checkmark \\
			\hline
			\cite{DBLP:journals/tmm/TangHH21} & \checkmark & \checkmark & {} & {} & {} & {} & {} & \checkmark & {} & \checkmark & \checkmark & {} \\
			\hline
			\cite{DBLP:journals/tvt/ZengHX0Z0L20} & \checkmark & \checkmark & \checkmark & {} & {} & {} & {} & \checkmark & \checkmark & {} & \checkmark & {} \\
			\hline
			\cite{DBLP:journals/tvt/ZhangC22} & \checkmark & \checkmark & \checkmark & {} & \checkmark & {} & {} & \checkmark & {} & \checkmark & \checkmark & {} \\
			\hline
			\cite{DBLP:journals/iotj/JiangYXSZ19} & \checkmark & \checkmark & {} & \checkmark & {} & {} & {} & \checkmark & {} & \checkmark & \checkmark & {} \\
			\hline
			\cite{DBLP:journals/tvt/MedeirosBC22} & \checkmark & {} & {} & {} & {} & \checkmark & {} & \checkmark & {} & \checkmark & {} & \checkmark \\
			\hline
			\cite{DBLP:journals/tvt/BurhanuddinLDCZ22} & \checkmark & \checkmark & {} & {} & {} & \checkmark & \checkmark & \checkmark & \checkmark & {} & {} & \checkmark \\
			\hline
   {\cite{DBLP:journals/twc/zhouTD24}} & 
            {\checkmark} & {} & {} & {} & {} & {\checkmark}  & {} & {\checkmark}  & {} & {\checkmark}  & {} & {\checkmark}  \\
			\hline
   {\cite{DBLP:journals/jsac/GuoTK23}} & 
            {\checkmark}  & {} & {\checkmark}  & {} & {\checkmark}  & {\checkmark}  & {} & {} & {\checkmark}  & {} & {} & {\checkmark}  \\
			\hline
   {\cite{DBLP:journals/jsac/GuoZWLB23}} & 
            {\checkmark}  & {} & {} & {} & {\checkmark}  & {\checkmark}  & {} & {} & {} & {\checkmark}  & {} & {\checkmark}  \\
			\hline
   {\cite{DBLP:journals/iotj/ShenLLCZCL22}} & 
            {\checkmark}  & {} & {} & {} & {} & {\checkmark}  & {} & {\checkmark}  & {} & {\checkmark}  & {} & {\checkmark}  \\
			\hline
   {\cite{DBLP:journals/jsac/YeWLDLWT24}} & 
            {\checkmark}  & {} & {} & {} & {} & {\checkmark}  & {} & {} & {} & {\checkmark}  & {} & {\checkmark}  \\
			\hline
		\end{tabular}
	\end{table*}

	{\textbf{Existing studies:}} This research topic has attracted a lot of attention from the academic community. %There have been limited methods for QoE-driven HAP deployment thus far. However, the research on QoE-driven UAV deployment has recently gained extensive attention. Table \ref{table_qoe_deployment} summarizes recent literature on QoE-driven UAV deployment.
	In terms of video transmission services, for example, the works in \cite{DBLP:journals/tnse/HuZXFJC21,DBLP:journals/tvt/BeraMC20,DBLP:conf/infocom/Shen22} proposed to minimize the video transmission latency by formulating and solving a UAV resource optimization problem. 
	In \cite{DBLP:journals/tnse/HuZXFJC21}, the authors addressed the issue of jointly optimizing computing offloading, UAV 3D trajectory, UAV transmission power control, and system bandwidth to maximize computation efficiency and guarantee users' video transmission latency requirements. 
	A latency-guided video delivery framework was proposed for cache-enabled multi-UAV networks in \cite{DBLP:journals/tvt/BeraMC20}. The authors studied the optimization problem of joint UAV and cache placement to minimize video delivery latency.
 
	Different from \cite{DBLP:journals/tnse/HuZXFJC21,DBLP:journals/tvt/BeraMC20,DBLP:conf/infocom/Shen22}, the works in \cite{DBLP:journals/tvt/ZhanH20,DBLP:conf/icc/HuZAW19,DBLP:journals/iotj/ZhouMHZCL22,DBLP:journals/tmm/ZhanHWFN20} proposed to optimize UAV resources to ensure smooth video playback. 
	In \cite{DBLP:journals/tvt/ZhanH20,DBLP:conf/icc/HuZAW19}, a logarithmic function was used to model smooth playback utility with respect to video transmission rate. The authors investigated joint optimization of UAV transmit power, UAV trajectory, and system bandwidth to maximize the minimum utility among all users, subject to constraints on UAV total power and video playback.
	The work in \cite{DBLP:journals/iotj/ZhouMHZCL22} utilized the mean opinion score (MOS) to measure the smoothness of playback video. The authors proposed to leverage a hybrid deep reinforcement learning (DRL) approach to adjust UAVs' moving directions and distances to achieve energy-efficient and smooth video playback. 
	
	Several works including \cite{DBLP:journals/tmm/TangHH21,DBLP:journals/tvt/ZengHX0Z0L20,DBLP:journals/tvt/ZhangC22,DBLP:journals/iotj/JiangYXSZ19} investigated the joint resource optimization problem in UAV networks to ensure high-quality video reception by users. In \cite{DBLP:journals/tmm/TangHH21}, the authors studied the joint optimization problem of UAV trajectory and transmit power to maximize the minimum PSNR of users' video reconstruction quality. An iterative optimization algorithm was developed, consisting of block-coordinated descent and successive convex approximation schemes, to solve the problem.
	The authors in \cite{DBLP:journals/tvt/ZengHX0Z0L20} proposed a joint optimization approach to ensure the quality of received videos by users by optimizing user scheduling, UAV trajectory, UAV transmit power, and bandwidth. They developed an alternative optimization approach to solve the mixed-integer and non-convex joint optimization problem.
	
	However, the above works \cite{DBLP:journals/tmm/TangHH21,DBLP:journals/tvt/ZengHX0Z0L20,DBLP:journals/tvt/ZhangC22,DBLP:journals/iotj/JiangYXSZ19} did not investigate the trade-off among QoE indicators. To tackle this issue, the works in \cite{DBLP:journals/tvt/MedeirosBC22,DBLP:journals/tvt/BurhanuddinLDCZ22,DBLP:journals/tnse/WuCYYWQ24} investigated the joint optimization of QoE indicators. 
	For instance, the work in \cite{DBLP:journals/tvt/MedeirosBC22} developed a UAV path planning scheme to deliver low-latency and high-quality videos to users. 	
	The authors in \cite{DBLP:journals/tvt/BurhanuddinLDCZ22} formulated the QoE by considering three factors, including video quality, jitter between video frames, and video latency. They jointly optimized the video resolution, UAV trajectory, and UAV transmit power using the deep Q-network (DQN) and actor-critic approaches to maximize the QoE of real-time video streaming.

 {In terms of DT and Metaverse services, for instance, the authors in \cite{ DBLP:journals/twc/zhouTD24} designed a DT-assisted tracking framework to perform accurate and real-time multiple target tracking (MTT) for a UAV swarm. It is time- and resource-consuming for the DT system to imitate the entire MTT process. Then, a tiered DT-assisted MTT framework was utilized. 
However, the developed tiered DT framework is not suitable for imitating higher-speed moving targets in real time. 
The authors in \cite{ DBLP:journals/jsac/GuoTK23} constructed a UAV-assisted mobile network to provide efficient communications for all mobile users in high-density and high-traffic environments. A digital twin-empowered dynamic bandwidth allocation and UAV trajectory optimization strategy based on online training was proposed to achieve the above goal. 
However, more realistic and large-scale scenarios such as varying movements of devices, multiple UAVs, and cross-cell interference are not considered in this paper. 
The work \cite{ DBLP:journals/jsac/GuoZWLB23} introduced DT into aerial computing networks and designed a DT-assisted UAV deployment strategy to ensure the service quality of ground users. 
However, this work studied the stationary task offloading scenario, and the proposed approach could not be applied to the more important mobile scenarios. 
The authors in \cite{ DBLP:journals/iotj/ShenLLCZCL22} introduced the DT technology to tackle two tricky issues encountered in the investigation of UAV swarm motion control using DRL approaches. That is the requirement for a high-fidelity simulation environment and an efficient data acquisition speed. Particularly, a DT-enabled DRL training framework for UAV swarm motion control was proposed. 
Nevertheless, the study on high-fidelity DT modeling of UAV swarms is still in its infancy. 
% The authors in \cite{ DBLP:journals/tgcn/LiLXZZ23} introduced the DT technology to real-time monitor the network state. Based on network state information, they designed a multi-agent DRL approach to optimize trajectories and transmit power of UAVs in a DT-empowered UAV network. 
% However, the optimization of UAVs’ deployment altitudes is not investigated in this work. At the same time, the impact of the changes in UAVs’ attitude on the design of beamforming scheme is not considered.
Besides, the work \cite{ DBLP:journals/jsac/YeWLDLWT24} designed a multi-agent DRL approach to optimize the trajectories of multiple UAVs, which moved around and collected data from mobile users wearing Metaverse devices.} 

 Table \ref{table_qoe_deployment} compares recent literature on QoE-driven UAV deployment. {We can observe from this table that controlling the movement of multiple UAVs using AI approaches (exactly, DRL) to satisfy users' QoE requirement has attracted a lot of attention. Nevertheless, how to optimize different QoE indicators simultaneously remains under-studied.}
	
	\subsection{Handoff Management}
	The handoff (or handover) means the shift of communication services for a particular user from one platform footprint (exactly, cell coverage of a platform on the ground) to another.

\begin{figure}[!t]
		\centering
		\includegraphics[height = 1.3 in]{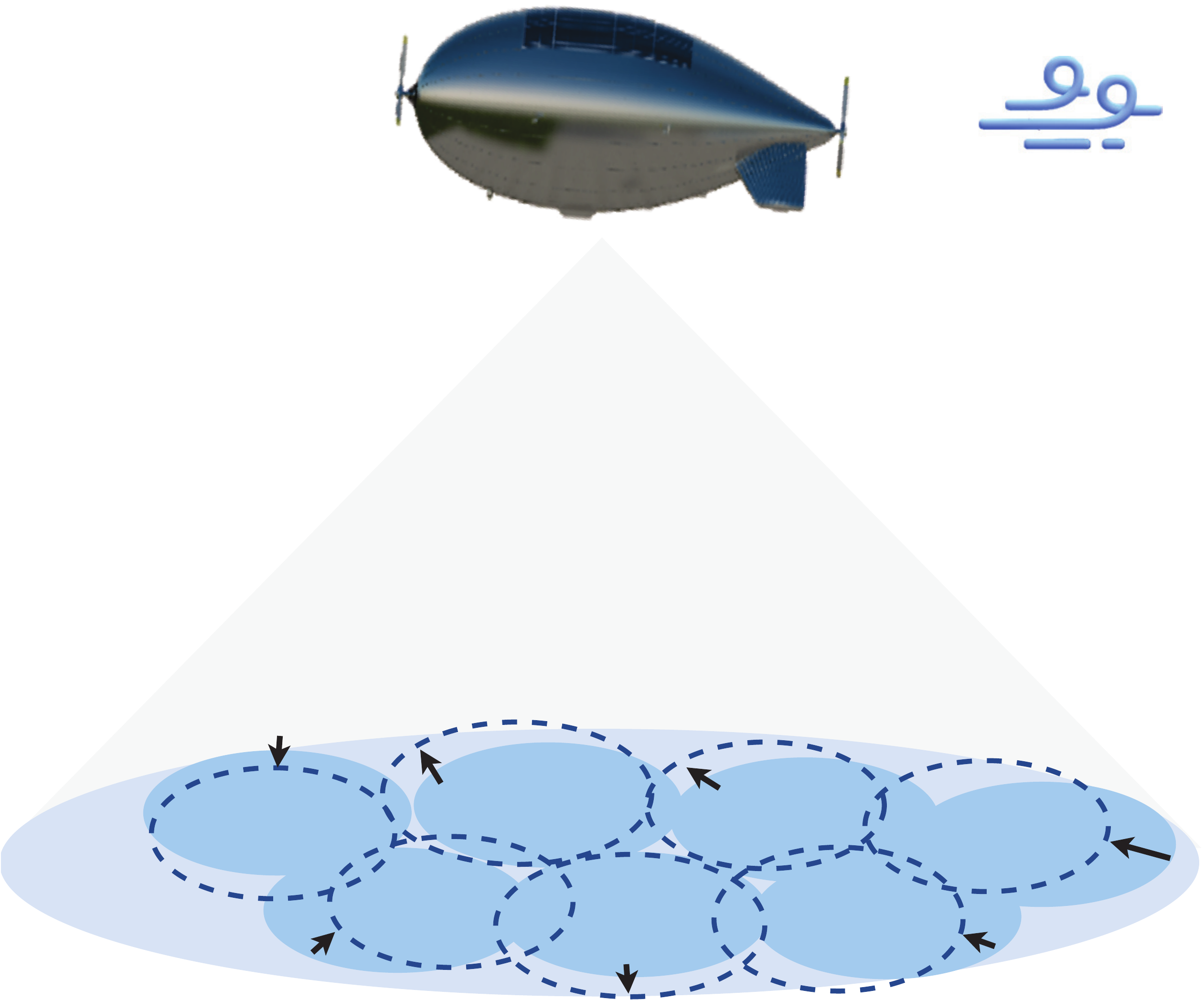}
		\caption{Changes in locations or shapes of footprints caused by airflow and mechanical vibration of platform.}
		\label{fig:footprint}
	\end{figure}

{\textbf{Discussion:}} As shown in Fig. \ref{fig:attitude_change}, the movement and mechanical vibrations of airborne platforms, as well as the airflow in the troposphere and stratosphere, can cause changes in attitudes and locations of airborne platforms. %\cite{grace2010low}. 
 %{We therefore conducted an experiment to test the realistic effects of these factors on an airship.}
 %Fig. \ref{fig:attitude_change} shows the attitude changes of an airship caused by mechanical vibrations and sudden airflow over two days. From this figure, we can observe that the pitch and the roll angles can vary by 3.3$^{\circ}$ and 1$^{\circ}$. Besides, Fig. \ref{fig:yaw_change} shows that sudden airflow has a significant affect on the yaw angle of airship, and the change is more than 5$^{\circ}$. %{The experimental results justify the necessity of investigating the phase delay of antennas mounted on a HAP.} 
 The locations or shapes of footprints projected by airborne platforms will then change frequently, as depicted in Figure \ref{fig:footprint}. 
 {Consequently, the handoff process of users in a footprint may be affected.% be triggered frequently. 
 %Therefore, both the mobility of serving users and the instability and mobility of airborne platforms will trigger the handoff process. 
 }
 % \begin{figure}[!t]
	% 	\centering
	% 	\includegraphics[width=3.2 in]{roll_angle.pdf}
	% 	\caption{{Attitude changes of an airship.}}
	% 	\label{fig:attitude_change}
	% \end{figure}
	% %The unstable platform movement caused by airflow includes the platform horizontal drift, vertical motion, rotation, and attitude angle disturbance (i.e., yaw, pitch, and roll/swing) \cite{grace2010low}.
 %  \begin{figure}[!t]
	% 	\centering
	% 	\includegraphics[width=3.0 in, height = 1.4 in]{yaw_angle.pdf}
	% 	\caption{{Change in the yaw angle of an airship.}}
	% 	\label{fig:yaw_change}
	% \end{figure}
	Nevertheless, there are some differences in terms of the handoff frequency for users in the footprints of airborne platforms. 
	The handoff frequency is influenced by various factors, including the distribution of users (e.g., inner-footprint users vs. footprint edge users), the speeds of both platforms and users (slow vs. fast), the coverage area of platforms (large vs. small), and the status of platforms (e.g., normal vs. abnormal operation).
	For instance, unstable platform movements will greatly increase the handoff probabilities of edge users in a footprint. 
	
	{\textbf{Classification:}} Handoff types can be categorized into two groups: hard handoff and soft handoff.
	In the case of a hard handoff, the session established by an old network is terminated before it is re-established by a new network. The situation is opposite in the case of a soft hand, where the session is maintained by the new network before the old network breaks the session. 
	
	Besides, the handoff can be horizontal or vertical. Horizontal handoff indicates the mitigation of a session from one footprint to another in the same network. On the contrary, vertical handoff refers to the seamless transition of a session between two heterogeneous networks.
	
	The handoff process results in the mitigation of state information and PHY layer connectivity from one footprint to another. Specifically, the handoff process consists of three phases: footprint discovery, trigger, and execution phases. During the footprint discovery phase, a user actively sends probes or passively listens to broadcast service advertisements to discover several footprints on assigned channels.
    According to the discovery, the user establishes a list of candidate footprints prioritized by handoff criteria (e.g., received signal strength (RSS), carrier-to-interference ratio (CIR), congestion status, and surveillance information).
	During the trigger phase, the decision to initiate a handoff is made based on various criteria, such as sub-network rental fees, SNR, handoff latency, elevation angle, and the availability of free channels.
	During the execution phase, the inter-footprint mitigation of sessions takes place, ensuring the seamless transfer of all contextual information related to the user. 
	
\begin{figure}[!t]
		\centering
		\includegraphics[width=3.4 in]{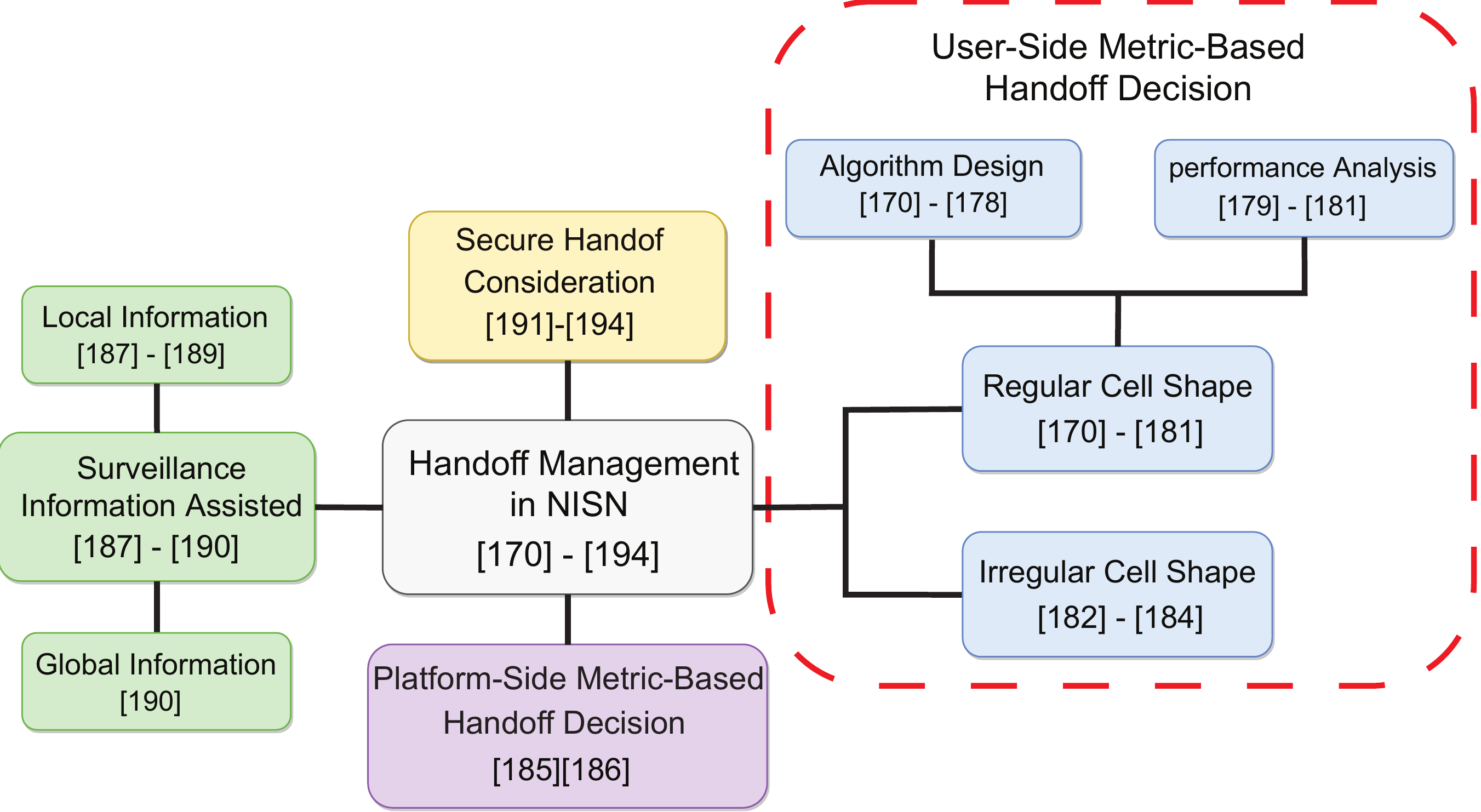}
		\caption{Classification diagram for the key handoff management strategies for NSIN in the literature.}
		\label{fig:handoff_management}
	\end{figure}

	A well-designed handoff algorithm is essential to reducing the overhead (e.g., session interruption, packet loss, handoff latency) of the handoff process \cite{Ren2023HandoffAwareDC}. 
Therefore, we highlight some of the key handoff management strategies developed for NSIN in the literature, the classification diagram of which is summarized in Fig. \ref{fig:handoff_management}.

     {\textbf{Existing studies:}} One of the significant goals of handoff algorithms is to ensure that the disruption experienced by users during handoff is imperceptible.
	Recently, a large number of methods have been introduced in the literature for reducing the overhead during the handoff process \cite{ grace2010low,Ren2023HandoffAwareDC,ren2022caching,DBLP:journals/iotj/ChenMLLC22,DBLP:journals/wc/KatoFTMTOL19, DBLP:journals/cm/ZhouFZCZH18,wei2022service,alsamhi2014performance,DBLP:conf/wcsp/HeCN16}. 
	For example, the work in \cite{ grace2010low} investigated the horizontal handoff between footprints served by the same HAP and the important problem of controlling which footprint to switch to at which time. With the aid of the measured RSS/CIR, a low-latency handoff algorithm was proposed. In this algorithm, a time-reuse time-division multiple/time-division multiple access (TDM/TDMA) frame structure was exploited. To maintain transparency in the handoff process for users, a single-frequency variant was proposed in which the HAP transmitted to or received from different spot beams (i.e., footprints) in various parts of the frame.
	
	Handoff algorithms with a fixed RSS/CIR threshold, however, could not effectively tackle the handoff issues in a HAP network. This is because the HAP is usually in a quasi-stationary state, and the RSS of cell edge users in the HAP will dynamically change. As a result, the fixed threshold setting will cause frequent handoffs between cells. To address this issue, the work in \cite{ DBLP:conf/wcsp/HeCN16} proposed an adaptive handoff algorithm to trigger the handoff according to predicted RSS by an improved least mean square (LMS) scheme. It was shown that the adaptive handoff algorithm could significantly reduce the unnecessary handoff time and keep the handoff failure probability at a lower level.
	
	Except for developing some schemes to manage the handoff, many researchers have paid much attention to the analysis of the impact of airborne platforms’ unstable movements on the handoff probability \cite{ he2016handover,siqing2011effect, albagory2014handover}. 
	For instance, the swing angle changes rapidly and unpredictably, making it challenging to adjust the parameters and orientations of airborne antennas in real-time to mitigate the impact caused by the swing.
	To this end, the work in \cite{ he2016handover} discussed the swing state of a HAP, analyzed the impact of the swing state on the handoff, and derived the average and maximum handoff probabilities.
	A novel HAP swinging model based on fixed angle beamforming was proposed in \cite{ siqing2011effect}. Owing to the platform swing, the so-called ``Ping-Pong handoff” would happen. Then, the authors analyzed the impact of platform swing in two cases: 1) all users executed the handoff immediately when the swing happened; 2) all users refused to handoff (if triggered) to avoid the “Ping-Pong handoff”. 
	Furthermore, in \cite{albagory2014handover}, the issue of yaw shift of HAP was addressed, and its influence on handoff was examined. A formula was derived to calculate the number of handoff calls, which indicated that various factors such as users' density and location distribution in the cell, HAP's yaw angle, cell geometry, and the number of active calling users could impact the number of handoff calls.
	
	In the above studies \cite{grace2010low,Ren2023HandoffAwareDC,ren2022caching,DBLP:journals/iotj/ChenMLLC22,DBLP:journals/wc/KatoFTMTOL19, DBLP:journals/cm/ZhouFZCZH18,wei2022service,alsamhi2014performance, DBLP:conf/wcsp/HeCN16,he2016handover,siqing2011effect,albagory2014handover}, a simple coverage model such as the circular and regular hexagonal coverage was introduced to mathematically analyze the handoff. While the mathematical analysis is simplified, it poses challenges to validating the derived conclusions in practical engineering scenarios. 
	Therefore, some handoff algorithms based on irregular or inconsistent cell shapes were developed \cite{ albagory2013smart,katzis2010inter,he2017adaptive}. 
	The work in \cite{ albagory2013smart} proposed a smart cell shape design for handoff management in a HAP network. Using user distribution and behavior data, the smart network design divided the coverage area into smaller spots and synthesized the desired cell by grouping some of these spots together. Consequently, the frequent handoffs from moving users could be significantly reduced. 
	To investigate various issues related to the horizontal HAP handoff process, the work in \cite{ katzis2010inter} designed inconsistent footprint shapes for users locating at different positions of the service area using directional antennas. 
    Based on the results, it was found that a wide beam-width antenna was suitable for serving users located in the center of the service area, whereas users located at the edge of the service area would benefit from a narrow beam-width antenna. 
	To address the frequent handoff issue caused by the user movement as well as the unstable HAP movement, an adaptive handoff scheme with an overlapped footprint design was developed in \cite{ he2017adaptive}. Particularly, this paper utilized the mechanism of cooperative transmission between two HAPs to guarantee the QoS requirements of users at the edge of the coverage area. To improve transmission reliability, the HAP with a higher channel gain was selected for cooperative transmission. Additionally, handoff decisions were made based on the user's movement direction and channel gain to minimize session interruption time caused by frequent handoffs.
	
	In the above studies \cite{albagory2013smart,katzis2010inter,he2017adaptive}, single-metric (e.g., RSS/CIR) or user-side multi-metric (e.g., user distribution and user movement direction) handoff decision algorithms were designed. 
	However, the metrics on the airborne platform side play a crucial role in the handoff decision-making process. For instance, when multiple users make handoff decisions simultaneously, solely relying on the RSS-based handoff algorithm can lead to HAPs with high RSS overload and HAPs with low RSS idle, potentially causing load imbalance within HAP networks. This, in turn, can result in network congestion and packet loss. 
	Therefore, some handoff algorithms considering metrics on the airborne platform side were designed \cite{an2013load, rouzbehani2011fuzzy}. 
	For instance, a load-balancing handoff algorithm considering both the RSS of users and the residual energy of HAPs was developed in \cite{an2013load}.
	To minimize handoff failures during changes in network traffic load, a fuzzy channel allocation scheme was introduced in \cite{rouzbehani2011fuzzy}. This scheme employed a guard channel policy and dynamically adjusted the optimal number of reserved channels based on QoS parameters. By efficiently allocating channel resources, the scheme prioritized ongoing calls over new calls.
	
	Diverse from the above handoff schemes, local and global surveillance information collected by airborne platforms was utilized to assist the handover decision-making in \cite{DBLP:journals/vtm/SunGSGA21, zhou2022handover, DBLP:conf/qrs/ZhouSYGGA21}. Particularly, to decrease handoff frequency, local information such as the broadcast location, velocity, and flying direction of the airborne platform was leveraged to predict its flight trajectory. A global information-supported machine learning (ML) method could effectively predict future traffic loads, and then a joint decision on handoff and resource allocation could be made.
	Additionally, in order to ensure uninterrupted UAV connectivity within an NSIN-integrated network, a predictive decision algorithm based on global air control information and a handoff strategy were devised in \cite{zeng2022predictive}. Leveraging the global air control information, the predictive algorithm determined available access points for UAVs to establish new links or recover from old links in future time slots. Complementing the predictive algorithm, the handoff strategy seamlessly maintained UAV connectivity by transitioning among soft handoff, tracking hard handoff, and non-tracking hardoff states based on the predictive results.
	
	However, the important issue of secure handoff was not investigated in the above studies. Some recent studies discussed the secure access authentication issue in the handoff process \cite{DBLP:conf/pimrc/DingZL20,DBLP:journals/wc/YaoGWXX20,DBLP:conf/icc/LaiC21, DBLP:conf/pst/HeLNY21}. 
	Ensuring fast and secure handover is crucial for maintaining seamless coverage of NSIN over a wide area.
    The work in \cite{DBLP:conf/pimrc/DingZL20} investigated the optimization of the handoff latency with a given security requirement in an NSIN-integrated network. A new handoff approach that utilizes pre-authentication and security context transfer was suggested.
	The proposed scheme distinguishes between two types of handoffs: intra-domain and cross-domain. In the case of intra-domain handoff, the user only needs to initiate an application to the controller. However, for cross-domain handoff, the user must initiate an application to the management center, which involves saving all relevant information related to the NSIN-integrated network.
	To reduce the handoff latency while ensuring security, some temporary proxy certificates were transferred in advance. 
	Meanwhile, to overcome the shortcomings of large handoff latency because of security requirements, the operations of platform trajectory prediction and pre-authentication were proactively performed based on periodically updated network status information. 
	The authors in \cite{DBLP:journals/wc/YaoGWXX20} developed a secure and lightweight access authentication scheme for handoff management in NSINs and space networks. To minimize authentication delay and signaling overhead during the handoff process, the access authentication scheme for re-authentication message transmission adopted a multicast communication mode. %The authentication process consisted of three phases: initialization, registration, and multicast authentication.
 Nevertheless, the security of a private key generator in the event of the failure of a single network node was not investigated. 
% Future work is to improve the security of private key generator to avoid a single point of failure, and the available methods include adopting distributed PKG or lightweight blockchain to enhance the reliability and security. 
 {To tackle this issue, blockchain techniques can be explored. As a distributed architecture, blockchain ensures the credibility of data and prevents tampering. However, the process of achieving consensus requires exchanging transaction data and authentication messages, which consumes a significant amount of network bandwidth and computing resources. An available scheme is to design a joint $\mu$Tesla and blockchain authentication protocol. $\mu$Tesla is leveraged to achieve lightweight authentication, and blockchain is utilized to manage the authentication \cite{Garcia2023TeslaBasedAF}.
 }
	
	\subsection{Network Management}
The recent studies on NSIN network management can be classified into network topology management and integrated network management. 
A summary of the corresponding management approaches is presented in Table \ref{table_network_management}. 
 \begin{table*}[!t]
	\renewcommand{\arraystretch}{1.2}
	\caption{A summary of goal, key technique, and key idea of network management approaches proposed in the literature about NSIN}
	\label{table_network_management}
	\newcommand{\tabincell}[2]{\begin{tabular}{@{}#1@{}}#2\end{tabular}}
	\centering
	\begin{tabular}{l}
%\begin{minipage}{0.18\textwidth}
			\centering
			\includegraphics[width=7.0 in]{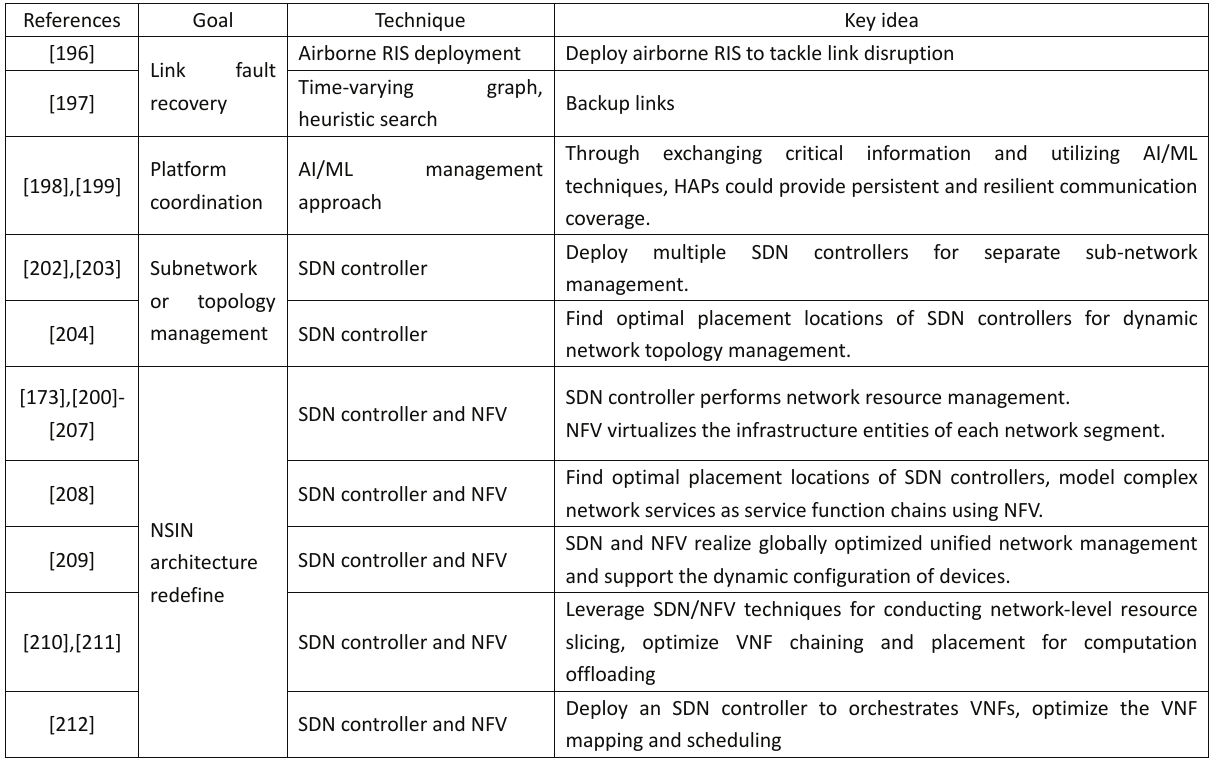}
		%\end{minipage} 
	\end{tabular}
\end{table*} 
\subsubsection{Network topology management}
The topology management of NSINs is essential for making NSINs more stable, efficient, and adaptable to meet changing application requirements and environmental conditions. 
%efficient network services and is a hot research topic. 
%To enhance the overall performance of the network to make it more stable, efficient and adaptable to meet 

 {\textbf{Discussion:}} Compared to space networks and TNs, NSINs possess distinctively dynamic connections and topologies. 
	For example, the flight trajectories of some platforms may be random in a 3D airspace, resulting in intermittent and unpredictable connections. The operational duration of flight platforms in NSIN is significantly shorter compared to network nodes in space networks and TN. 
	In this case, NSIN should be resilient to the dynamical topology to sustain service provision. 
    Besides, as en essential component of 6G networks, NSIN should be designed to efficiently support multiple types of traffic concurrently. 
	Adopting self-organization mechanisms holds promise for NSIN to be resilient to topology changes and adapt to distinct service requests.
    These mechanisms encompass self-configuration, self-healing, self-optimization, and self-protection functions. 
	Self-configuration indicates the automatic recognition and registration of new neighbor nodes. 
    Self-healing mechanisms in NSIN automatically address node failures and connection interruptions, ensuring uninterrupted services or mitigating service degradation.
    Additionally, self-optimization mechanisms automatically optimize the technical parameters of nodes to achieve specific objectives.
	Finally, to ensure network security and data confidentiality, the self-protection automatically protects NSIN from penetration by unregistered users.
	Let's illustrate the aforementioned process with an example. When a new node emerges or an existing node fails, neighboring nodes undergo a neighbor discovery process to acknowledge this change. Network modifications trigger a series of message exchanges among nodes to reorganize the network.
	The next step is the connectivity establishment of nodes during self-organization by executing connection mechanisms in the PHY and upper layers. 
	After establishing the connections, the service recovery process will be performed to recover from local service disruptions.
	Meanwhile, some optimization schemes, such as energy optimization, will be implemented to implement network and service optimization. 
	
	However, the task of designing dependable and effective self-organization functionalities in NSIN is challenging. 
	The processes of self-organization require frequent measurements and probing. Then, the important issue of the trade-off between the optimality of self-organization and the signaling cost should be investigated. 
	Incomplete, delayed, or even erroneous feedback may happen, which may significantly impact the efficiency of self-optimization processes and the service performance of the re-organized NSIN. 
	Meanwhile, ultra-reliable self-organization functionalities in NSIN should be designed to minimize or avoid human intervention. 
	
	{\textbf{Existing studies:}} Some endeavors have been made to ease the design of self-organization functionalities in NSIN \cite{bariah2021ris,qu2022link,anicho2019comparative,anicho2019autonomously}. For example, considering the unprecedented abilities of RIS in terms of enhancing spectral efficiency, coverage expansion, and security enhancement, the work in \cite{bariah2021ris} proposed to utilize RIS technologies in NSIN-integrated air-space networks. With the utilization of RIS, the quality of inter- and intra-layer communications was improved, and the job of heterogeneous network self-organization was eased \cite{bariah2021ris}. 
 
 By utilizing RIS to program the dynamic RF environment, the traditional concept of cell boundaries could also be disrupted.
    Airborne platforms can be configured as airborne-RISs to extend service coverage. Link disruption caused by interference can also be mitigated by airborne-RISs. 
    By analyzing a time sequence link weight graph, the authors of \cite{qu2022link} developed a link fault recovery approach that targeted the unstable inter-domain neighbor relationships, frequent routing updates, and slow routing convergence issues in an NSIN-integrated space-ground network.The key idea of this approach was to dynamically choose backup links and search for a recovery path with the shortest length and the maximum path weight based on backup links when the link failure occurred.
    
	AI/ML techniques can also be explored to effectively tackle the important issues in the self-organized NSIN. 
	Deploying multiple HAPs as a network can greatly boost communication coverage. 
	However, the construction of HAP networks is technically and economically challenging. The proposal for network self-organization aims to eliminate the need for direct human intervention and reduce operational costs.
    The work in \cite{anicho2019comparative} then applied reinforcement learning (RL) and swarm intelligence approaches to solve the self-organization and coordination problem of multiple HAPs for communication coverage. 
	The work in \cite{anicho2019autonomously} investigated the coordination of multiple self-organizing HAPs in a volcanic cloud emergency scenario for communication coverage. 
	To coordinate a swarm of HAPs, a swarm intelligence-based algorithm was developed. This algorithm comprised different modes, including scouting mode, exploitation mode, decision loop, and main exploitation mode. Similar to a foraging behavior, the participating HAPs exchanged vital information as they explored the environment. The HAPs achieved persistent and resilient communication coverage by exchanging critical information and utilizing swarm techniques. The swarm's resilience was tested by simulating the failure of a participating HAP.
	
\subsubsection{Integrated network management}	
 As mentioned above, NSIN are a type of heterogeneous, stereoscopic, multi-tier, and time-varying networks. Different frequency bands may be allocated to diverse sub-networks (or network segments), and the capabilities (e.g., communication, computing, and storage) of airborne platforms in different sub-networks are heterogeneous; thus, the communication system of a specific sub-network may be separate from other sub-networks. The separation of sub-networks will significantly limit the reconfigurability and interoperability of NSIN and greatly increase network management costs and complexity.
	
	\textbf{Discussion:} Exploring the concepts of software and virtualization, software-defined networking (SDN) and network function virtualization (NFV) techniques can redefine the NSIN architecture to tackle some challenges in NSIN, including inconsistent air interfaces and complicated heterogeneity, as well as reduce the management costs and complexity of NSIN. 
	Furthermore, the flexible integration of sub-networks in NSIN can be greatly enhanced by leveraging SDN and NFV techniques. These techniques enable advanced network management strategies, promote business agility, and facilitate service innovation.

    On one hand, SDN facilitates centralized control in the data plane by decoupling the controller from the switch.
    This grants the controller a comprehensive overview of NSIN and enables the global management capability. 
	In the data plane, SDN-enabled platforms will receive instructions from the centralized controller and do not need to understand diverse protocols. In this way, the flexibility of managing these platforms can be significantly improved.
	
	On the other hand, NFV can implement various network functions through visualization technologies and decouple network functions from proprietary hardware devices. Thus, the dispatching of network functions as instances of common software to service providers breaks through the barriers created by dedicated hardware devices.
	
\begin{figure*}[!t]
		\centering
		\includegraphics[width=6.9 in]{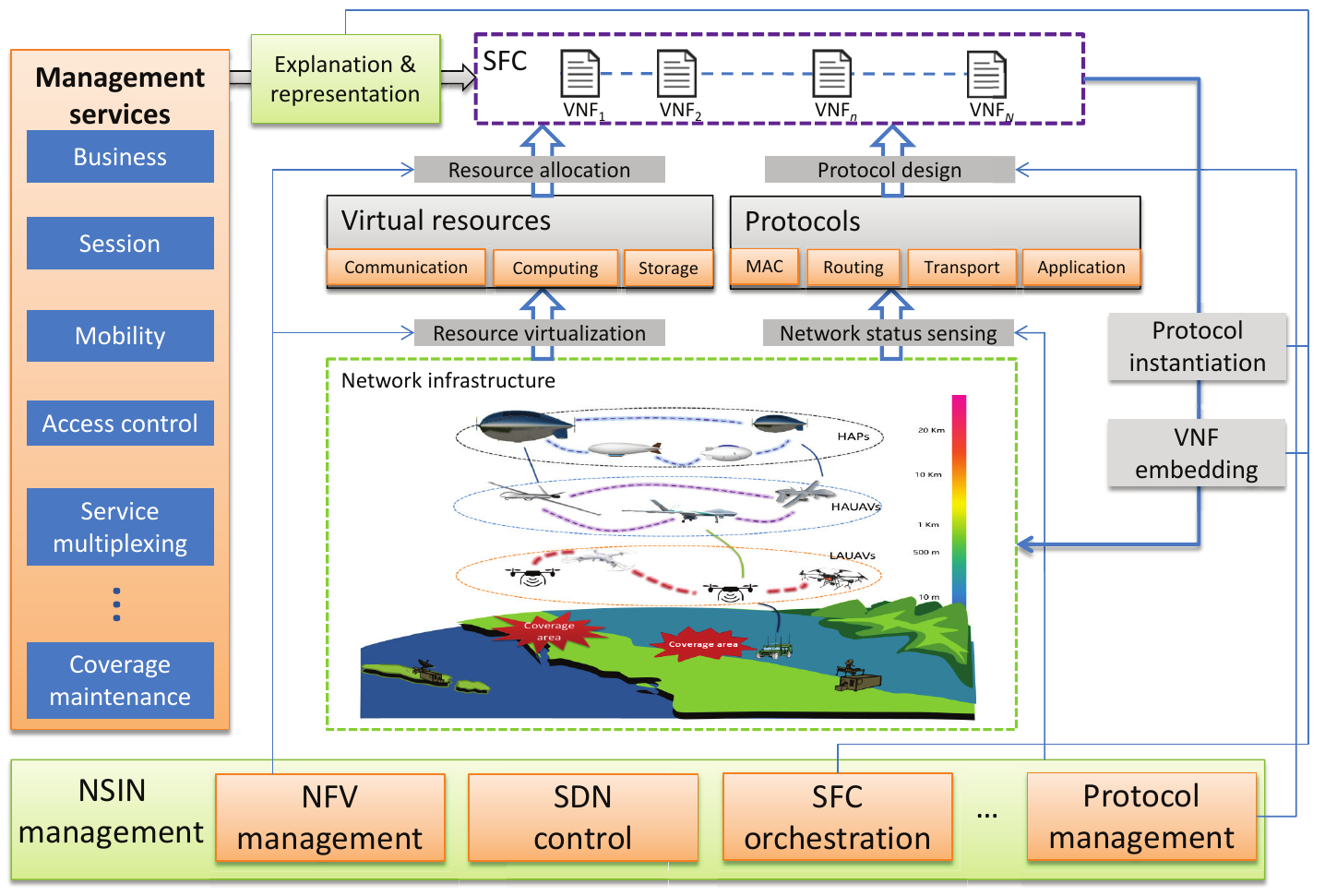}
		\caption{SDN/NFV-enabled NSIN management architecture.}
		\label{fig:fig_SDN_NFV_Mag_Archi}
	\end{figure*}

	Fig. \ref{fig:fig_SDN_NFV_Mag_Archi} illustrates an SDN- and NFV-based network management framework of NSIN for providing flexible network services. In this framework, the data flow needs to flow through a sequence of virtualized network functions (VNFs) to implement a specific service. Different network services such as business management, session management, mobility management, access control, service multiplexing, and coverage maintenance can be implemented by diverse service function chains (i.e., sets of ordered virtual functions with logical dependencies). 
	
	By exploring the concept of NFV, heterogeneous resources from diverse sub-networks can be extracted into unified virtual resource pools for flexible resource management, orchestration, and allocation. 
	Separating the data plane from the control plane enables the SDN controller to design efficient network protocols and issue instructions for SDN-enabled platforms. 
	This type of framework enables flexible network reconfiguration, dynamically adapting network service provisions to changing service requirements. 
	When service requests change or the capability of network service provisions changes (due to dynamic network topology), reconfigured services can be easily implemented by updating software or the sequence of VNFs. 
	
	\textbf{Existing studies:} Recently, many promising SDN and NFV solutions have been developed to manage the integrated NSIN \cite{DBLP:journals/iotj/ChenMLLC22,Yuan2022SoftwareDI,DBLP:journals/access/RuanLZCL20,DBLP:journals/cn/ZhangZW20,DBLP:journals/vtm/ZhangXMXDLKF19,yang2020research,sheng2021space,DBLP:journals/comsur/LiuSFK18,DBLP:journals/comsur/GuoLLTK22,torkzaban2021joint,hong2020space,DBLP:journals/pieee/ZhuangYLCR20,DBLP:journals/wc/WuCZSCXZS20,DBLP:journals/iotj/LiSWZS22}.
	In \cite{Yuan2022SoftwareDI}, an SDN- and NFV-enabled system architecture consisting of infrastructure, control, and application layers for NSIN-integrated network management was designed. 
	In the infrastructure layer, some airborne platforms acted as SDN-enabled switches or gateways to forward data packets based on flow tables. The infrastructure entities in each sub-network were virtualized, and VNFs were deployed on demand. 
	In the control layer, both master SDN controllers and slave SDN controllers were configured to effectively manage the network. 
	The master SDN controller collaborated with its slave counterparts to gather the global network status and oversee the entire network topology for resource management at a high level, while the slave SDN controllers undertook fine-grained resource management tasks. 
	The application layer would provide diverse programmable services and management modules through some interfaces. %The interfaces will also act as an interpretation bridge between 
	The resource request issues of the application layer would be interpreted by the interfaces into the instructions of SDN controllers. 
	In \cite{torkzaban2021joint}, an SDN/NFV-enabled NSIN-integrated network management architecture was developed. To realize the SDN/NFV-enabled architecture effectively, the crucial issues of the design of gateway deployment policies, the optimal placement of SDN controllers, the embedding of service function chains, and the establishment of traffic routes between these chains were investigated. 
	In \cite{qiu2019air}, an integrated NSIN hierarchical network architecture was proposed for beyond 5G (B5G) wireless networks. In this architecture, a centralized SDN controller was embedded to manage the operation of the whole NSIN and incorporated the cross-tier sub-networks at the upper level. Three separate SDN controllers in the lower layer were embedded into sub-networks for local network management (or fine-grained management). 
	In \cite{qu2020sdn}, an SDN-enabled NSIN-integrated network architecture was developed for network management. Particularly, the optimal placement strategy of an SDN controller for dynamic network topology management was studied.
	Besides, an SDN/NFV-enabled NSIN-integrated network architecture was proposed in \cite{hong2020space} to implement unified network management and dynamic device configuration. 
	In the NSIN-integrated network, SDN controllers were deployed to create SDN tunnels in the core network, and VNFs were deployed in the clouds to enhance network flexibility and reliability.

	\subsection{Summary and {Lessons Learned}}
{\textbf{Summary:}} In this section, we discuss the existing works on QoS- and QoE-driven NSIN deployment in detail, which may shed light on the comprehensive modeling of NSIN deployment problems.
Handoff has a significant impact on the service performance of NSINs, especially for users located near the boundaries of NSIN footprints. 
We explore the primary reasons for handoff among NSIN-covered users, summarize the impact of handoff frequency, and survey recent advances in addressing handoff issues. 
We discuss the network management issue of NSINs from the perspectives of network topology management and integrated network management. 
Specifically, we provide detailed explanations of the procedures involved in network self-organization mechanisms for topology management, and we provide an overview of network architectures for integrated network management based on SDN and NFV {techniques}.

 {\textbf{Lessons learned:}}
 The deployment of NSINs is task-driven, and therefore, NSINs should be deployed to satisfy the QoS and QoE requirements of tasks. 
It is essential to formulate the NSIN deployment problem as an optimization problem. 
However, this type of optimization problem is hard to formulate due to the multi-layered and heterogeneous network structure as well as {the existence of} some restrictions on the flight airspace and the platform itself. 
Therefore, great efforts should be spent {in the process of modeling the problem constraints with a} consideration of network characteristics. 
{For example, in order to characterize the heterogeneity of aerial platforms, a viable method is to constrain the service capabilities of platforms from the perspective of limiting available computing, storage, and network resources \cite{zhang2022impacts}.}
%Novel approaches are also desired to solve this type of problem effectively. 
%	Based on our research on NSINs deployment, handoff management, and network management, we can conclude that 
{Constructing large-scale NSINs is an important trend in NSIN development, where an efficient dynamic networking design is indispensable. 
The dynamic networking of large-scale NSINs needs to achieve efficient interference management. Yet, both the heterogeneity in terms of platform capability and deployment altitude and the large number of network platforms bring in complexly overlapped coverage. As a result, the interference experienced by network nodes or terminals served by NSINs becomes incredibly complex. To tackle this issue, one is encouraged to model the impact of the heterogeneity of network platforms on coverage and utilize the stochastic geometric theory to analyze the statistical average interference.}
 {Additionally, situational information, which includes security situations and network situations, is crucial for designing effective networking approaches.
	Yet, the research on situation-assisted NSIN networking is still in its initial stage, and the accurate perception of situations in NSINs is challenging. For instance, there are a large number of small threat targets without obvious features, which make it difficult to detect them in time and accurately.} 
{A viable method includes cooperative situation perception of heterogeneous aerial platforms. HAPs that have a wide-area sensing capability can send coarse perception information to UAVs. With the support of coarse information, UAVs approaching targets can generate accurate perception information \cite{Yang2022NetworkingOI}.}
	
	\section{Transmission via NSINs}
	In this section, we will present some key issues encountered in experiments, which should be considered when investigating the transmission topic of NSINs. 
 %Some PHY layer techniques and upper layer communication protocols will be surveyed. 
 %we overview the design of key air interface techniques that aim to solve the effective transmission issue in NSINs. 	
	%\subsection{PHY Layer}
 
\begin{figure}[!t]
		\centering
		\includegraphics[width=3.4 in]{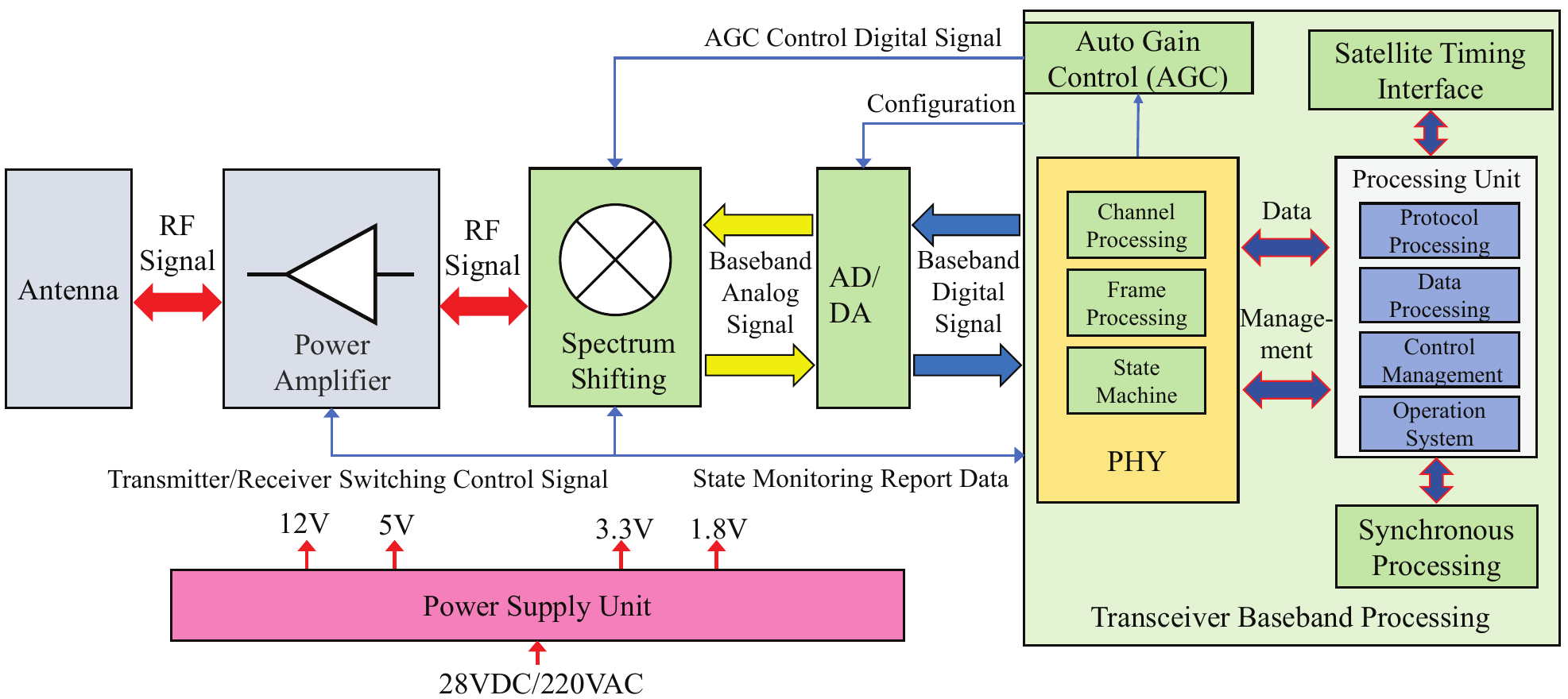}
		\caption{{A logical diagram of the digital signal processing module of Beihang airship.}}
		\label{fig:digital_PHY}
	\end{figure}

\begin{figure}[!t]
		\centering
		\includegraphics[width=3.4 in]{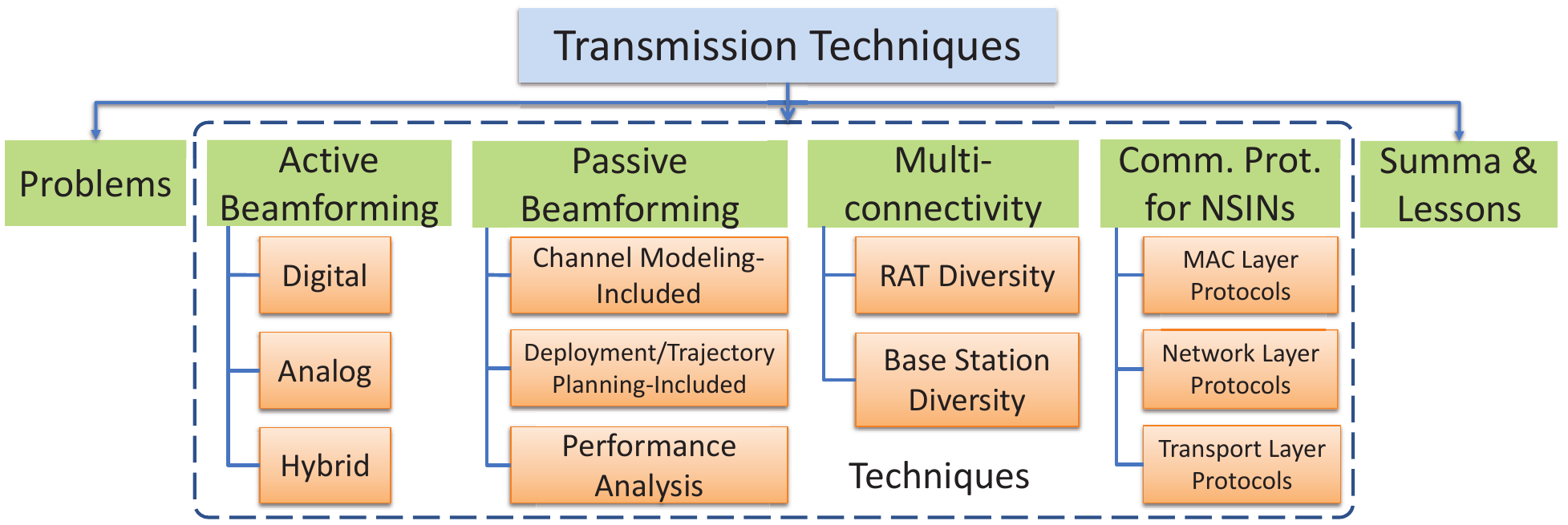}
		\caption{{Categorization of the transmission techniques in this section.}}
		\label{fig:multiconnectivity_PHY}
	\end{figure}

 Fig. \ref{fig:digital_PHY} shows a logical diagram of the digital signal processing module designed for communication systems of Beihang airship. In this PHY layer module, many techniques (e.g., beamforming and multi-connectivity) have to be studied. %Beamforming is one of the most significant techniques.
	The purpose of beamforming is to implement efficient coverage by electronically controlling the amplitudes and phases of the weights of antenna array elements.
	It can be divided into active beamforming and passive beamforming. 
 {Multi-connectivity is a key technology for achieving reliable and low-latency transmissions. 
 Besides, it is essential to overview the upper layer communication protocols of NSINs, which enable efficient end-to-end transmission.
%  Table \ref{table_beamforming} describes the main ideas, pros, and cons of different types of beamforming techniques. 
The technical categories of this section are illustrated in Fig. \ref{fig:multiconnectivity_PHY}.}
% 	\begin{table*}[!t]
% 	\renewcommand{\arraystretch}{1.2}
% 	\caption{Comparison of different types of beamforming techniques}
% 	\label{table_beamforming}
% 	\newcommand{\tabincell}[2]{\begin{tabular}{@{}#1@{}}#2\end{tabular}}
% 	\centering
% 	\begin{tabular}{l}
% %\begin{minipage}{0.18\textwidth}
% 			\centering
% 			\includegraphics[width=7.0 in]{Table_beamforming.pdf}
% 		%\end{minipage} 
% 	\end{tabular}
% \end{table*} 
% 	%Summarize the goal of the physical layer architecture design. The research directions in the aspect of phy layer design. e.g., Beamforming, interference management, resource optimization; Next, give some examples, 

\subsection{{Problems Encountered in Experiments}}
\begin{figure*}[!t]
\centering
    \includegraphics[width=5.5 in, height = 2.8 in]{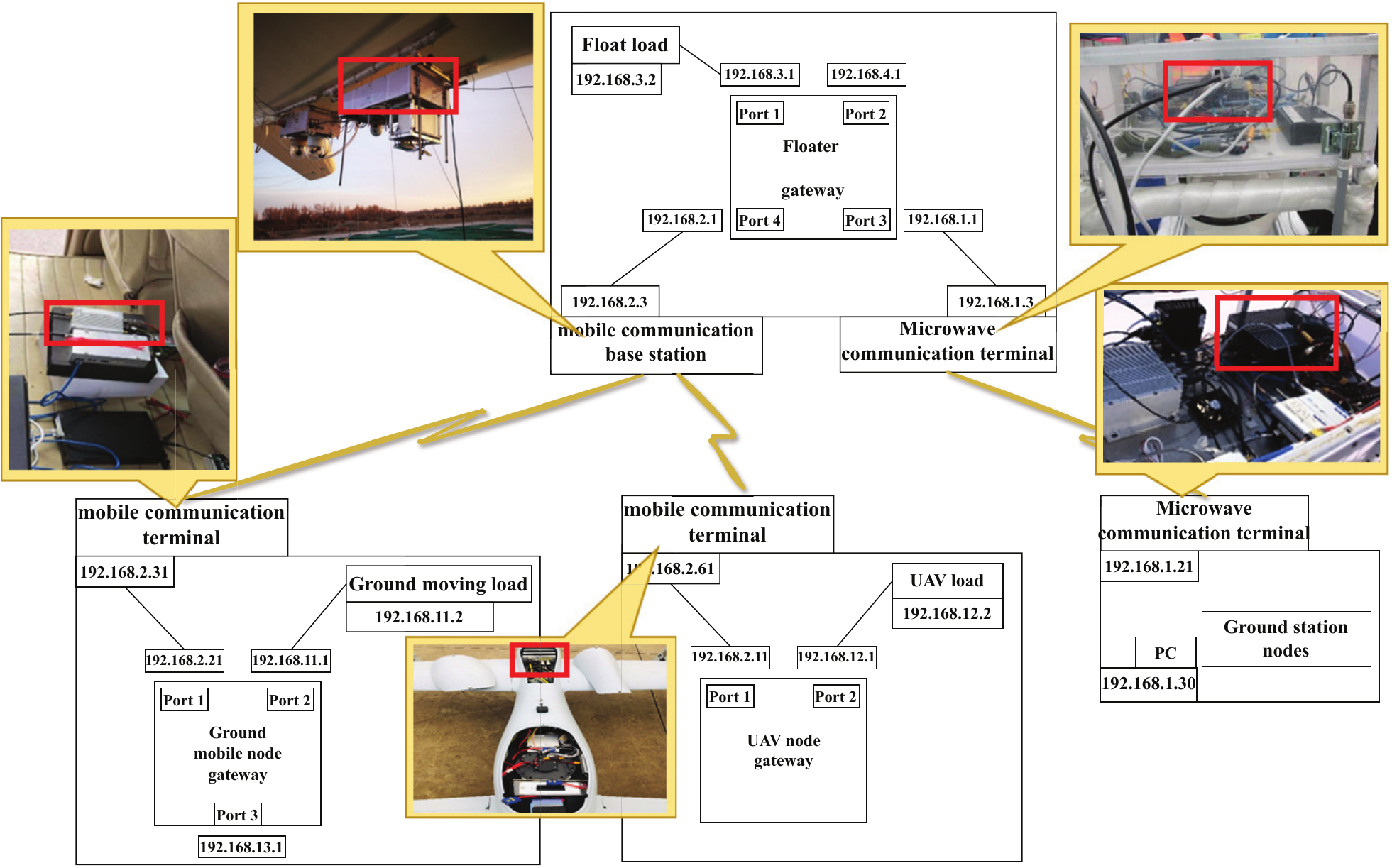}
    \caption{{The network topology of the NSINs utilized to transmit real-time video streams.}}
    % An experiment of conducting live steaming transmission of the front window view of a moving car via our NSINs
    \label{fig:topology}
\end{figure*}

{The authors in this article conducted an experiment of transmitting real-time video streaming via NSINs. 
Fig. \ref{fig:topology} illustrates the topology of the actual NSINs. As can be observed, the NSINs adopt two types of communication systems, including mobile communication and microwave communication systems. 
These two types of communication systems work at different frequency bands and adopt different frame structures and MAC protocols as well. 
The mobile communication system is designed to support communication access for wide-area mobile users. The microwave communication system is designed to achieve ultra-high speed information transmission. 
Therefore, heterogeneity is a significant characteristic of NSINs and should be taken into account when investigating the transmission techniques of NSINs.
}
%for real-time video transmission and the effects of video transmission. integrated communication protocols design Therefore, the heterogeneity in the adopted communication systems in NSINs should be taken into account when investigating the transmission techniques of NSINs.
	\subsection{Active beamforming}
Active beamforming array is an architecture where RF amplifiers are utilized as internal elements of an antenna array.
	According to the hardware structure of the antenna array, the active beamforming architecture can be classified into digital beamforming, analog beamforming, and hybrid beamforming.
	
	\subsubsection{Digital beamforming:} 
\begin{figure}[!t]
		\centering
		\includegraphics[width=2.6 in]{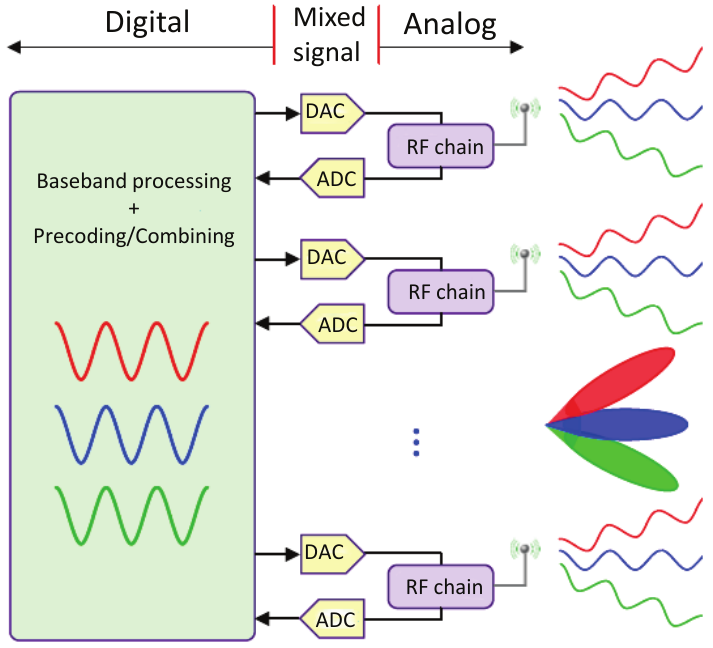}
		\caption{Architecture of the digital beamforming \cite{Qasim2023what}.}
		\label{fig:Digital_beamforming}
	\end{figure}

	For the digital beamforming architecture, as shown in Fig. \ref{fig:Digital_beamforming}, all operations (e.g., precoding, multiplexing, signal weighting, phase shifting) are performed in the digital domain.
	At the transmit end, amplitudes and phases of digital signals can be simultaneously and flexibly adjusted.
	%In digital beamforming, amplitude/phase variation (i.e. wk) is applied to digital signal after ADC/DDC conversion at the transmit end. In the receiver, received signals are first passed from ADC converters and digital down converters before summation operation. 
	%In this architecture, all the operations such as precoding, multiplexing, signal weighting, phase shifting etc. occurs in the digital domain which has many benefits. It is used in frequency selective beamforming applications.
	 In digital beamforming architecture, each antenna element has a dedicated RF chain, enabling a variable number of beams based on the number of elements forming the beam. Additionally, all beams can be steered, and algorithms can synthesize a drive signal to optimize the desired characteristics of each beam.
  
\begin{figure}[!ht]
		\centering
		\subfigure[{Experiment setup and configuration}]{
    \label{fig:subfig:f_setup} %% label for first subfigure
    \includegraphics[height = 2.3 in]{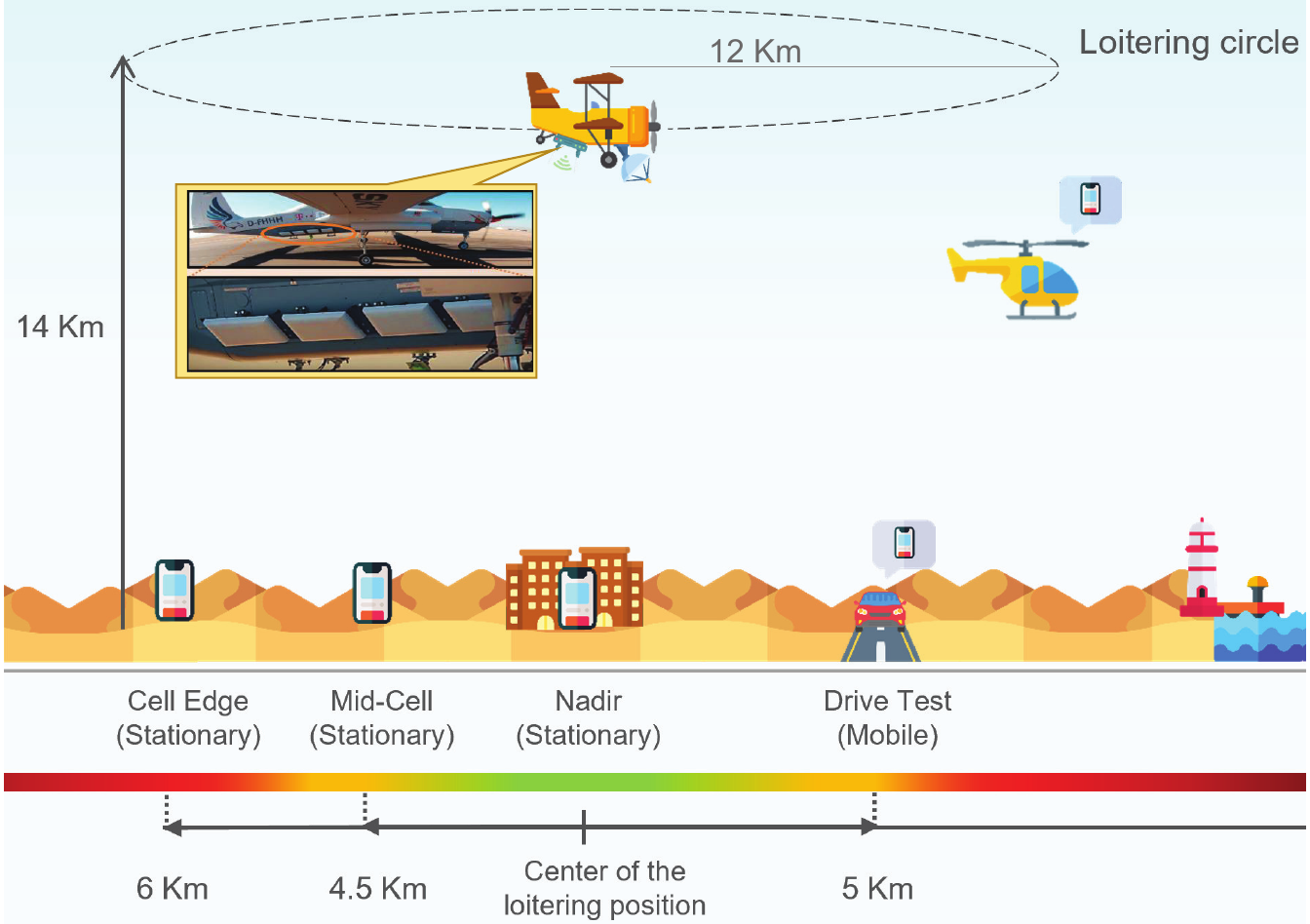}}
    \hspace{3pt}
    \subfigure[{The trend of user’s receiving data rates with a constant user-HAP distance 18.2 km}]{
    \label{fig:subfig:f_constant} %% label for first subfigure
    \includegraphics[width=2.1 in]{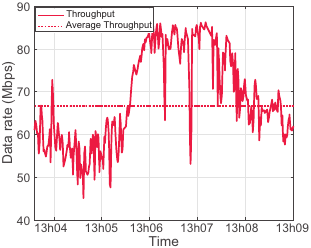}}
    \hspace{3pt}
    \subfigure[{The trend of user’s receiving data rates with varying the user-HAP distance}]{
    \label{fig:subfig:g_varing} %% label for first subfigure
    \includegraphics[width=2.1 in]{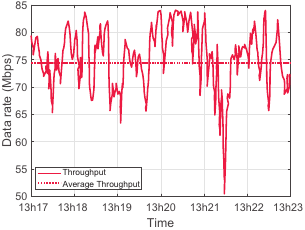}}
		\caption{{A digital beamforming experiment \cite{Belmekki2024CellularNF}.}}
		\label{fig:channel_measurement}
	\end{figure}

\textbf{Existing studies:}	Owing to its high flexibility in producing beams, many digital beamforming methods for NSIN have been proposed in recent years \cite{DBLP:conf/vtc/HoshinoSO19,DBLP:journals/wcl/NaPKA22,DBLP:journals/wcl/LianJHH19,DBLP:journals/icl/SudheeshMMSM18,DBLP:conf/apcc/ZakiaTIK13,DBLP:journals/tsp/LinLHCZ19,DBLP:conf/wpmc/MiuraMOMST21,DBLP:journals/wcl/NaPKA22,DBLP:conf/wpmc/OuchiKC21}. For example, \cite{DBLP:conf/vtc/HoshinoSO19} introduced a digital beamforming method that combined horizontal and vertical adjustments to address the displacement of HAP cells projected on the ground due to variations in stratospheric wind or turning flight.
	The work in \cite{DBLP:journals/wcl/NaPKA22} investigated the interference mitigation problem in NSIN by optimizing beamforming vectors of HAPs and formulated this optimization problem as second-order cone programming, which was solved by an interior-point method. The work in \cite{DBLP:journals/wcl/LianJHH19} studied the digital beamforming for HAP multiple-input multiple-output (MIMO) systems and derived a beamformer matrix according to the critical observation that signal power concentrated on a statistical eigenmode instead of a statistical correlation matrix. However, the above beamforming schemes \cite{DBLP:journals/wcl/NaPKA22,DBLP:journals/wcl/LianJHH19} were developed for stationary HAP networks, and the impact of the movement of HAP was not considered.
	%However, the digital beamforming architecture increases digital chip content and the corresponding cost and power consumption. %Therefore, 
 %It consumes highest DC power.	It offers I/Q signal routing complexity.	It offers LO signal routing complexity.	It offers highest hardware complexity i.e. full RF chain per element or column in the array.	Hardware can not fit within the lattice at high frequencies (no 2D scan capability for planar arrays).
	%Fig. 7 (a) illustrates the basic digital beamforming architecture at the transmitter, where each AE is connected to an independent radio-frequency (RF) chain. Beamforming is performed in the baseband via digital signal processing, which yields a high flexibility with sufficient DoFs to implement efficient precoding algorithms. Thus, in theory, digital beamforming achieves a higher performance compared to other architectures [74]. It can accommodate multi-stream transmission, and can distinguish signals simultaneously received from different directions. However, the digital beamforming architecture requires a dedicated RF chain for each AE. The corresponding hardware components, including ADCs, DACs, data converters, and mixers, entail a high hardware complexity and a large energy consumption. 
	The authors in \cite{Belmekki2024CellularNF} conducted an experiment using a moving HAP to provide high data rates and ubiquitous 5G coverage via digital beamforming. 
 Fig. \ref{fig:subfig:f_setup} illustrates the experiment setup and configuration. The communication architecture consists of a 5G gNB and a backhaul radio on-board a HAP. 
 From Fig. \ref{fig:subfig:f_constant} and \ref{fig:subfig:g_varing}, it can be observed that the peak downlink data rate was 86.5 Mbps, and the average downlink data rate was 68.3 Mbps, when a user is located at the center point of the center-cell. The data rate fluctuation is caused by the changes in the position and shape of the cell. When the user is located at the mid-cell, the peak downlink data rate was 84 Mbps, and the average downlink data rate was 75.5 Mbps. The data rate fluctuation is caused by the varying distance between the user and HAP. Overall, the experimental results illustrate that a HAP is capable of supporting high-bandwidth applications (e.g., 4K resolution video transmission) using the beamforming technique \cite{ Belmekki2024CellularNF}. 
	However, the digital beamforming architecture increases digital chip content and the corresponding cost and power consumption. 
 
	\subsubsection{Analog beamforming:} 
\begin{figure}[!t]
		\centering
		\includegraphics[width=2.6 in]{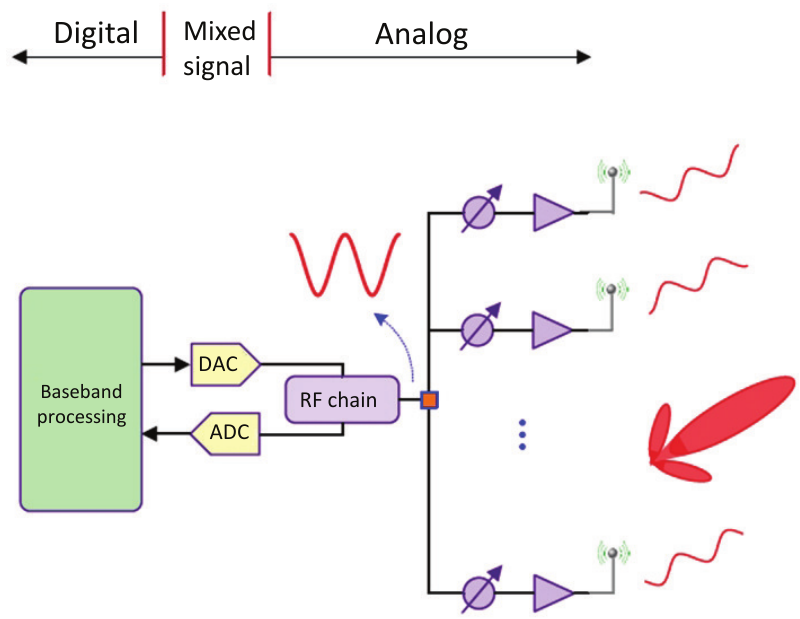}
		\caption{Architecture of the analog beamforming \cite{Qasim2023what}.}
		\label{fig:Analog_beamforming}
	\end{figure}

  As illustrated in Figure \ref{fig:Analog_beamforming}, the analog beamforming architecture incorporates a phase shifter at each antenna element, allowing for flexible and adaptive beams that can be directed in any direction.
  All phase shifters share a common RF chain in this type of architecture; as a result, analog beamforming is power- and cost-efficient. 
  
  \textbf{Existing studies:} The authors in \cite{DBLP:conf/iccchina/TongLZCW18} investigated the problem of analog beamforming to achieve low power consumption and exploited the benefits of beamforming in HAP-based multi-user MIMO communication systems. They formulated the analog beamforming problem as quadratically constrained quadratic programming and proposed a semi-definite relaxation approach to solve it.
    In \cite{ DBLP:conf/vtc/MatsuuraO23}, the authors investigated how to control the beam directions of the antenna onboard a HAP with consideration of the position of a ground gateway and attitude changes of a HAP. 
    In \cite{ DBLP:conf/vtc/ShibataTHNO23}, the authors proposed to optimize beamwidth and beam directions to prevent existing TN from being interfered with by a HAP.
    Besides, the authors in \cite{DBLP:conf/globecom/XiLHJSD17} studied the issue of HAP interference alleviation in HAP massive MIMO communication systems through user equipment beamforming. They proposed a semi-definite relaxation method to obtain the user equipment beamforming vector by maximizing its signal-to-pilot contamination ratio (SPR). 
    However, the fine tuning of the beams in this type of beamforming architecture is limited due to the low resolution of quantized phase shifts.
 
	\subsubsection{Hybrid beamforming:} 
	This beamforming architecture utilizes the advantages of both analog beamforming and digital beamforming architectures. 
 %Here precoding is applied to both analog domain and digital domains i.e. 
 \begin{figure}[!t]
		\centering
		\includegraphics[width=2.6 in]{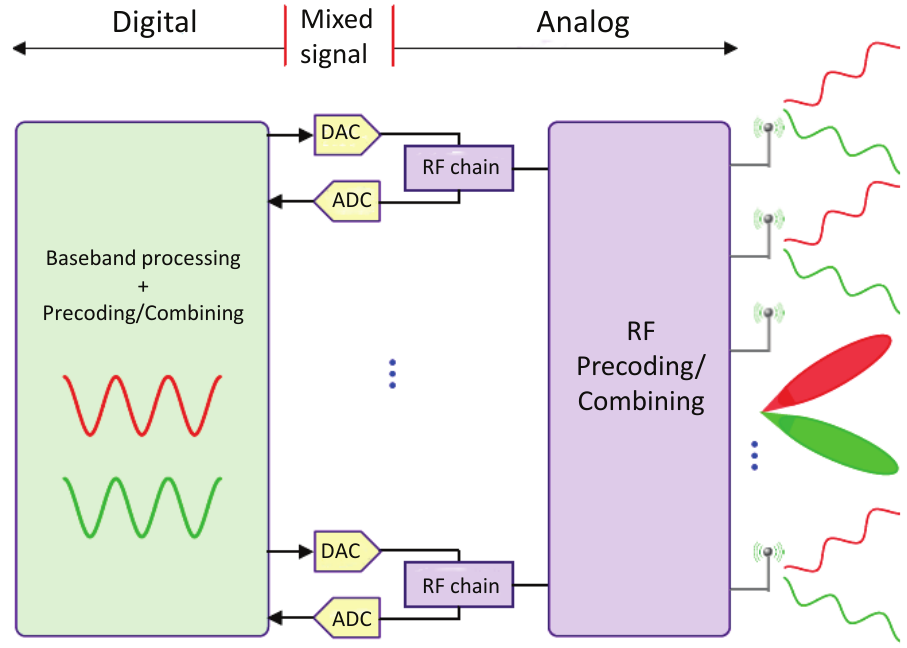}
		\caption{Architecture of the hybrid beamforming \cite{Qasim2023what}.}
		\label{fig:hybrid_beamforming}
	\end{figure}
 It employs precoding/beamforming at both RF and baseband, where the antenna elements are still driven by analog phase shifters, as shown in Fig. \ref{fig:hybrid_beamforming}. Yet, the number of RF chains remains smaller compared to the number of antennas.
 The reduction in the number of RF chains and corresponding data converters results in less cost, computational load, and power consumption. Overall, this type of beamforming architecture strikes a trade-off between complexity and flexibility.
As a result, it has garnered increasing attention from the research community. 

\textbf{Existing studies:} For example, the authors in \cite{DBLP:conf/sinc/JiJHH19} designed a hybrid beamforming scheme for HAP massive MIMO systems. Particularly, they derived the relationship between RF beamforming and statistical channel state information according to the duality theory and utilized the zero-forcing method to conduct baseband beamforming. 
 The authors in \cite{DBLP:conf/iscas/ZhangJJHHW20} investigated the hybrid beamforming for HAP massive MIMO systems and formulated the RF beamforming as an average signal-to-leakage-plus-noise ratio maximization problem. A greedy algorithm was then designed to obtain the RF beamforming matrix. Given the RF beamforming matrix, the baseband beamforming matrix was obtained by a regularized zero-forcing method.
 The authors in \cite{Hassan2023MachineLH} proposed to solve the high power consumption, high cost, and high hardware complexity issue of conducting hybrid beamforming for HAP massive MIMO systems. To this end, they designed an adaptive cross-entropy-based optimization method to update the RF beamforming matrix and employed a zero-forcing method to obtain a digital beamforming matrix.
Besides, the authors in \cite{ DBLP:conf/vtc/TashiroHN23} developed a two-stage precoding scheme to alleviate the interference between HAPs. Particularly, the proposed scheme separately controlled the interference between HAPs and user-specific beams by factorizing the overall precoding weight matrix into an outer precoder for interference reduction and an inner precoder for user-specific beam generation.

	\subsection{Passive beamforming}
	Beamforming architecture in which amplifiers or radios are used as external elements of an antenna array is known as a passive beamforming array. 
 
	\textbf{Discussion:} A typical passive beamforming pattern is RIS or IRS.
	In some scenarios, many aerial platforms of NSIN are expected to act as aerial relays. To complete the task of relaying signals, aerial platforms have two selections: 1) act as radio relays that mount active beamforming payloads; 2) integrate with RIS/IRS. 
	When acting as radio relays, aerial platforms have to mount some payloads, such as antennas, analog-to-digital converters (ADCs), digital-to-analog converters (DACs), data converters, mixers, power amplifiers, and filters, entailing high hardware complexity and large energy consumption. 
	On the contrary, RIS/IRS-mounted relays can be designed by coating the surfaces of aerial platforms with a thin layer of meta-surfaces, entailing low hardware complexity.
	Moreover, several studies have demonstrated that the capacity attained through RIS/IRS-mounted relays is comparable to that achieved through radio relays \cite{Huang2018EnergyEM,DiRenzo2019ReconfigurableIS}.
    Further, for NSIN, the energy consumption of using RIS/IRS is much lower than that of using radio relays \cite{DBLP:journals/wcl/AlfattaniYYY23,DBLP:journals/vtm/AlfattaniJYY23}. For example, using RIS/IRS reduces the communication energy consumption of NSIN to the required power for the RIS/IRS controller. In a recent experiment, it was found that the configuration of each reflector unit consumed 0.33 mW \cite{Tang2019WirelessCW}.
Studies have indicated that an RIS/IRS-assisted communication system could achieve a remarkable $40$\% increase in energy efficiency compared to a radio relay-assisted system \cite{Huang2018EnergyEM}. Therefore, RIS/IRS-mounted relays will be a promising solution for NSIN to build multi-hop links in a cost-efficient way. 
	%(aerial-RIS)
	
	Actually, RIS/IRS are a planar array composed of a large number of passive reflecting elements. By precisely manipulating the phases and directions of incident signals in a controlled manner, the planar array enables passive beamforming, which intelligently reflects signals towards targeted directions using passive reflecting elements \cite{Airbus2021Zephyr}. Mathematically, denote $x_i$ as the incident wave of the $i$-th element of RIS/IRS. Then, the reflected signal can be written as $y_i = I_i{\rm e}^{j\alpha_i}x_i$, $\forall i = 1, \ldots, M$, where $I_i \in \{[0, 1]\}$ and $\alpha_i \in [0, 2\pi)$ control the amplitude and phase shift of the reflected signal, respectively, and $M$ denotes the total number of array elements. 
	
	\textbf{Existing studies:} Recently, the investigation of integrating RIS/IRS on different aerial platforms has attracted lots of attention in the research community. Table \ref{table_aerial_RIS} compares the recent studies on aerial-RIS (ARIS). Summarily, the recent researches can be classified into three groups: channel modeling of ARIS \cite{xu2022reconfigurable,Azizi2023RISMA,DBLP:conf/icc/TekbiyikKHEY21,DBLP:journals/icl/IacovelliCG21,DBLP:journals/tsp/ChenCZW22}, deployment or trajectory planning of ARIS \cite{DBLP:journals/access/LiYDMD21,DBLP:journals/jcin/JiaoFZZ20,DBLP:conf/globecom/GeZW21,DBLP:journals/corr/abs-2111-11650,yu2022fair,DBLP:journals/wcl/GaoJLM21,ai2021joint,DBLP:conf/globecom/Xu0ZL21,DBLP:conf/globecom/LongCYLWXS20,DBLP:journals/twc/KhaliliMZJYJ22,DBLP:journals/jstsp/YuYC21}, and performance analysis of ARIS \cite{solanki2022ambient,DBLP:journals/wcl/Al-JarrahAAS21,DBLP:journals/tgcn/MahmoudMSADY21,DBLP:journals/ojcs/AlfattaniJHYY21,DBLP:journals/tcom/Al-JarrahAAIA21,DBLP:journals/wcl/WangTN21}. 
	
	\begin{table*}[!t]
		\newcommand{\tabincell}[2]{\begin{tabular}{@{}#1@{}}#2\end{tabular}}
	%\renewcommand{\arraystretch}{1.3}
	% if using array.sty, it might be a good idea to tweak the value of
%	 \extrarowheight as needed to properly center the text within the cells
	\caption{Comparison of most recent studies on ARIS}
	\label{table_aerial_RIS}
	\centering
	% Some packages, such as MDW tools, offer better commands for making tables
	% than the plain LaTeX2e tabular which is used here.
	\begin{tabular}{|c|c|c|c|c|c|c|c|c|c|c|c|c|}
	\hline
	Refs. & Amplitude & \tabincell{l}{Phase \\ shift}	& \tabincell{l}{Phase \\ error} &	\tabincell{l}{Channel \\ modeling}	& \tabincell{l}{ARIS \\ deploy- \\ ment}	& \tabincell{l}{ARIS \\ trajectory}	& \tabincell{l}{Aerial \\ platform \\ features} & \tabincell{l}{Single \\ ARIS} &	\tabincell{l}{Multi- \\ ARIS} & \tabincell{l}{Optimi- \\ zation} & \tabincell{l}{Lea- \\ rn} & \tabincell{l}{Perfor- \\ mance \\ analysis} \\
	\hline
	\cite{xu2022reconfigurable,Azizi2023RISMA} & {} & \checkmark & {} & \checkmark & {} & {} & {} & \checkmark & {} & \checkmark & {} & {} \\
	\hline
 \cite{DBLP:conf/icc/TekbiyikKHEY21} & {} & \checkmark & {} & \checkmark & {} & {} & {} & \checkmark & {} & {} & \checkmark & {} \\
	\hline
	\cite{DBLP:journals/icl/IacovelliCG21} & \checkmark & \checkmark & {} & \checkmark & {} & {} & {} & \checkmark & {} & {} & {} & {} \\
	\hline
	\cite{DBLP:journals/tsp/ChenCZW22} & \checkmark & \checkmark & {} & \checkmark & {} & {} & \checkmark & {} & \checkmark & {} & {} & {} \\
	\hline
	\cite{DBLP:journals/access/LiYDMD21} & {} & \checkmark & {} & {} & \checkmark & {} & {} & {} & \checkmark & {} & \checkmark & {} \\
	\hline
	\cite{DBLP:journals/jcin/JiaoFZZ20} & {} & \checkmark & {} & {} & \checkmark & {} & {} & \checkmark & {} & \checkmark & {} & {} \\
	\hline
	\cite{DBLP:conf/globecom/GeZW21} & {} & \checkmark & {} & {} & \checkmark & {} & {} & {} & \checkmark & \checkmark & {} & {} \\
	\hline
	\cite{DBLP:journals/corr/abs-2111-11650} & \checkmark & \checkmark & {} & {} & {} & \checkmark & {} & \checkmark & {} & \checkmark & {} & {} \\
	\hline
	\cite{DBLP:journals/wcl/GaoJLM21} & {} & \checkmark & {} & {} & {} & \checkmark & {} & \checkmark & {} & \checkmark & \checkmark & {} \\
	\hline
	\cite{yu2022fair,ai2021joint} & {} & \checkmark & {} & {} & {} & \checkmark & {} & \checkmark & {} & \checkmark & {} & {} \\
	\hline
	\cite{DBLP:conf/globecom/Xu0ZL21} & {} & \checkmark & {} & {} & {} & \checkmark & {} & \checkmark & {} & {} & \checkmark & {} \\
	\hline
	\cite{DBLP:conf/globecom/LongCYLWXS20} & {} & \checkmark & {} & {} & {} & \checkmark & {} & \checkmark & {} & \checkmark & {} & {} \\
	\hline
	\cite{DBLP:journals/twc/KhaliliMZJYJ22} & {} & \checkmark & {} & {} & {} & \checkmark & {} & {} & \checkmark & {} & \checkmark & {} \\
	\hline
	\cite{DBLP:journals/jstsp/YuYC21} & {} & \checkmark & {} & {} & {} & \checkmark & {} & {} & \checkmark & \checkmark & {} & {} \\
	\hline
	\cite{solanki2022ambient,DBLP:journals/wcl/Al-JarrahAAS21} & {} & {} & \checkmark & {} & {} & {} & {} & \checkmark & {} & {} & {} & \checkmark \\
	\hline
	\cite{DBLP:journals/tgcn/MahmoudMSADY21,DBLP:journals/wcl/WangTN21} & {} & {} & {} & {} & {} & {} & {} & \checkmark & {} & {} & {} & \checkmark \\
	\hline
	\cite{DBLP:journals/ojcs/AlfattaniJHYY21} & {} & {} & {} & {} & {} & {} & \checkmark & \checkmark & {} & {} & {} & \checkmark \\
	\hline
	\cite{DBLP:journals/tcom/Al-JarrahAAIA21} & {} & {} & \checkmark & {} & {} & {} & \checkmark & \checkmark & {} & {} & {} & \checkmark \\
	\hline
	\end{tabular}
	\end{table*}
	
	In terms of channel modeling of ARIS, for example, under the condition of considering the correlations of fading channels in the spatial, time, and frequency domains imposed by the AoA/AoD, the Doppler, and the orthogonal frequency division multiplexing (OFDM) operations, the channel modeling of high-dimensional and high-mobility ARIS was investigated in \cite{xu2022reconfigurable}. In \cite{DBLP:conf/icc/TekbiyikKHEY21}, the cascaded channel estimation in full-duplex ARIS communications under the condition of varying channel characteristics was studied. For the channel estimation, a graph attention network was utilized. Under Rician fading conditions, a composite channel gain expression was proposed in \cite{DBLP:journals/icl/IacovelliCG21}. Besides, considering the position perturbation of ARIS, a novel atomic norm minimization method was developed to estimate the direction of arrival (DoA) in \cite{DBLP:journals/tsp/ChenCZW22}. 
	
  In terms of the deployment or trajectory planning of ARIS, for instance, an optimization framework was formulated and solved in \cite{DBLP:journals/access/LiYDMD21}, which considered ARIS's deployment, power allocation at a macro base station, phase shift of ARIS, and blocklength of ultra-reliable low latency communications (URLLC). The work presented in \cite{DBLP:journals/jcin/JiaoFZZ20} investigated the optimization of ARIS's horizontal position and phase shift of ARIS to maximize the rate of a strong terrestrial user while guaranteeing the target rate of a weak terrestrial user.
 In \cite{DBLP:journals/corr/abs-2111-11650}, the problem of optimizing ARIS beamforming and trajectory to enable reliable data transmissions from ARIS to an intended terrestrial user was explored. Furthermore, in \cite{yu2022fair}, the objective was to maximize the minimum throughput among all mobile vehicles by jointly optimizing ARIS passive beamforming and trajectory, with the aim of achieving communication fairness.
	
	In terms of performance analysis of ARIS, for example, the outage probability and ergodic spectral efficiency expressions of ARIS communications were derived by the probability theory in \cite{solanki2022ambient}. The impact of imperfect phase knowledge on the system capacity of ARIS communications was investigated in \cite{DBLP:journals/wcl/Al-JarrahAAS21}, where the phase error was modeled as a von Mises random variable. On the other hand, \cite{DBLP:journals/tgcn/MahmoudMSADY21} derived tractable analytic expressions for the achievable symbol error rate (SER), ergodic capacity, and outage probability of ARIS communications.
    The tight upper and lower bounds of the average SNR were also derived. Besides, the work in \cite{DBLP:journals/ojcs/AlfattaniJHYY21} theoretically analyzed the link budget of ARIS communications for ARIS `specular' and `scattering' reflection paradigms. 
	
	\subsection{Multi-connectivity}
    {Multi-connectivity by introducing link/path diversity can significantly improve reliability and capacity performance. There are various architectures for multi-connectivity with diverse ways of diversity, such as technology diversity and base station diversity \cite{Salehi2023ReliabilityAD}.} 

{\textbf{Existing studies:} In terms of improving transmission performance via technology diversity, the authors in \cite{Salehi2023ReliabilityAD} considered radio access technology (RAT) diversity of air-to-ground (A2G), A2A, HAP, and LEO satellites to provide seamless connectivity for remote piloting of aerial vehicles.} 

{CoMP is a typical technique utilizing the principle of base station diversity. 
It was introduced by the 3GPP in Release 11 to mitigate cell-edge interference and convert the unwanted signals into useful signals. As a result, the capacity of cell-edge users could be significantly improved. Besides, CoMP can dramatically enhance transmission reliability by simultaneously transmitting the same data to an intended user. 
The authors in \cite{DBLP:journals/sensors/DongLGLG15} theoretically analyzed the diversity of receiving performance of a wireless sensor network from a constellation of inter-connected HAPs, where multiple HAPs coordinately transmitted signals to a receiving sensor at the same frequency. Nevertheless, the synchronization overhead for achieving coordination among HAPs was not considered in this work. The work \cite{DBLP:conf/softcom/ZakariaGMS18} investigated how joint transmission CoMP could be extended to a HAP system architecture by exploiting a phased array antenna. The phased array antenna generated multiple beams that formed cells. Each cell could map on to pooled virtual base stations, thereby making it possible to achieve base station diversity. Then, multiple virtual base stations cooperated to send the same data simultaneously to an intended user, which improved the performance of users in the system. However, the calculation of the channel capacity was based on the free space path loss model, which might make the simulation result significantly different from the actual results. 
}

 	\subsection{{Communication Protocols for NSINs}}
    {In this subsection, we elaborate on the communication protocols for NSINs, including the MAC layer protocols, network layer protocols, and transport layer protocols.} 
    
 \subsubsection{MAC layer protocols}
	The primary goal of MAC protocols is to improve network capacity and reduce packet transmission latency. MAC protocols greatly impact the performance of wireless networks, including network throughput, latency, and reliability. 
	There are three categories of MAC protocols, which include fixed allocation protocols, random contention protocols, and reservation protocols.
	
	Fixed allocation protocols needing to reserve network resources are suitable for wireless networks with stable topologies or associations, a delay-sensitive constraint, and regular traffic flows. 
	Random contention protocols are suitable for wireless networks with random traffic flows, a delay-tolerant constraint, and time-varying topologies or associations. 
	Reservation protocols are suitable for wireless networks with irregular and highly fluctuating traffic flows and time-varying topologies or associations.
	
	\textbf{Fixed allocation protocols:} Channel resources in this kind of MAC protocol are allocated or reserved to each node in wireless networks according to a certain allocation algorithm in advance. 
	Many types of multiple access techniques, including TDMA, code division multiple access (CDMA), frequency division multiple access (FDMA), spatial division multiple access (SDMA), and NOMA, can be explored in fixed allocation protocols \cite{Zhou2023AerospaceIN}. 
	It is appropriate to apply fixed allocation protocols to the HAP sub-network in NSIN as a HAP is quasi-stationary and has a large footprint. Numerous relevant studies have been presented in the literature.
    For instance, the authors in \cite{DBLP:conf/tsp/TacuriT19} investigated the interference issue in a HAP and fixed wireless access (FWA) coexistence system. Under the TDMA assumption, they introduced a probabilistic model to analyze the aggregate interference caused by FWA transmitters for users served by the HAP.
 %The paper \cite{DBLP:journals/tvt/CaoYYH21} proposed a HAP-reserved ground-HAP-space (GHS) transmission scheme using TDMA to support ground-to-space (G2S) communications. It designed a transmission control strategy enabling ground users to switch to the GHS link with a certain probability and formulated an optimization problem aiming at maximizing users’ throughput. 
	The paper \cite{hidayat2015pilot} considered a HAP uplink communication scenario using a single-carrier FDMA (SC-FDMA) scheme. The investigation on the performance of a pilot-based uplink channel estimation method was conducted. The impact of some crucial coefficients, such as a user’s elevation angle, modulation scheme, LTE channel bandwidth, and Doppler shift, on a HAP uplink channel model was evaluated. 
	The paper \cite{kurniawan2014closed} designed a joint power control and space diversity algorithm to improve the performance of CDMA-enabled HAP communications at a low elevation angle. The authors in \cite{DBLP:journals/adt/AhmedG13} studied the wideband CDMA (WCDMA) uplink capacity and the interference statistics of HAP macrocells using statistical geometry and probability theories. 
	Exploiting beamforming to design SDMA-enabled HAP systems is currently under active research \cite{liu2022joint,DBLP:conf/vtc/PopoolaGC20,DBLP:journals/twc/DingWZLL22,DBLP:conf/iccchina/TongLZCW18,DBLP:journals/iotj/XuLRK22,DBLP:journals/icl/SudheeshMMSM18,DBLP:journals/wcl/LianJHH19}. The work in \cite{liu2022joint} investigated the problem of augmenting ground communications through deploying an SDMA-enabled HAP. The authors in \cite{DBLP:conf/vtc/PopoolaGC20} considered a scenario of deploying a HAP to provide broadband infotainment services for vehicular users in rural areas and proposed to utilize steerable beam directional antennas to realize SDMA for vehicular users. 
	
	The investigation into the NOMA-enabled HAP networks has attracted much attention from the research community. 
	The papers \cite{he2022noma,DBLP:conf/iccchina/HeWHTZ21} proposed the deployment of NSIN to provide reliable communication services to ground users, with the adoption of the NOMA technique to enhance the quality of downlink links for users located at the edge of NSIN.
	The paper \cite{DBLP:journals/tii/WangHYSNZ22} designed a novel secure transmission framework to provide latency-sensitive medical-care services for ground users by deploying a HAP and many low-altitude platforms. The authors proposed to deliver medical-care information to multiple hotspots via a NOMA transmission scheme and formulated an optimization problem for minimizing transmission latency, subject to power and spectrum constraints. An alternating optimization framework was designed to solve the optimization problem. The authors of \cite{DBLP:journals/icl/JiJHLH21} investigated energy-efficient beamforming for HAP-NOMA systems operating over Rician fading channels. Particularly, they derived the expression of the NLoS channel of a HAP equipped with a uniform planar array (UPA). Based on the obtained channel expression, a HAP-NOMA scheme, including user grouping and power optimization, was designed. The paper \cite{DBLP:journals/tvt/QinZZFLZ21} considered clustered-NOMA NSIN and investigated the joint UAV location optimization and resource allocation problem to maximize the overall network uplink achievable rates. To solve this problem, UAV location optimization and resource allocation were iteratively performed. A simulation of HAP backbone networks utilizing the NOMA scheme was studied in \cite{azzahra2019noma}. 
	The paper \cite{DBLP:journals/wcl/ShuaiGAHZ22} considered an integrated HAP space information network using a NOMA scheme. The authors derived the closed-form expressions of the outage probabilities of the integrated network in the presence of imperfect channel state information (CSI) and successive interference cancellation (SIC). 
In \cite{ DBLP:journals/tvt/CumaliOKY23}, the authors discussed the issue of exploiting NOMA-based HAP communications and multiple antennas with user selection to meet the connectivity, reliability, and high achievable data rate requirements of users. 
 In \cite{ Javed2022AnIA}, the authors proposed to deploy a HAP communication system to enable NOMA for ground multi-cell users. 
 The work in \cite{ Guo2023TwoWaySN} theoretically analyzed the outage probability and ergodic capacity of a two-way satellite-HAP-terrestrial network with NOMA. 
 Besides, in \cite{Qin2023JointTP}, the authors investigated the issue of deploying clustered-NOMA-enabled NSIN to cover remote areas where the clustered-NOMA technology was leveraged to achieve channel multiplexing and reduce the complexity of SIC.
	
	\textbf{Random contention protocols:} In this type of MAC protocol, each node in wireless networks accesses channels in a competitive way. Once obtaining the opportunity to access a channel, a node immediately accesses the channel or randomly accesses the channel with a certain probability after detecting that the channel is idle. 
	If a packet collision occurs, the transmission fails, and the node chooses to back off and wait for the next transmission. The greater the number of transmission failures, the lower the random access probability.
	There are many famous random contention protocols, such as additive links on-line hawaii area (ALOHA), carrier sense multiple access (CSMA), and CSMA with collision avoidance (CSMA/CA). 
	The ALOHA protocol utilizes the simplest random competition mechanism. Users access a channel immediately when they have data to send. 
	In order to reduce collisions, users should avoid accessing the channel when the channel is not idle. For this purpose, the CSMA protocol has been developed. In this type of protocol, a user first monitors the channel before sending data, which thus reduces packet collisions to a certain extent. However, packet collisions may still happen due to propagation delays. 
	To further reduce possible collisions, the CSMA/CA protocol is introduced. It adopts collision avoidance methods to reduce possible collisions. Specifically, it utilizes request-to-send (RTS) and allow-to-send control messages to implement channel reservation and collision avoidance, which thus improves the success probability of data transmission. 
	For example, the authors in \cite{Ruan2020AnAC} adopted the CSMA/CA and RTS/clear to send (CTS) handshake mechanisms to solve the efficient access problem among a HAP and multiple UAVs when sharing channel resources.
 
	\textbf{Reservation protocols:} 
 In this type of protocol, according to the amount of network traffic in a node, the node first competitively sends several short packets to reserve a channel in advance. This type of protocol mandates the exchange of reservation control information between nodes. As a result, a great deal of overhead will be generated when the payload in the networks is limited or the number of nodes varies. The impact of channel reservation information on channel utilization is a major concern for this type of protocol. 
	The polling MAC protocol is a typical reservation protocol that maintains a central control node (or master) to centrally control each node's access to the channel without contention. 
	
	Summarily, the random contention protocols and reservation protocols for dynamic networks can be applied to HAUAV and LAUAV sub-networks in NSIN.The HAUAV and LAUAV sub-networks are characterized by high mobility, which leads to fluctuating link quality and time-varying topologies. Nevertheless, the packet latency constraint is difficult to be satisfied in the dynamic HAUAV and LAUAV sub-networks.
	During the past few years, the problem of designing effective MAC protocols for UAV networks has been extensively studied in the literature. Table \ref{table_mac-surveys} summarizes the surveys on UAV MAC protocols.
	
	\begin{table*}[!t]
		\newcommand{\tabincell}[2]{\begin{tabular}{@{}#1@{}}#2\end{tabular}}
		%\renewcommand{\arraystretch}{1.3}
		% if using array.sty, it might be a good idea to tweak the value of
		%	 \extrarowheight as needed to properly center the text within the cells
		\caption{Relevant Surveys On UAV MAC Protocols.}
		\label{table_mac-surveys}
		\centering
		% Some packages, such as MDW tools, offer better commands for making tables
		% than the plain LaTeX2e tabular which is used here.
		\begin{tabular}{|l|l|}
			\hline
			Refs. & Focus \\
			\hline
			\cite{DBLP:journals/adhoc/BekmezciST13} & \tabincell{l}{Promising technological advancements, including directional antenna and full-duplex radio circuits \\ with multi-packet reception, which can be adopted by MAC protocols of FANET.}   \\
			\hline
			\cite{batthsurvey,DBLP:journals/comcom/VashishtJA20} &  \tabincell{l}{Design of MAC protocols for UAV networks with \\ Omni-directional and directional antennas.} \\
			\hline
			\cite{DBLP:journals/csi/HentatiF20} & \tabincell{l}{Classify the MAC protocols for UAV networks into two categories: \\ antenna category-based MAC protocols and access mechanism-based MAC protocols.} \\
			 \hline
			 \cite{DBLP:journals/sensors/KhisaM20} & \tabincell{l}{Discus the design considerations for MAC protocols of the Internet of Things based on UAV (UIoT), \\ and summarize the contention-based, contention-free, and AI-based MAC protocols for UIoT.} \\
			 \hline
			 \cite{DBLP:journals/access/PoudelM19} & \tabincell{l}{Summarize the contention-based, contention-free, and hybrid MAC protocols \\ for UAV-aided wireless sensor networks and compare the MAC protocols \\ in terms of the features and focal ideas. } \\
			 \hline
			 \cite{DBLP:journals/comsur/GuptaJV16,bashir2019green} & \tabincell{l}{Discuss the protocols that help save energy in the MAC layer of UAV networks. } \\
			 \hline
		\end{tabular}
	\end{table*}
	
	\subsubsection{Network Layer}
	A routing protocol aimed at tackling the issue of obtaining the most connected E2E path to reliably deliver traffic to one or more destination(s) with a reduced E2E latency. 
	
 {\textbf{Discussion:}} Considering the high heterogeneity and complexity of NSIN, it is highly challenging to design an efficient routing protocol to enable packet transmission across different network segments. 
	Most existing routing protocols for UAV networks (partially) cannot be applied to NSIN. As noted in numerous surveys of UAV routing protocols, these protocols can be broadly categorized into topology-based routing protocols (such as flat, cluster-based, and hybrid routing protocols) and geographical routing protocols.
 
	The topology-based routing protocol (exactly, flat routing protocol), including some famous protocols such as proactive, reactive, and hybrid routing protocols, cannot not be directly applied to route packets in NSIN. 
 %A type of topology-based routing protocol (exactly, a flat routing protocol), including some famous protocols such as proactive, reactive, and hybrid routing protocols, cannot be directly applied to route packets in NSIN. 
 This is because all of these routing protocols assume that the nodes in the network have equal roles  \cite{DBLP:journals/comsur/LakewSDNC20}, which is not suitable for NSIN.
	The idea of adopting a greedy scheme and recovery scheme in UAV geographical routing protocols may be applicable for NSIN.  However, the entire routing protocol cannot be directly applied to NSIN, mainly due to the large footprint of HAP as well as the long distance towards the destination. For instance, owing to its large footprint, a destination may always be in the coverage range of a HAP. Yet, owing to the high deployment altitude, the transmission distance of a UAV-HAP link is greater than that of a UAV-UAV link.
	Besides, for the geographic information-based routing protocols, they routed packets using the location information of airborne platforms and adopted a greedy scheme to forward packets. An important procedure in this type of routing protocol is that an airborne platform will deliver packets to a neighboring platform closer to the destination, and the routing process may fail if an invalid route occurs. 

 \textbf{Existing studies:} Therefore, routing protocols for NSIN should be separately designed.
 Existing routing protocols for NSIN can be categorized into two groups: HAP-assisted routing protocols and unified routing protocols.
 Table \ref{table_routing} compares the main ideas and some key performance indicators, e.g., signal overhead, communication latency, and packet delivery ratio, of these routing protocols. 
\begin{table*}[!t]
	\renewcommand{\arraystretch}{1.2}
	\caption{Classification, main idea, and comparison of routing protocols for NSIN}
	\label{table_routing}
	\newcommand{\tabincell}[2]{\begin{tabular}{@{}#1@{}}#2\end{tabular}}
	\centering
	\begin{tabular}{l}
%\begin{minipage}{0.18\textwidth}
			\centering
			\includegraphics[width=7.0 in]{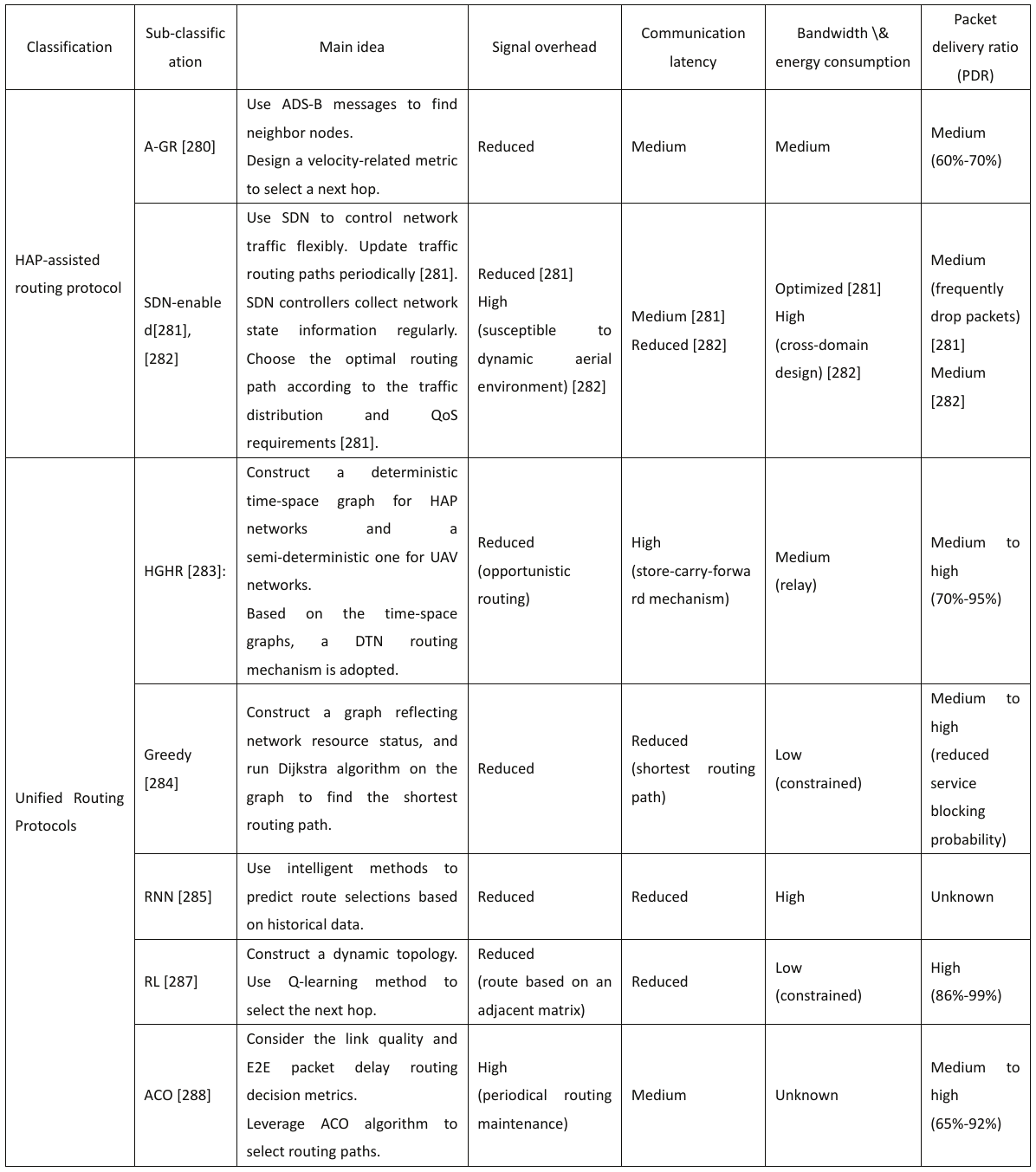}
		%\end{minipage} 
	\end{tabular}
\end{table*} 
 
	\emph{{A. HAP-assisted routing protocol:}}
In this type of routing protocol, HAP networks are consistently utilized as a control information delivery layer. Specifically, HAPs can broadcast control information (e.g., automatic dependent surveillance-broadcast (ADS-B) and SDN information) that will be useful for the efficient protocol design of UAV networks. 
	This type of protocol includes: 
 
	\textbf{ADS-B system-aided Geographic Routing (A-GR) \cite{DBLP:journals/jsa/WangFDGSY13}:} It is an ADS-B system-aided geographic routing protocol. 
	ADS-B is a cooperative surveillance system for air traffic management (ATM). 
 An ADS-B-enabled aircraft will broadcast a state vector including identification, position, velocity, altitude, and other flight-related information. 
	Owing to their large footprints, HAPs are leveraged to expand the coverage of the ADS-B system.
 HAPs will collect ADS-B messages from different aircrafts and then broadcast this type of message.
	
A-GR differs from traditional routing protocols in that it builds a neighbor table using ADS-B messages instead of sending hello beacons, which reduces the beaconing overhead. Further, as the ADS-B system and the payload data transmission system adopt different antennas and frequency bands, ADS-B messages will interfere with payload packets. 
 %explores aircraft positions and velocities in ADS-B message to avoid the traditional routing beaconing and builds a neighbor table using ADS-B message.  design a velocity-based metric for the selection of the next hop. 	The routing procedures of A-GR routing protocol consisted of neighbor discovery, next hop selection, and data forwarding. It builds a neighbor table based on ADS-B messages
  %received from the neighbor aircraft, which include the position and vector state of aircraft. The beaconing overhead of traditional geographical routing protocol is totally eliminated, when ADS-B messages instead of traditional hello beacons are used to find neighbor nodes. As ADS-B system and AANET use their own physical layer antenna, ADS-B messages do not interfere with data packets for routing. 
 The update frequency of the neighbor table is determined by the ADS-B message cycle.
	A-GR designs a routing decision metric (i.e., instantaneous flight time, IFT) that considers the distance towards the destination and the relative velocities of airborne platforms. 
	% \begin{equation}\label{eq_IFT}
	% 	IFT = \mathop {\min }\limits_{k \in {N_i}} \left\{ {\frac{{\Delta {d_{i,k}}(t,D)}}{{{v_i}(t) - {v_k}(t)}}} \right\}
	% \end{equation}
	% where $N_i$ is the neighbor set of platform $i$, ${\Delta {d_{i,k}}(t,D)}$ denotes the geographic distance between $i$ and $j$ to destination $D$, $v_i(t)$ and $ v_k(t)$ are the velocities of platform $i$ and $k$ at moment $t$, respectively.
	During the packet forwarding phase, a typical platform searches its neighbor table to determine the routing path for the packets. If multiple platforms are within its transmission range, the one with the best IFT is selected as the next hop. In the absence of platforms within the transmission range of the platform, the packets are buffered for a certain period of time until suitable platforms appear. If no suitable platform is found, the packets are dropped.

 \textbf{SDN-enabled approach \cite{DBLP:journals/jfi/QiSKG19,DBLP:journals/network/ShiCLK18}:} 
%	\textbf{Traffic-differentiated routing (TDR) \cite{}:} 
By providing an overview of NSIN and decoupling the control plan and data plan of the network, SDN can be deployed to design efficient routing protocols.
For example, the authors in \cite{DBLP:journals/jfi/QiSKG19} proposed to deploy HAP-SDN controllers in NSIN and designed a traffic-differentiated routing (TDR) protocol according to information provided by these controllers. 
Specifically, in this protocol, HAP-SDN controllers were deployed to acquire a comprehensive abstract network view, facilitate unified scheduling of resources, and guide efficient data processing and delivery.
Supported by controller-collected network state information (e.g., positions and speeds of UAVs, the number of successfully/unsuccessfully delivered packets, packet queuing delay, and transmission delay), an estimation of the availability of links in the near future was performed.
The forwarding abilities of UAVs as well as packet delay on transmission links were predicted.
Meanwhile, to address specific delay and reliability requirements of traffic flows, this protocol assigned different weights to flows based on their sensitivity to delay and level of importance. 
An optimization problem aiming to minimize the total cost of all traffic flows was then formulated and solved to update traffic routing paths.
 In SDN-enabled NSIN, however, the fast movement of UAVs may lead to long configuration update times in the data plane. 
	In \cite{DBLP:journals/network/ShiCLK18}, the authors proposed to deploy a HAP-SDN controller to assist the design of a routing protocol for NSIN. 
  Due to its wide coverage and relative stability, the configuration updating time could be reduced by HAP-SDN.
	The main functionalities of the controller include: 1) topology discovery, which involves perceiving and updating network topology in real-time as airborne platforms are added or removed from the data plane. 
	2) routing decision, which involves selecting the optimal routing path based on traffic distributions and QoS requirements within the integrated network.

\begin{figure*}[!t]
\centering
%\begin{tabular}{ccc}
    \subfigure[{Prototype of the gateway routing protocol}]{
    \label{fig:subfig:f} %% label for first subfigure
    \includegraphics[width=3.7 in, height = 1.6 in]{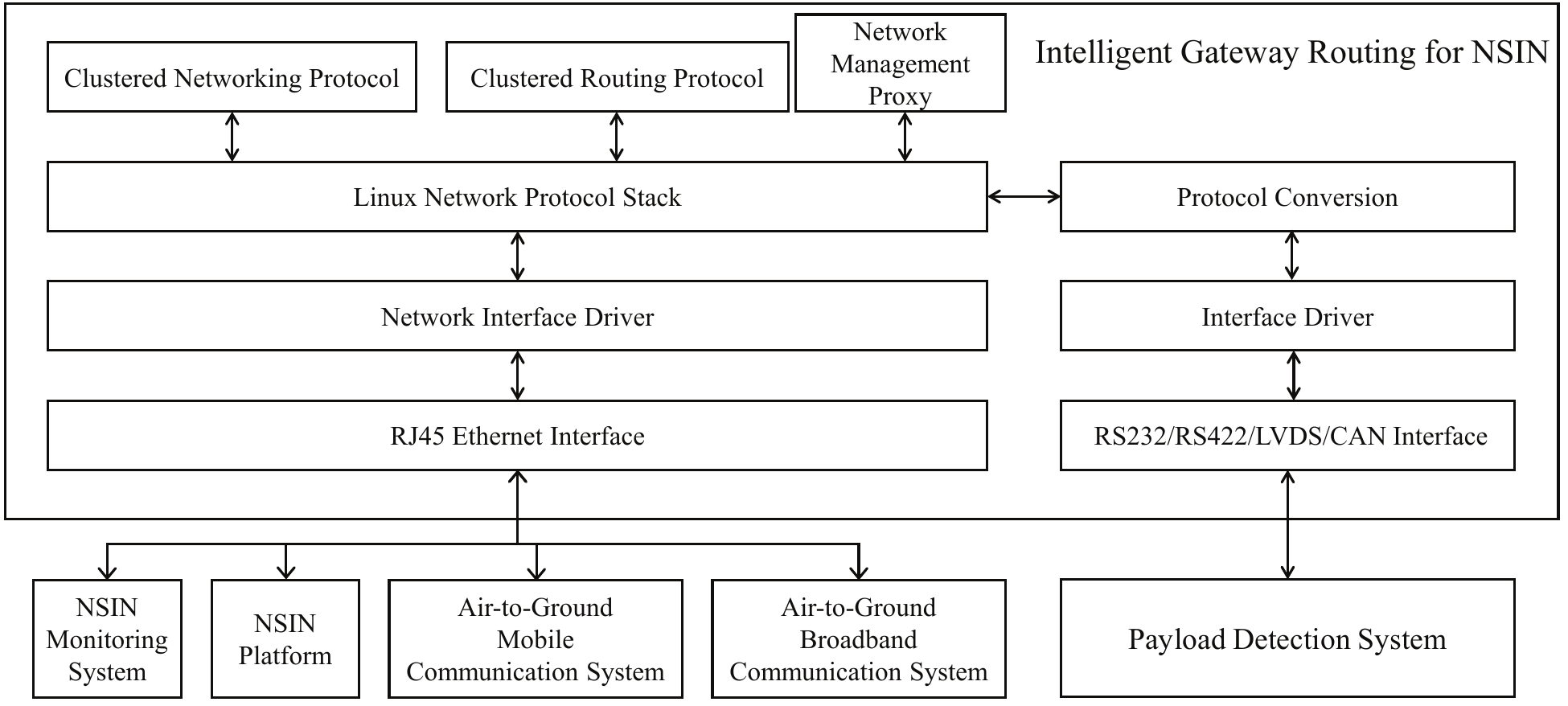}}
    \hspace{3pt}
    \subfigure[{Hardware device}]{
    \label{fig:subfig:g} %% label for first subfigure
    \includegraphics[width=3.0 in, height = 1.6 in]{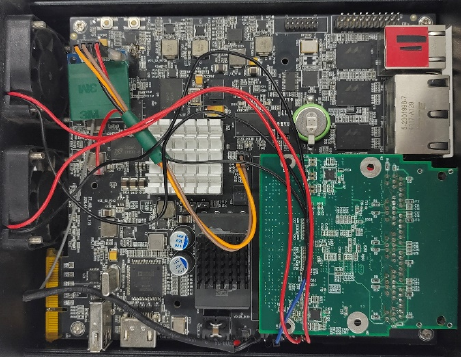}}
    \caption{{A prototype and hardware device of the designed adaptive gateway routing protocol.}}
    \label{fig:gateway_routing}
%\end{tabular}
\end{figure*}

\begin{figure}[!t]
\centering
%\begin{tabular}{ccc}
    %\subfigure[{Live video broadcast}]{
    %\label{fig:subfig:g_v} %% label for first subfigure
    \includegraphics[width=2.8 in, height = 1.4 in]{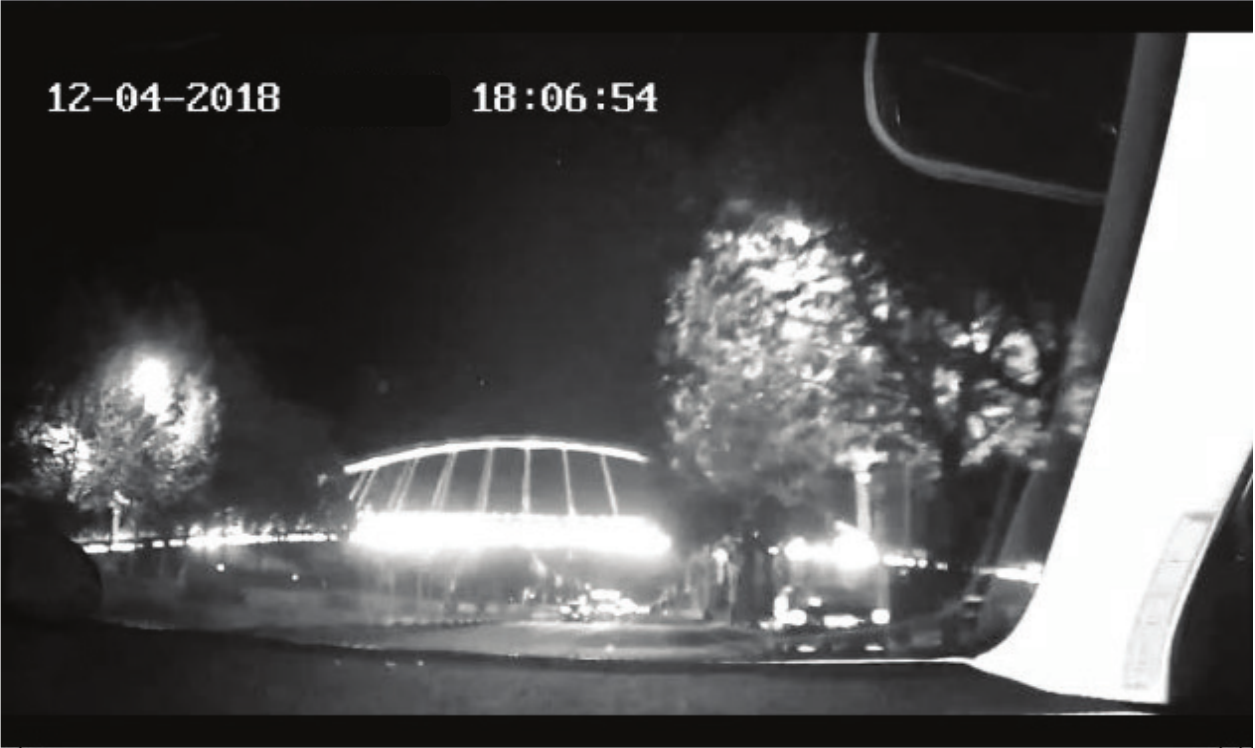}
    \caption{{A subjective result of transmitting front window videos of a moving car utilizing practical NSINs.}}
    \label{fig:subjective_result}
%\end{tabular}
\end{figure}

	\emph{{B. Unified routing protocols:}}
 {Different subnetworks in NSINs may have diverse transmission capabilities and adopt different communication protocol formats and even communication system architectures. To achieve high efficient information transmission over NSINs,} unified routing protocols are desired.
 %to selecting an E2E path in NSINs for achieving uniform deployment and centralized management. 
 When different communication protocol formats or system architectures are adopted, applying a number of gateways to interconnect subnetworks will be a feasible proposal. For example, the authors in this article designed an intelligent gateway routing protocol, its prototype and hardware device are shown in Fig. \ref{fig:gateway_routing}.
 It intelligently adapts the communication performance of subnetworks by dynamically adjusting the communication parameters of gateway routers to guarantee smooth video transmission in NSINs. 
 {The authors in this article also conducted an experiment to verify the performance of the designed gateway routing protocol by implementing it on actual NSINs, the topology of which is shown in Fig. \ref{fig:topology}. %illustrates the network topology of NSINs.
 In this experiment, the authors utilized a moving car to shoot its front window video. The car transmitted the shot video streams to an airship acting as a relay. The airship delivered the received video streams back to a ground terminal. 
 Fig. \ref{fig:subjective_result} shows a subjective result of video transmission via the NSINs. From this figure, it can be observed that there is low distortion in the received image, and the image details are clear.} 
 %An experiment of conducting live streaming transmission of the front window view of a moving car via our NSINs.
 %\caption{ and }
 % \begin{figure}[!t]
	% 	\centering
	% 	\includegraphics[width=3.2in]{adaptive_routing_arc.pdf}
	% 	\caption{An architecture of our designed adaptive gateway routing protocol.}
	% 	\label{fig:gateway_routing}
	% \end{figure}
% \begin{figure}[!t]
% \centering
% %\begin{tabular}{ccc}
%     \subfigure[{Network topology of the adopted NSINs}]{
%     \label{fig:subfig:f_v} %% label for first subfigure
%     \includegraphics[width=3.5 in, height = 2.6 in]{Network_topology.pdf}}
%     \hspace{3pt}
%     \subfigure[{Live video broadcast}]{
%     \label{fig:subfig:g_v} %% label for first subfigure
%     \includegraphics[width=2.8 in, height = 1.4 in]{Vehicle_video.pdf}}
%     \caption{{An experiment of conducting live steaming transmission of the front window view of a moving car via our NSINs.}}
%     \label{fig:topology}
% %\end{tabular}
% \end{figure} 
%\textcolor{blue}{Nevertheless, as mentioned above, different subnetworks may adopt different communication protocol formats, carrier frequencies, sub-network transmission capabilities, and even different communication system architectures, then a gateway that can connect diverse subnetworks will be essential for promoting the unified routing protocol design. For example, we Utilizing gateway is a traditional and feasible way of achieving the integration design of heterogeneous airborne networks. For example, we constructed an integrated NSINs by designing intelligent gateways to integrate NSINs The topology of the testing network, in the experiments, we transmit real-time videos collected by cars to Except for gateway, some advanced strategies can be explored to achieve the above goal. Specifically, except for intelligent gateway routing protocols, many other different types of unified routing protocol have been proposed, including: }

{Except for gateway routing protocols, many other types of unified routing protocols have been designed with a consideration of the diverse transmission capabilities among different subnetworks, including} 
%hybrid time-space graph supporting hierarchical routing (HGHR) \cite{DBLP:journals/access/QiHGSJ16}, greedy routing protocols \cite{DBLP:journals/jsac/WangZZNS20}, recurrent neural network (RNN)-based routing protocols \cite{DBLP:journals/wc/GuZX20}, RL-based routing protocols \cite{DBLP:conf/wcsp/ShiRD21}, and ant colony optimization (ACO)-based routing protocols \cite{DBLP:journals/ppna/ZhengZY21}.}
 
	\textbf{Hybrid time-space Graph supporting Hierarchical Routing (HGHR) \cite{DBLP:journals/access/QiHGSJ16}:} 
 HGHR is a unified routing protocol designed for space and UAV networks. The work in \cite{DBLP:journals/access/QiHGSJ16} designed a unified routing protocol, HGHR, for space and UAV networks. Yet, HGHR can be deployed in NSIN. Similar to space networks, the topology of HAP networks is predictable. 
	
 HGHR is designed based on a hybrid time-space graph containing two subgraphs, i.e., a semi-deterministic subgraph for UAV networks and a deterministic subgraph for space networks. 
 In the semi-deterministic subgraph, a UAV represents a node, and the weight of an edge is measured by UAV contact time and contact probability. The prediction of UAV contact time and contact probability is accomplished by utilizing a discrete time-homogeneous semi-Markov process with the state transition probability and sojourn time probability of each UAV as inputs. 
 A deterministic subgraph for space networks can be constructed due to the predictable topology of space networks. 
	%we first introduce a discrete time homogeneous semi-Markov model for the air network in order to predict the contact probability and contact time between UAV pairs. 	1) Prediction model: 	The prediction model is based on a discrete time-homogeneous semi-Markov process, which takes each UAV's historical information including state (air-mark) transition probability and sojourn time probability distribution as inputs. And then the future moving trajectory of each UAV to other air-marks will be predicted so that the contact time and contact probability between all UAV pairs can be obtained.	Based on the prediction results, a semi-deterministic time-space subgraph is available. 
	By combining two subgraphs, a hybrid graph for the integrated networks is obtained. 
	Under the hybrid graph, HGHR explores the store-carry-forward mechanism in a delay-tolerant networks (DTN) architecture to implement packet forwarding. 
 %It delivers packets across the path with low energy consumption and E2E latency. 
	%The data is forwarded to its neighboring node in proper time, the state-space graph is updated and a new path is selected. Data is transferred hop by hop in the direction of closing to the destination.
 
	%Next, the hybrid time-space graph is transformed into a state-space one by setting a time window and updating weights of sub-graphs within the time window, in order to remove time dimension from graph edges. 
	
	%Under the store-carry-forward mechanism in DTN architecture which introduces an overlay above regional lower layer protocols, a data forwarding rule is designed based on the state-space graph.

	\textbf{Greedy routing protocol \cite{DBLP:journals/jsac/WangZZNS20}:} 
 The work in \cite{DBLP:journals/jsac/WangZZNS20} proposed a greedy routing protocol for packet forwarding in NSIN-ground integrated networks. 
 During the process of finding routes, this protocol allocated a higher priority to TN to fully utilize its computation resources and reduce resource utilization costs. 
 When the E2E latency could not be guaranteed solely by TN, platforms in NSIN were considered to reduce the number of transmission hops. 
 The protocol employed a greedy search approach to find the optimal route between a source node and a destination node, taking into account factors such as available bandwidth resources, transmission latency, and the level of function sharing requirements.
 %A feasible route should satisfy the bandwidth resource and transmission latency requirements. according to  using the Dijkstra algorithm.  However,  Then, the optimal path is searched in a greedy manner according to the QoS requirement, the level of function sharing and bandwidth capacity. 
	%the ground and aerial networks are provided with different priorities for resource utilization. 
 %The priority mode for resource utilization can make the best use of the advantages of both ground and aerial networks to reduce the resource costs. 
 %Algorithm 2 is a greedy algorithm to find the optimal path from source node to destination node for each service request based on Dijkstra algorithm. A feasible routing path for the traffic should satisfy the bandwidth resource requirement, and the total hopping delay cannot exceed the deadline.
 % Specifically, for any physical links $(n,m)$ in the integrated networks, their link weight takes the following form
	% \begin{equation}\label{eq_weight}
	% 	Weigh{t_q}(n,m) = \frac{{{D_{n,m}}I\left( {B_{n,m}^R - {B_q}} \right)}}{{\exp \left( {\rho \left( {{W_{n,q}} + {W_{m,q}}} \right)} \right)}}
	% \end{equation}
	% where $D_{n,m}$ denotes the delay of the link, $I(\cdot)$ is an indicator function, $B_{n,m}^{R}$ the remaining bandwidth resources of the link after serving the last service request, $B_q$ the bandwidth requirement of service request $q$, $\rho$ is the aggregation factor, which is a non-negative parameter controlling the aggregation ratio, and $W_{n,q}$ the function sharing factor. 
 Next, the route with the minimal total weights was selected.
Finally, the protocol selected the most suitable nodes along the chosen route based on the available computation resources and the level of function sharing required.
 
 %At the start point of SFC planning, the network status, including network topology, available physical resources, and channel conditions, are known in advance. Besides, the service requests are arrived with different deadlines. Then, each service request is served by the stand-alone ground network first, using Algorithm 2 to find the best routing path for the traffic. This is because the physical resources in aerial network are much more scarce with higher cost weights. Thus, the service requests are preferred to be served by ground network to reduce the resource cost. 
%	However, if the service request is blocked by the standalone ground network, the aerial resources are leveraged. 

	\textbf{Recurrent neural network (RNN)-based routing protocols \cite{DBLP:journals/wc/GuZX20}:} 
	The authors in \cite{DBLP:journals/wc/GuZX20} investigated the challenging issue of finding effective paths to deliver packets across a variety of network segments, including space networks, NSIN, and TN, using RNNs. 
 %[proposed] data packets in SAGINs need to be transported in a variety of network segments (i.e., space, air, and ground), which complicates the issue of route planning. 
 Particularly, the long short-term memory network (LSTM) was explored to predict routes based on historical data. However, the training of neural networks is time-consuming. To address this problem, they investigated a coded technique that could alleviate the computational burden. By employing a coded computation model and selecting suitable machine learning network models, the training process could be accelerated through coding and computation offloading \cite{DBLP:journals/tit/LeeLPPR18}.

	\textbf{Reinforcement Learning-based Routing \cite{DBLP:conf/wcsp/ShiRD21}:} 
 Considering that TN and NTN had different characteristics concerning bandwidth resources, energy resources, and transmission delay, the authors in \cite{DBLP:conf/wcsp/ShiRD21} proposed an integrated space-air-ground routing protocol based on a RL approach. 
 Specifically, they formulated a path selection optimization problem aimed at minimizing the transmission latency while considering the constraints on the remaining network energy and bandwidth. 
 A Q-learning method was explored to solve this problem, and a penalty mechanism was activated when a platform had little residual energy and/or its bandwidth utilization was too high.

	\textbf{Ant Colony Optimization (ACO)-based routing protocol \cite{DBLP:journals/ppna/ZhengZY21}:} 
 The authors in \cite{DBLP:journals/ppna/ZhengZY21} designed an ACO-based routing protocol for space-air-ground networks. The routing protocol designed a routing metric by exploring information, including link quality, E2E latency, and queue length, and consisted of three phases: 
	%based on which the heterogeneous network framework is described as a vector weighted topology. Instead of a scale, the weighted parameter of the topology is a vector with elements of signal-to-noise ratio (SNR), variation of SNR, end-to-end delay and queuing length. To meet the time-varying requirements, a Wiener predictor is adopted for obtaining the estimated channel information, the expectation of queuing delay is also acquired by modeling the process of packets waiting the transmitting buffer as a M/M/1 queuing system. 
 %Considering the Ant Colony Optimization (ACO) algorithm sharing the common decentralized feature with routing algorithm in SAGINs, a novel ACO-based cross-layer routing algorithm for SAGINs is proposed. 
	%paper includes two phases: the routing establishment phase and the routing maintenance phase. In the algorithm, nodes do not need to maintain the routing table all the time, and only start to establish routes and maintain the corresponding routing table when data transmission is needed. The pheromone of each path is stored in the routing table. The algorithm can be subdivided into three phases: 
 route search, link selection, and route maintenance. 
	
In the phase of route search, a source node will broadcast forward ants to find new routes toward a destination if a route cannot be found in its routing table. 
Once the forward ants reach the destination, the destination will dispatch backward ants to inform the source node. During the process of route searching, ants will find multiple routes.
%if a maintained routing table.	When the source node wants to communicate with the destination node,  the source node first checks whether there is a route to the destination node in the routing table.  If so, the next hop can be selected according to the routing table.  Otherwise, if there is no routing to the destination node, the forward ants are broadcast by the source node to search for new paths.  When the forward ants arrive at the destination node, the destination node sends backward ants. The backward ants return to the source node along the path that the forward ants have traveled through, and the nodes on the path update the routing information.  During the path search process, ants can search for multiple paths for route selection in the second phase. 
In the phase of link selection, nodes on established routes will randomly select an effective link to deliver packets based on a link selection probability related to SNR and remaining pheromone concentration. %, which is computed by
	%When the routes are established, nodes will randomly select	a path to transmit the data packet according to the selection	probability, and monitor and maintain all path information	in real-time.	The path selection probability is computed by 
	% \begin{equation}\label{eq_p_ij}
	% 	{p_{ij}}(t) = \frac{{{{\left[ {{\tau _{ij}}(t)} \right]}^\alpha }{{\left[ {SN{R_{ij}}(t){e^{\Delta SN{R_{ij}}(t)}}} \right]}^\beta }}}{{\sum\limits_{s \in {N_i}} {{{\left[ {{\tau _{ij}}(t)} \right]}^\alpha }{{\left[ {SN{R_{is}}(t){e^{\Delta SN{R_{is}}(t)}}} \right]}^\beta }} }}
	% \end{equation}
	% where ${\tau _{ij}}(t) = f({D_j},{T_{ij}})$ denotes the pheromone concentration left by ants across the link $(i,j)$ at time $t$, $D_j$ the queue length of $j$, $T_{ij}$ the transmission latency between nodes $i$ and $j$, $SNR_{ij} (t)$ the SNR that may be experienced by $j$ at time $t$, $\Delta SNR_{ij} (t)$ the SNR variation at time $t$,
 % $N_i$ the neighboring node set of $i$, $\alpha$ the pheromone influencing factor, and $\beta$ the heuristic information influencing factor.
During the routing maintenance phase, the source node regularly sends forward ants to the destination to assess the route quality.
If a failure link is detected during monitoring, the forward ants will promptly return, and the source node will update the routing table.

	\subsubsection{Transport layer protocols}
	The efficient design of transport layer protocols is a crucial and hot research topic in the research area of NSINs. 
 
 {\textbf{Discussion:}} Reliable transmission, congestion control, and flow control are three main responsibilities of the transport layer protocol of NSIN. 
	Particularly, a transport layer protocol in NSIN must be designed to ensure reliable transmission. 
	Certainly, as different types of applications may have diverse requirements for transmission reliability and latency, transport layer protocols should support diverse reliability levels for diverse vertical applications. 
	Congestion control is a key function of the transport layer protocol. A congested NSIN will result in a decrease in the packet delivery ratio and an increase in the E2E transmission latency. Therefore, efficient congestion control mechanisms should be designed for NSIN to avoid data congestion. 
	Besides, flow control is a crucial function of the transport layer protocol to tackle the possible overload issue at a receiver. 
	
	TCP is the most widely used transport layer protocol; however, the application of conventional TCP to NSIN faces many challenges: 1) TCP's flow control relies on a framing mechanism, and the size of the congestion control window fluctuates continuously, making it unable to adapt to the changing network topology. Additionally, accurately estimating the round-trip time (RTT) remains a challenging issue. 2) TCP will suffer from significant performance degradation mainly due to the frequent handover in NSIN; 3) besides, complex network structures, inconsistent communication protocols, and incompatible network equipment greatly degrade the performance of conventional TCP when applying in NSIN.
	
	\textbf{Existing studies:} During the past few years, researchers had attempted to improve TCP or design new transmission control protocols following different design routes. Generally, the existing transmission control protocols in NSIN can be classified into three groups: improved TCP-based, scalable traffic control-based, and resource optimization-based protocols. Table \ref{table_transport} summarizes the main ideas and main pros and cons of these protocols. 
 \begin{table*}[!t]
	\renewcommand{\arraystretch}{1.2}
	\caption{Classification, main idea, main pros and cons of transport layer protocols for NSIN}
	\label{table_transport}
	\newcommand{\tabincell}[2]{\begin{tabular}{@{}#1@{}}#2\end{tabular}}
	\centering
	\begin{tabular}{l}
%\begin{minipage}{0.18\textwidth}
			\centering
			\includegraphics[width=7.0 in]{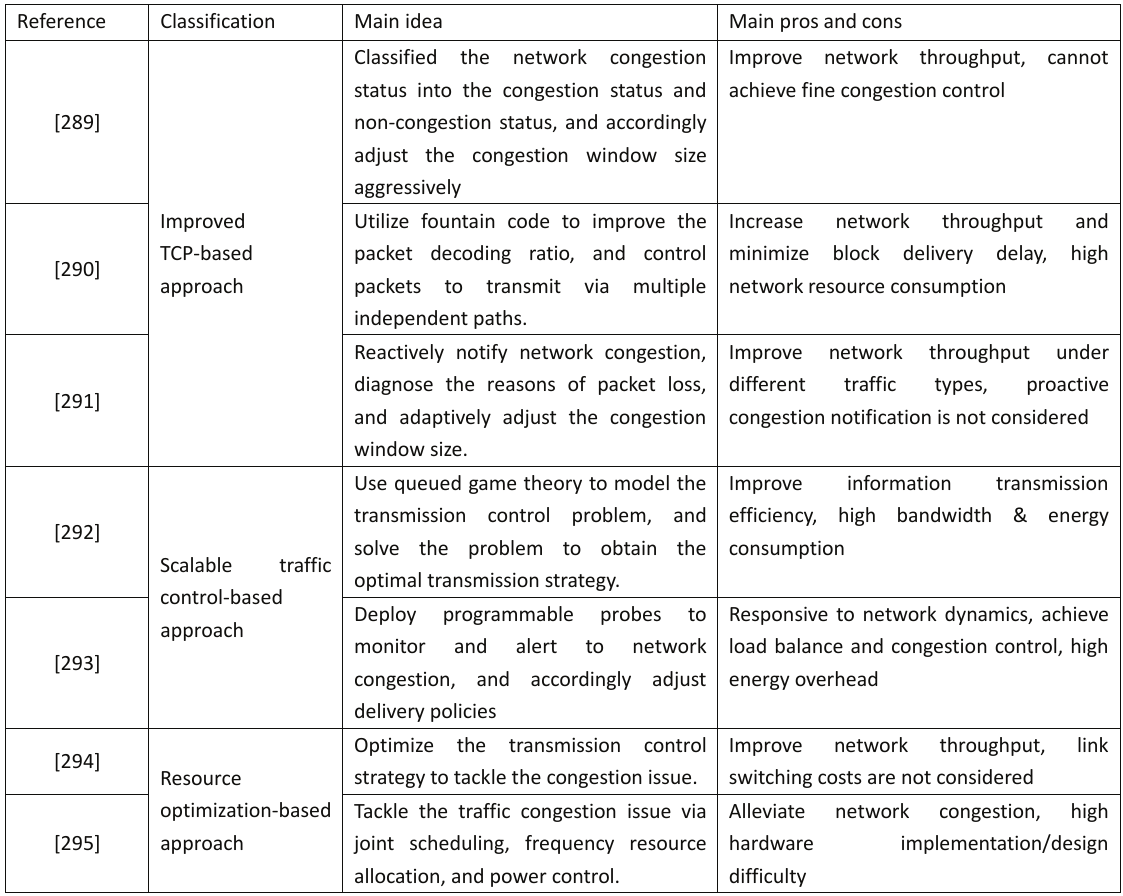}
		%\end{minipage} 
	\end{tabular}
\end{table*} 
	
	\emph{{A. Improved TCP-based protocol:}} 
  To tackle the issues of applying a TCP in NSIN, researchers have made some improvements by considering the distinctive features of NSIN, which makes it more appropriate for NSIN applications.
 
	For example, the authors in \cite{DBLP:conf/cwsn/WangSXYW12a} proposed a new TCP for satellite-HAP networks to support high-speed data transmission. Unlike the conventional TCP congestion control mechanism that utilized the standard additive increase multiplicative decrease approach to adjust the size of the congestion window, they classified the network congestion status into congestion status and non-congestion status. Based on different network congestion statuses, the authors designed a more aggressive congestion window size increment or decrease mechanism. 
The work in \cite{Li2015AeroMTPAF} developed a fountain code-based multipath transport protocol for NSIN. The protocol utilized the fountain code to improve the probability of successfully decoding packets. Considering the possibility of establishing multiple independent paths via diverse network interfaces onboard NSIN platforms, the authors proposed to deploy an independent TCP congestion control mechanism for each path. Through transmitting valid packets and ACKs via different paths, the congestion issue caused by the link capacity asymmetry was well addressed.
	The improvement of TCP for a HAP network was also investigated in \cite{weiqiang2011optimization}. In particular, the improved TCP involved three new mechanisms: 1) a relay node detecting the network congestion would send congestion messages to the source to achieve rapid congestion notification; 2) distinguished the reasons that resulted in the packet loss to avoid decreasing TCP throughput by mistake; 3) designed different congestion window adjustment schemes to adapt to the types of traffic.
	
	\emph{{B. Scalable traffic control-based protocol:}}
	By deploying programmable network devices or using the SDN technique, this type of control approach aims to improve the flexibility of traffic transmission control in complex and heterogeneous NSIN.
	
	To adapt to the transmission requirements of large-scale space-air-ground integrated networks (SAGINs), the authors in \cite{DBLP:journals/twc/GuoGXZH22} proposed a dynamic handover software-defined transmission control that supported heterogeneous users over SAGINs. Particularly, they explored the SDN technique to increase the data forwarding efficiency and implement the network platforms’ versatility by separating the control and switch functionalities of SAGINs. They also utilized a queued game theory to model the SDSAGINs' transmission control problem and determined the transmission strategy by maximizing the social welfare of the queued game theory.
	
	The work in \cite{DBLP:journals/network/PanYMZL21} designed a traffic control-empowered satellite-HAP network architecture. In this architecture, the deployment of programmable satellite-HAP network nodes allowed for the utilization of telemetry functions and dynamic probes to gather network status information, such as link utilization and network congestion, and monitor network events, including congestion information and handover errors. 
	Upon the arrival of a probe at a destination node or when the transmission hops of the probe reached a limit, feedback signals carrying local congestion information or other network statuses were generated and transmitted back to the source node. These feedback signals were then utilized to adjust network policies, such as sending rates or sub-network switching.
	
	\emph{{C. Resource optimization-based protocol:}}
	The primary concept behind this type of control protocol is to address congestion alleviation by formulating and resolving a network resource optimization problem.

	The work in \cite{DBLP:journals/tvt/CaoYYH21} studied the congestion control issue in a HAP-reserved SAGIN. Initially, the authors formulated an optimization problem for transmission control with the objective of maximizing the overall throughput of ground users while taking into account congestion control constraints. These constraints included the transmission probability of the ground-air-space link for a user and the number of users selecting the ground-air-space links. Subsequently, they presented a transmission control strategy to assist ground users in determining the probability of switching between a ground-air-space link and a ground-space link. 
	
	The work in \cite{zhang2018matching} investigated the applicability of NOMA in supporting the NSIN to alleviate transmission congestion. In this work, a HAP acting as a centralized decision maker was utilized to perform semi-persistent scheduling and non-orthogonally allocate time-frequency resources, and massive UAVs were deployed to execute power control in a distributed manner. Through solving the joint scheduling, frequency resource allocation, and power control problem, the congestion issue of data traffic in NSIN could be alleviated.

	%\subsection{Cross layer}
	%What's the meaning of cross layer protocol design. 
	
	\subsection{Summary and Lessons Learned}
{\textbf{Summary:}}
	%Beamforming is an important and attractive PHY layer research topic in the research field of NSINs, which can significantly improve the coverage efficiency of NSINs. 
In this section, we distinguish two important beamforming techniques, i.e., active beamforming and passive beamforming. Existing beamforming techniques in recent ten years for NSINs have been comprehensively surveyed. 
RIS/IRS, a typical and emerging passive beamforming technique, has attracted widespread interest in both academia and industry. We then discuss a categorization of recent research on ARIS.
Three types of MAC protocols for NSINs, i.e., fixed allocation protocol, random contention protocol, and reservation protocol, are surveyed. As HAP networks have quasi-stationary network topologies, fixed allocation protocols (e.g., TDMA, CDMA, FDMA, SDMA, and NOMA) for HAP networks have been extensively studied. Considering the highly dynamic network topologies, researchers have designed random contention and reservation protocols for UAV networks. 
Routing protocol design is a popular topic in the research area of UAV networks, with various {delivery} schemes {being} proposed to ensure efficient packet delivery. 
Nonetheless, existing UAV routing protocols may not be fully applicable to NSINs. We explain some reasons in this section and present some routing protocols designed separately for NSINs. 
Besides, we have comprehensively surveyed the endeavors in the design of transport layer protocols for NSINs over the past ten years. 

{\textbf{Lessons learned:}}
{
%The active beamforming technique includes three types of beamforming architectures. 
It is now widely accepted that the selection of beamforming architecture should take into account various factors such as applications, hardware and software complexities, techniques, and so on. Nevertheless, things are changing rapidly in this area. With advances in data conversion techniques and an increased level of on-chip integration, the era of all-digital beamforming is on the horizon.
In addition to beamforming and multi-connectivity, many other key techniques for implementing efficient transmission via NSINs are worthy of in-depth study, for example, waveform design and modulation and coding schemes. 
The waveform structure serves as the foundation of wireless communications. However, there are no standards that describe a specific waveform structure for NSINs. The issue of designing the waveform structure for NSINs has also not been investigated so far \cite{DBLP:journals/comsur/KurtKAIDAYY21}.
High mobility is a significant characteristic of NSINs. However, in high-mobility scenarios, the performance of some widely adopted modulation schemes (e.g., OFDM) may degrade \cite{DBLP:journals/wc/WeiYLYBHH21}. 
Orthogonal time-frequency space (OTFS) modulation may be a viable modulation scheme for NSINs. OTFS modulates information in the delay-Doppler domain, providing strong delay-resilience and Doppler-resilience \cite{DBLP:journals/wc/WeiYLYBHH21}.}
%Thus, OTFS achieves better performance than OFDM in high mobility scenarios [10]. It is worth noting that the development of OTFS systems is still in its infancy. Whether OTFS can ensure other KPIs of TI applications remains unclear and deserves further study.
{Besides, NSINs have some specific characteristics, such as high mobility, frequently-changed topology, mechanical and aerodynamic constraints, and a harsh communication environment. These characteristics will result in the performance degradation of some widely utilized communication protocols when {applied to} NSINs. 
As a result, the design of NSIN communication protocols should holistically consider the characteristics (e.g., heterogeneity, integration, and platform constraints) of NSINs to achieve an integrated design.}
%In summary, the design of NSINs protocols should comprehensively consider the characteristics (i.e., heterogeneity, ) of NSINs to achieve an integrated design. 

	\section{Open issues and promising directions}
 In this section, we present open issues and promising directions of NSINs deserved for future study and discuss the corresponding challenges.
	\subsection{Generalized Channel Model} 
	Generalization capability is an important performance evaluation metric for channel models. 
 It is expected that the standardized NSIN channel models can produce a generalized channel model framework with varying parameter sets tailored to different scenarios. 
	However, due to the heterogeneity of scenarios and different scales over the wavelengths, it is highly challenging to develop a general channel model for NSINs. 
	The general channel model is likely to have a complex, non-linear relationship between the channel model parameters and the channel characteristics. These parameters encompass antenna patterns, large-scale model parameters such as various types of path loss and shadowing, as well as small-scale model parameters including multipath amplitudes and delays and Doppler shifts. The channel characteristics comprise the time autocorrelation function, spatial cross-correlation function, root mean square (RMS) delay spread, RMS angle spread, Doppler power spectral density, and stationary interval.
	Then, how to design a general channel model capturing the model parameters and characteristics requires careful investigations.  
{A viable method is to design a general channel model in two stages, including model initialization and time evolution \cite{ Chang2023AG3}. In the 1$^{\rm st}$ stage, generate scenario-related scattering cluster parameters and large-scale fading parameters. In the 2$^{\rm nd}$ stage, generate or update scenario-related small-scale fading parameters periodically.}

	\subsection{{Quantum Communication-Empowered NSINs}}
	With the increasing requirement for providing services via NSINs, the feasibility of the traditional centralized application service architecture of NSINs is threatened. In a traditional centralized NSIN architecture, a single application service data center controlled by a subject or a third-party certification agency manages the credibility, security, and integrity of all network entities, network behaviors, and network data. As a result, once the data center or certification agency was attacked, NSINs would become unbelievable. 
 {Besides, due to the long-distance transmission in NSINs, malicious nodes have higher chances of obtaining users’ private information.} 
 
  {To address these key issues, quantum communication-empowered NSINs should be investigated. Quantum communication harnesses the distinct attributes of quantum mechanics like superposition and entanglement to enable secure transmission of information, ensuring inherent safeguards against tampering and interception \cite{Xu2022WhenQI}. In particular, the quantum cryptography technique can be utilized to protect the data center or certification agency from being attacked. A quantum teleportation, which is a quantum communication protocol for quantum information transfer, can be leveraged to establish secure transmissions in NSINs.}
 % To address this key issue, blockchain-enabled NSINs should be investigated.
%	As a distributed architecture, blockchain ensures the credibility of data and prevents tampering. However, the process of achieving consensus requires exchanging transaction data and authentication messages, which consumes a significant amount of network bandwidth and computing resources. {The amount of required resources} becomes even greater when deploying NSINs with a larger scale. 
% Therefore, it is important to research and develop approaches to orchestrate and manage resources effectively in order to provide sufficient storage, bandwidth, and computing resources for deploying blockchain-enable NSINs.
	
	\subsection{Exploration of All Frequency Spectra}
	To fully unlock the performance benefits of NSINs, it is essential to explore all frequency spectra, including sub-6 GHz, mmWave, Terahertz (THz), and optical frequency bands. 
 
 While sub-6 GHz bands have been the primary working frequency bands in TNs and are indispensable in NSINs, mmWave communications show promise in providing ultra-high capacity.
	Several open issues remain to be tackled before applying mmWave communications {to} NSINs. 
	Firstly, the primary challenge posed by current mmWave systems is their design for small cells with radii of a few hundred meters or less. However, the ground coverage radius can {be extended} to tens of kilometers or more in NSINs. This calls for a revolutionary redesign of mmWave systems to cater to the requirements of NSINs. 
 Secondly, a flexible channel sounder design becomes imperative for {mmWave-enabled NSINs}. 
	Channel sounding data is essential for understanding near-space mmWave channel characteristics. However, conducting channel measurement campaigns for NSINs proves to be costly and challenging.
	Furthermore, diverse NSINs {deployment} scenarios need to utilize different channel sounders \cite{DBLP:journals/chinaf/YouWHGZWHZJWZSW21}. Therefore, flexible channel sounders are {essential} to support sufficiently high measurement speeds and multiple measurement scenarios \cite{DBLP:journals/chinaf/YouWHGZWHZJWZSW21}. 
 
 In addition, THz communications are envisioned to achieve high transmission rates ranging from hundreds of Gbps to several Tbps. The practical implementation of THz communications in NSINs is attractive and challenging. 
	For instance, the inhomogeneous atmospheric propagation medium in the near-space environment results in the undesirable refraction effect, making it difficult to achieve THz beam alignment for ultra-long distance THz propagation links. 
	In addition, significant research effort is needed to develop high-performance THz devices and packaging techniques to make THz communications commercially viable for NSINs.
 
	The optical communication system offers sufficient unlicensed bandwidth for robust, fast, safe, and efficient transmission, with optical bandwidth being three orders of magnitude larger than the spectrum resources available in the RF bands \cite{DBLP:journals/chinaf/YouWHGZWHZJWZSW21}.
	Owing to the unique deployment environment, HAP networks adopting optical communication techniques will be immune to atmospheric turbulence and absorption effects. 
	However, the limited transmit power under sky radiance and background shot noise significantly affects the practical application of optical communication techniques in NSINs. Advanced transceiver design should be considered to tackle the above issue. 	
   Moreover, in order to address the challenges of implementing optical communications in a near-space environment, it is crucial to conduct comprehensive studies on {some} issues like the analysis of beam wandering and pointing errors and {how to mitigate} their adverse effects.

 \subsection{{Generative AI-Empowered Large-Scale NSINs}}
	We know that NSINs have a multi-layered network architecture consisting of HAP networks and UAV networks. The definition of NSINs does not constrain the number of platforms in NSINs. Unlike NSINs, there are a large number of aerial platforms in large-scale NSINs. The utilization of large-scale NSINs will significantly improve the efficiency of accomplishing various types of vertical applications. 
{Although conventional AI methods have been widely applied to solve many key issues (e.g., channel modeling and network management) in NSINs, it becomes highly difficult to leverage them to learn and make decisions rapidly for large-scale NSINs in complex and dynamic environments.} 

{Studying the generative AI-empowered large-scale NSINs is just at that time. 
Generative AI, with its capabilities in causal reasoning and a deep understanding of data patterns and structures, offers great potential for solving many communication problems \cite{DBLP:journals/comsur/YangDLLXNCSM23}.} 

Take the network reconfiguration of large-scale NSINs as an example. 
%For instance, generative AI approaches can be explored to achieve effective network reconfiguration for large-scale {NSINs}. 
The network topology of large-scale NSINs is highly dynamic, with many network nodes joining (e.g., new nodes or recharged old nodes) and leaving (e.g., due to malfunction or energy exhaustion) the network continuously. To maintain network connectivity, the exchange of a tremendous amount of data among large-scale network nodes and the process of searching for the optimal {route} may be triggered continuously. Further, the impact of increasing or decreasing the number of different types of network nodes will be quite diverse. The complexity of network reconfiguration will increase when deploying HAPs with different functionalities and mobility features. 
Overall, the network topology and data distribution inevitably vary with time and space in large-scale NSINs. We cannot utilize already-trained network reconfiguration models to reconfigure networks for continuously changing scenarios. 
{Generative AI approaches can capture the similarity between existing knowledge (e.g., resource allocation and topology reconfiguration schemes) and new knowledge. Through transferring the similarity, generative AI approaches can help to improve the learning and decision-making efficiency in network reconfiguration for NSINs.} 

Take the situation sensing and sharing as another example. 
	Given the deployment of HAPs at altitudes ranging from 17 km to 22 km, NSINs need to produce situations for an expansive and complex airspace. 
%{Nevertheless, it is highly challenging to investigate} and {produce} effective situations in such a complex airspace.
In such an airspace, a large amount of data (especially videos, images, and radar data) will be sensed, stored, and analyzed, which poses a huge pressure on improving the utilization of the limited network resources. The scenes in such a vast airspace are susceptible to sudden changes, which creates significant challenges for accurate sensing. Meanwhile, the issue of the explosion of network data, network scale, and scenarios in large-scale NSINs places immense strain on the transmission capability of UAVs and HAPs, making it challenging for them to share sensed situations. 

To this end, a natural question arises: how do we acquire an accurate network situation without directly delivering and sharing the massive amount of raw data in large-scale NSINs?
{A viable solution is to integrate generative AI into large-scale NSINs. 
Firstly, generative AI can generate a variety of new and meaningful content, including text, images, audio, and 3-D models. Secondly, generative AI can utilize deep latent space generative models to significantly reduce the dimensionality of multimodal data. Sharing compressed data will alleviate the strain of data transmission on the limited communication resources of large-scale NSINs. Receivers can then recover the original data through decoders and mine situational information from the recovered data.}

	\section{Conclusion}
	This article highlighted the characteristics and potential of near-space information networks (NSIN). With the potential to achieve sensing, seamless communications, and ubiquitous computing in a wide area, NSIN had emerged as an essential component of future wireless communication networks. However, the specific characteristics of NSIN make the existing technologies designed for terrestrial networks (TN), unmanned aerial vehicle (UAV) networks, and satellite networks not completely suitable for NSIN. This article comprehensively surveyed the latest advancements in the channel modeling, networking, and transmission aspects of NSIN. A wide range of research topics in these three aspects were discussed from a forward-looking, comparative, and technological evolutionary perspective. The impact of unstable airborne platforms on the phase delays of diverse types of onboard antenna arrays was discussed. The recent advancements in high-altitude platform (HAP) channel modeling were provided, and the differences between HAP channel modeling and UAV channel modeling were revealed. Along with the integration of promising passive beamforming, softwarized, and machine learning (ML) or artificial intelligence (AI) techniques, a comprehensive review of the networking strategies of NSIN in network deployment, handoff management, and network management aspects was provided. Besides, the promising technologies and communication protocols of the physical layer, medium access control (MAC) layer, network layer, and transport layer of NSIN for achieving efficient transmission over NSIN were overviewed. 
Overall, the goal of this article is to characterize NSIN, present prospective use-cases of NSIN in the near future and beyond, and disseminate the significant contributions in the crucial research area of NSIN. By presenting a comprehensive overview, we expect that more attention from the research community could be paid to this emerging and promising research field.

	%\appendix

	\ifCLASSOPTIONcaptionsoff
	\newpage
	\fi

	
	
	
	\bibliographystyle{IEEEtran}
	\bibliography{Study_NSIN}

\end{document}